\documentclass[modern]{aastex61}

\usepackage{natbib} 
\usepackage{gensymb} 

\definecolor{Amalthea}{HTML}{FF00FF} 
\definecolor{Galilians}{HTML}{0000FF} 

\definecolor{Themisto}{HTML}{00FF00} 
\definecolor{Himalia}{HTML}{008000} 
\definecolor{Carpo}{HTML}{00FF7F} 

\definecolor{Carme}{HTML}{8B4513} 
\definecolor{Ananke}{HTML}{FFA500} 
\definecolor{Pasiphae}{HTML}{FF0000} 

\definecolor{MainRing}{HTML}{800080} 

\definecolor{IcySats}{HTML}{00008B} 
\definecolor{Trojans}{HTML}{4682B4} 
\definecolor{Alkanoids}{HTML}{00BFFF} 

\definecolor{Inuit}{HTML}{006400} 
\definecolor{Gallic}{HTML}{FF8C00} 
\definecolor{Norse}{HTML}{8B0000} 

\received{June 8th, 2017}
\revised{April 9th, 2018}
\accepted{April 10th, 2018}
\published{MM DD, 2018}
\submitjournal{The Astrophysical Journal}

\begin{document}


\title{Cladistical analysis of the Jovian and Saturnian satellite systems}
\shorttitle{Cladistics of Satellite Systems}

\author{Timothy. R. Holt}
\affiliation{University of Southern Queensland, Computational Engineering and Science Research Centre, Queensland, Australia}
\affiliation{Swinburne University of Technology, Center for Astrophysics and Supercomputing, Victoria, Australia.}

\author{Adrian. J. Brown}
\affiliation{Plancius Research LLC, Severna Park, MD. USA.}

\author{David Nesvorn{\'y}}
\affiliation{Southwest Research Institute, Department of Space Studies, Boulder, CO. USA.}

\author{Jonathan Horner}
\affiliation{University of Southern Queensland, Computational Engineering and Science Research Centre, Queensland, Australia}

\author{Brad Carter}
\affiliation{University of Southern Queensland, Computational Engineering and Science Research Centre, Queensland, Australia}

\correspondingauthor{Timothy. R. Holt}
\email{timothy.holt@usq.edu.au}

\begin{abstract}

Jupiter and Saturn each have complex systems of satellites and rings. These satellites can be classified into dynamical groups, \replaced{indicative of}{implying}  similar formation scenarios. Recently, a larger number of additional irregular satellites have been discovered around both gas giants that have yet to be classified. 
The aim of this paper is to examine the relationships between the satellites and rings of the gas giants, using an analytical technique called \textit{cladistics}. 
Cladistics is traditionally used to examine relationships between living organisms, the `tree of life'. In this work, we perform the first cladistical study of objects in a planetary science context. Our method uses the orbital, physical and compositional characteristics of satellites to classify the objects in the Jovian and Saturnian systems.
We find that the major relationships between the satellites in the two systems, such as families, as presented in previous studies, are broadly preserved. In addition, based on our analysis of the Jovian system, we identify a new retrograde irregular family, the Iocaste family, and suggest that the Phoebe family of the Saturnian system can be further divided into two subfamilies. We also propose that the Saturnian irregular families be renamed, to be consistent with the convention used in Jovian families. Using cladistics, we are also able to assign the new unclassified irregular satellites into families. 
Taken together, the results of this study demonstrate the potential use of the cladistical technique in the investigation of relationships between orbital bodies.

\end{abstract}



\section{Introduction}

The two gas giants of the \replaced{solar}{Solar} system, Jupiter and Saturn, are host to a large number of satellites and rings. The satellites of both planets follow a similar progression pattern. The inner region of each system consists of small icy satellites, with an accompanying ring system \citep{Thomas1998InnerSatJup, Throop2004JupiterRings, Porco2005CassiniRingSat,Thomas2013InnerSatSatu}. Further out, there are larger icy/silicate satellites \citep{Thomas2010SaturnSatsCassiniProps, Deienno2014OrbitGalileanSat}. In the outer system, both planets have a series of irregular satellites, small satellites with high eccentricities and inclinations \citep{Nesvorny2003IrrSatEvol, Sheppard2003IrrSatNature, Jewitt2007IrregularSats}. It is thought that these satellites were captured from other populations of small \replaced{solar}{Solar} system bodies \citep{Colombo1971JupSatsForm, Heppenheimer1977Capture, Pollack1979SatGasDrag, Sheppard2003IrrSatNature, Nesvorny2004IrrSatFamilyOrigin, Johnson2005PhoebeKuiper, Nesvorny2007IrrSatCap, Nesvorny2014IrrCapture}. This is in contrast to the inner satellites, which are thought to have accreted in a circumplanetary disk \citep[e.g.][]{Canup2002GalSatAcc, Canup2010SaturnSatOrigin}. Such a formation mechanism is thought to resemble the accretion of planets in a protoplanetary disk around a young star \citep{Lissauer1987PlanetAccretion}, a conclusion that is supported by the recent discovery of the TRAPPIST-1 planetary system \citep{Gillon2016Trapist1}. That system features at least seven Earth-mass planets orbiting a very low mass star. The star itself, TRAPPIST-1, is within two orders of magnitude more massive than Jupiter, and similar in size. The seven planets span an area comparable to that of Jupiter’s regular satellite system. Studying and understanding the gas giant systems in our own \replaced{solar}{Solar} system, can therefore provide context for future exploration of low-mass exoplanetary systems. 

\subsection{The Jovian System}
Historically, \citet{Galileo1610SidereusNuncius} discovered the first satellites in the Jovian system, the large Galileans, Io, Europa, Ganymede and Callisto. Our knowledge of these satellites has increased greatly, as a result of both improved ground-based instrumentation \citep[e.g.][]{Sparks2016EuropaHST,Vasundhara2017GalileansGround} and spacecraft visitations \citep[e.g.][]{Smith1979Jupiter, Grundy2007NewHorizonsJupiterSats, Greenberg2010IcyJovian}\deleted{have given us a more detailed understanding of these objects}. 

Amalthea, one of the inner set of Jovian satellites, was discovered by \citet{Barnard1892Amalthea}. A few years later, the first two small irregular satellites, Himalia \citep{Perrine1905Himalia} and Elara \citep{Perrine1905Elara}, were discovered in inclined, prograde orbits. The discovery of Pasiphae three years later by \citet{Melotte1908Pasiphae} is significant as this was only the second satellite in the Solar system to be found on a retrograde orbit, and the first such object found in the Jovian system. Several other irregular satellites were discovered in the first half of the 20th century, Sinope \citep{Nicholson1914Sinope}, Lysithea \citep{Nicholson1938LysitheaCarme}, Carme \citep{Nicholson1938LysitheaCarme} and Ananke \citep{Nicholson1951Ananke}. Leda, another small prograde irregular, was discovered 20 years later by \citet{Kowal1975Leda}. Themisto, the first Jovian satellite smaller than 10km to be discovered, was found that same year \citep{Kowal1975Themisto} and subsequently lost. Themisto was rediscovered by \citet{Sheppard2000ThemistoReDis} nearly 20 years later. The Voyager visitations of Jupiter discovered the remaining three inner satellites, Metis \citep{Synnott1981MetisDiscov}, Adrastea \citep{Jewitt1979AdrasteaDiscov} and Thebe \citep{Synnott1980ThebeDiscov}, along with a ring system \citep{Smith1979Jupiter}. These three satellites, Amalthea and the ring system, would be imaged again by the Galileo \citep{OckertBell1999JupiterRing} and Cassini \citep{Porco2005CassiniRingSat} spacecraft during their missions. 

The irregular Jovian satellites orbit the planet with semi-major axes an order of magnitude greater than the Galilean moons, and have large eccentricities and inclinations. In the early years of the 21st century, extensive surveys were carried out to search for the Jovian irregular satellites \citep{Scotti2000Callirrhoe, Sheppard2001SatsJupDiscovery,
Sheppard2002JupSatDisc, Gladman2003JupIAU1, Gladman2003JupIAU2,
Sheppard2003JupIAU1, Sheppard2003JupIAU2, Sheppard2003JupIAU3,
Sheppard2003IrrSatNature,
Sheppard2004JupIAU1,Sheppard2004JupIAU2, Beauge2007IrregSatRsonance,
Jacobson2011NewJupSats, Sheppard2012JupIAU}. These surveys increased the number of known Jovian satellites from 14 after Voyager, to the 67 known today.  The inner five irregular satellites, Leda, Himalia, Lystea, Elara and Dia, have prograde orbits and have previously been classified into the Himalia group \citep{Nesvorny2003IrrSatEvol, Sheppard2003IrrSatNature}. Themisto and Carpo were proposed as single members of their own  groups by \citet{Sheppard2003IrrSatNature}. The remainder of the irregular satellites have retrograde orbits. Based on similarities in semi-major axis, inclination and eccentricity, these satellites have been grouped into families by \citet{Sheppard2003IrrSatNature} and \citet{Nesvorny2003IrrSatEvol}. These dynamical families are typified by their largest member, Himalia representing the inner prograde satellites, with the retrograde ones being broken down into the Ananke, Pasiphae and Carme families. Recently, several additional small irregular satellites have been discovered \citep{Jacobson2011NewJupSats, Sheppard2012JupIAU} which are yet to be named or classified. With the discovery of new satellites \citep{Scotti2000Callirrhoe, Sheppard2001SatsJupDiscovery, Beauge2007IrregSatRsonance,
Jacobson2011NewJupSats, Sheppard2012JupIAU} and additional information from the Cassini spacecraft \citep{Porco2005CassiniRingSat}, a revisitation of the classification of the Jovian irregular satellites \citep{Nesvorny2003IrrSatEvol, Sheppard2003IrrSatNature, Jewitt2007IrregularSats} is warranted. 

\subsection{The Saturnian System}
The Saturnian system is broadly similar to that of Jupiter, but exhibits greater complexity. One of the most striking features, visible to even the most modest telescope, is Saturn's ring system. First observed by Galileo in 1610, it was \citet{Huygens1659systema} that observed that the objects surrounding Saturn were in fact rings. The rings themselves are composed of individual particles, from micrometer to meter size \citep{Zebker1985SaturnRingParticle}.  Embedded within several of the main rings are a series of small moonlets \citep{Tiscareno2006SaturnAringMoonlets} and several shepherd satellites \citep{Showalter1991Pan, Porco2007SaturnSmallSats, Cuzzi2014FringPromethius}. The co-orbitals Janus and Epimetheus \citep{Yoder1983SaturnCoorobiting, Yoder1989JanusEpiMassOrbit, Nicholson1992CoorbitalSaturn, Treffenstdt2015JanusEpiFormation, ElMoutamid2016JansSwapAring}, and their associated faint ring system \citep{Winter2016JanusEpiRing} are unique to the Saturn system. Just beyond the Janus/Epimetheus orbit, there is a diffuse G-ring, the source of which is the satellite Aegaeon \citep{Hedman2007Gring}.

\citet{Huygens1659systema} also discovered Saturn's largest satellite, Titan. Earth-based observations highlighted the methane based atmosphere of Titan \citep{Kuiper1944TitanAtmos, Karkoschka1994TitanESO}, with further characterization by the Cassini spacecraft \citep{Niemann2005TitanAtmos} and Huygens lander \citep{Lebreton2005HuygensTiten}. The bulk composition of Titan is analogous to that of the other icy satellites, with an icy shell, subsurface water ocean and silicate core \citep{Hemingway2013TitanIceshell}. There are seven other mid-sized icy satellites, with semi-major axes on a similar order of magnitude to \added{that of} Titan. The five largest, Mimas, Enceladus, Tethys, Dione and Rhea are large enough to be in hydrostatic equilibrium. All of the mid-sized satellites are thought to be predominantly composed of water ice, with some contribution from silicate rock, and may contain subsurface liquid oceans \citep{Matson2009SaturnSat, Filacchione2012VIMS3}. Those satellites closer to Saturn than Titan, Mimas, Enceladus, Tethys, Dione and Rhea, are embedded in the E-ring \citep{Feibelman1967SatEring, Baum1981SatEring, Hillier2007EringComp, Hedman2012EringStruc}. The Cassini mission identified the source of this ring as the southern cryo-plumes of Enceladus \citep{Sphan2006EnceladusEring}. 

In addition to the larger icy satellites, there are four small Trojan satellites \citep{Porco2005CassiniRingSat}, situated at the leading and trailing Lagrange points, 60\degree \ ahead or behind the parent satellites in their orbit. Tethys has Telesto and Calypso as Trojan satellites, while Helene and Polydeuces are Trojan satellites of Dione. So far, these Trojan satellites are unique to the Saturnian system. Between the orbits of Mimas and Enceladus, there are the Alkyonides, Methone, Anthe and Pallene, recently discovered by the Cassini spacecraft \citep{Porco2005CassiniRingSat}. Each of the Alkyonides have their own faint ring arcs \citep{Hedman2009SatRingArcs} comprised of similar material to the satellite. Dynamical modeling by \citet{Sun2017MethoneDust} supports the theory of \citet{Hedman2009SatRingArcs}, that the parent satellite is the source of the rings.

In the outer Saturnian system there are a large number of smaller irregular satellites, with 38 known to date. The first of these irregular satellites to be discovered was Phoebe, which was the first planetary satellite to be discovered photographically \citep{Pickering1899Phoebe}. Phoebe was also the first satellite to be discovered moving on a retrograde orbit \citep{Pickering1905Phoebe, Ross1905Phoebe}. Phoebe is the best studied irregular satellite and the only one for which in-situ observations have been obtained \citep{Clark2005Phoebe}. Recently, a large outer ring associated with Phoebe and the other irregular satellites has been discovered \citep{Verbiscer2009SatLarRing}. It has been suggested that Phoebe may have originated in the Edgeworth-Kuiper Belt and captured into orbit around Saturn \citep{Johnson2005PhoebeKuiper}. The other Saturnian irregular satellites were discovered in extensive surveys during the early 21st century \citep{Gladman200112Sat, Sheppard2003IAUJupSat, Jewitt2005IAUCSat,
Sheppard2006SatIAUC, Sheppard2007SatIAUC}. \replaced{With}{Due to} the small size of the majority of these satellites, only their orbital information is available. There are nine prograde and 29 retrograde outer satellites, of which attempts have been made to place into families based on dynamical \citep{Gladman200112Sat, Jewitt2007IrregularSats, Turrini2008IrregularSatsSaturn} and photometric \citep{Grav2003IrregSatPhoto, Grav2007IrregSatCol} information. In the traditional naming convention \citep{Grav2003IrregSatPhoto}, the Inuit family, Ijiraq, Kiviuq, Paaliaq, Siarnaq and Tarqeq, are small prograde satellites, whose inclination is between 45\degree and 50\degree . The Gallic family, Albiorix, Bebhionn, Erriapus and Tarvos, is a similar, prograde group, but with inclinations between 35\degree and 40\degree . The retrograde satellites are all grouped into the Norse family, including Phoebe. There is a possibility that the Norse family could be further split into subfamilies, based on photometric studies \citep{Grav2003IrregSatPhoto, Grav2007IrregSatCol}. The convention of using names from respective mythologies for the satellite clusters \citep{Jewitt2007IrregularSats}, has become the default standard for the irregular satellite families of Saturn. 

\subsection{Formation Theories}

The purpose of taxonomy and classification, beyond simple grouping, is to investigate the origin of objects. The origin of the irregular satellites is a major topic of ongoing study \citep{Nesvorny2012JumpingJupiter, Nesvorny2014IrrCapture}. Here we present an overview for context. There are three main theories in the formation of the Jovian satellites: formation via disk accretion \citep{Canup2002GalSatAcc}; via nebula drag \citep{Pollack1979SatGasDrag}; or via dynamic capture \citep{Nesvorny2003IrrSatEvol, Nesvorny2007IrrSatCap}. The satellites that are captured either by nebula drag or through dynamical means, are thought to be from \replaced{solar}{Solar} system debris, such as asteroids and comets. 

The disk accretion theory has generally been accepted as the mechanism for the formation of the inner prograde satellites of Jupiter \citep{Canup2002GalSatAcc}. The satellites form from dust surrounding proto-Jupiter in a process analogous to the formation of planets around a star \citep{Lissauer1987PlanetAccretion}. This surrounding disk would have lain in the equatorial plane of Jupiter, with material being accreted to the planet itself through the disk. This would explain both the prograde, coplanar orbits of the regular satellites and their near circular orbits.

The second theory requires satellites to be captured in the original Jovian nebula \citep{Pollack1979SatGasDrag, Cuk2004HimaliaGasDrag}. Before it coalesced into a planet, Jupiter is proposed to have had a greater radius, and lower density than now. There was a `nebula' surrounding this proto-Jupiter. As other pieces of \replaced{solar}{Solar} system debris crossed into the Hill sphere of this nebula, they would be slowed down by friction and be captured as a satellite. Related to this is the concept of a pull down mechanism \citep{Heppenheimer1977Capture}. As a gas giant increases in mass from accretion \citep{Pollack1996GiantPlanetAccretion}, the hills sphere increases. As a subsequent effect, small \replaced{solar}{Solar} system bodies can possibly be captured as irregular satellites.

Dynamical capture can explain the retrograde orbits of the Jovian satellites \citep{Nesvorny2003IrrSatEvol}. The Hill sphere of a planet dictates the limit of its gravitational influence over other bodies. The theory \citep{Nesvorny2003IrrSatEvol, Nesvorny2007IrrSatCap} states that is it impossible for a satellite to be captured in a three body system (Sun, planet and satellite). The Nice model of the \replaced{solar}{Solar} system \citep{Tsiganis2005NICEplanets, Nesvorny2007IrrSatCap, Nesvorny2014IrrCapture} has a fourth body interaction placing the satellite into a stable orbit inside the Hill sphere of the gas giant. Recently the Nice model was updated to include a fifth giant planet \citep{Nesvorny2012JumpingJupiter}. This updated theory has the new planet interacting with Jupiter and allowing for the capture of the satellites, before the fifth giant planet is ejected from the \replaced{solar}{Solar} system. Collisions between objects could also play a part in the dynamical capture of the irregular satellites \citep{Colombo1971JupSatsForm}.

The formation of the Saturnian satellite system is thought to be similarly complex. The inner satellites are possibly formed from accretion within the ring system \citep{Charnoz2010SaturnMooletsfromMainRings} or from the breakup of a large, lost satellite \citep{Canup2010SaturnSatOrigin}. Modeling of the Saturnian system by \citet{Salmon2017SaturnMidAccretion} has \replaced{indicated}{shown} that the mid-sized satellites could have formed from a large ice-dominated ring, with contamination of \replaced{asteroids}{cometary material} during the Late Heavy Bombardment, delivering the requisite silicate rock. Being the largest satellite in the Saturnian system, Titan is thought to have formed from accretion of proto-satellites \citep{Asphaug2013SatMerger}. The Saturnian irregular satellites are predicted to be captured objects \citep{Jewitt2007IrregularSats}, though their origins are still in dispute. Collisions are thought to have played a part in the capture of the irregular satellites of Saturn \citep{Turrini2009IrregularSatsSaturn}. The cratering data provided by the Cassini spacecraft \citep{Giese2006PhoebeTopo} supports this hypothesis.

\subsection{This Project}
With the discovery of several new irregular satellites \citep{Scotti2000Callirrhoe, Gladman200112Sat, Sheppard2001SatsJupDiscovery,
Sheppard2002JupSatDisc, Gladman2003JupIAU1,Gladman2003JupIAU2,
Sheppard2003IAUJupSat, Sheppard2003JupIAU1,Sheppard2003JupIAU2,Sheppard2003JupIAU3,
Sheppard2003IrrSatNature,
Sheppard2004JupIAU1,Sheppard2004JupIAU2, Jewitt2005IAUCSat, Sheppard2006SatIAUC,Sheppard2007SatIAUC, 
Jacobson2011NewJupSats, Sheppard2012JupIAU}, along with the detailed examination of the Jovian and Saturnian system by the Cassini spacecraft \citep{Brown2003CassiniJupiter, 
Porco2005CassiniRingSat, Cooper2006CassiniAmaltheaThebe, Giese2006PhoebeTopo, Porco2006EnceladusPlume, Sphan2006EnceladusEring, Filacchione2007VIMS1, Nicholson2008VIMSRings, Matson2009SaturnSat, Buratti2010SatInnerSat, Filacchione2010VIMS2, Thomas2010SaturnSatsCassiniProps, 
Clark2012VIMSIapetus, Filacchione2012VIMS3, Spitale2012s2009s1, Tosi2010IapetusDark, Hirtzig2013VIMSTitan, Brown2014Rayleigh, Filacchione2014VIMSrings, Filacchione2016VIMS4}, there is an opportunity to revisit the classification of the satellite systems of the gas giants. We apply a technique called \textit{cladistics} to characteristics of the Jovian and Saturnian satellites, in order to examine the relationships between objects in the systems. The purpose of this is two fold. First, due to their \replaced{well established}{well-established} classification systems, the Jovian and Saturnian satellite systems offer an opportunity to test the cladistical technique in a planetary science context. This project is an extension of \citet{Holt2016JovSatCald} and together they form the first use of cladistics for planetary bodies. The second aim of the project is to classify recently discovered satellites, as well as providing context for future work.

In Section \ref{Methods}, we introduce the cladistical technique, and how it is used in this paper. The resulting taxonomic trees for the Jovian and Saturnian systems, along with their implications for the taxonomy of the satellites, are presented in Sections \ref{JupiterTax} and \ref{SaturnTax} respectively. Section \ref{Discussion} discusses the implications of cladistics in a planetary science context, along with some remarks on origins of the gas giant satellites and possible future work.

\section{Methods}
\label{Methods}
In this section, we present an overview of the cladistical method and how it is applied to the Jovian and Saturnian satellite systems. Following a general overview of cladistics, the section progresses into the specifics of this study, including characteristics used in the paper. The section concludes with an explanation on the specific matrices of the Jovian and Saturnian satellites and how they are applied to the cladistical method. 
\subsection{Cladistics}
\label{cladistics}

Cladistics is an analytical technique, originally developed to examine the relationships between living organisms \citep{Hennig1965PhylogeneticSystem}. A \textit{clade} is the term used for a cluster of objects or \textit{taxa}, that are related to each other at some level. In astronomy/astrophysics, the technique has been used to look at the relationships between stars \citep{FraixBurnet2015StarClads,Jofre2017StarsClads}, gamma-ray bursts \citep{Cardone2013GRBClads}, globular clusters \citep{FraixBurnet2009GlobularClusters} and galaxies \citep{FraixBurnet2006DwarfGalaxies, FraixBurnet2010EarlyGalx, FraixBurnet2012SixPermGal, FraixBurnet2015GalClad}. These works, along with this study, form a  body of work in the new field of `Astrocladistics' \citep{FraixBurnet2015GalClad}. There are good reasons to believe that cladistics can provide sensible groupings in a planetary science context. Objects that have similar formation mechanisms should have comparable characteristics. Daughter objects that are formed by breaking pieces off a larger object should also have similar characteristics. The advantage of this method over other multivariate analysis systems is the inclusion of a larger number of characteristics, enabling us to infer more detailed relationships. 

The vast majority of work in cladistics and phylogenetics has been undertaken in the Biological and Paleontological sciences. Biologists and Paleontologists use cladistics as a method to investigate the common origins\replaced{ or `tree', of life}{, or `tree' of life} \citep{Darwin1859Origin, Hennig1965PhylogeneticSystem, Hug2016TreeLife}, and how different species are related to one another \citep[e.g.][]{Van1993new, Salisbury2006originCrocs, Vrivcan2011two, Smith2017new, Aria2017burgess}. Historically, the investigation into relationships between different organisms reaches back to \citet{Darwin1859Origin}. Early attempts at using tree analysis techniques occurred in the early 20th century \citep{Mitchell1901BridsClads, Tillyard1926insects, Zimmermann1931arbeitsweise}. \citet{Hennig1965PhylogeneticSystem} is regarded as one of the first to propose `phylogenetic systematics', the technique that would become modern cladistical/phylogenetic analysis. The technique was quickly adopted by the biological community and used to analyze every form of life, from Bacteria \citep[e.g.][]{Olsen1994winds} to Dinosauria \citep[e.g.][]{Bakker1974dinosaur} and our own ancestors \citep[e.g.][]{Chamberlain1987EarlyHominid}. Recently, the use of DNA led to the expansion of the technique to become molecular phylogenetics \citep{Suarez2008HistoryPhylo}. As computing power improves, larger datasets can be examined, and our understanding of the `Tree of Life' improves \citep{Hug2016TreeLife}. For a detailed examination of the history of cladistics and pyholgenetics, we refer the interested reader to \citet{Hamilton2014EvolSystem}.

The cladisitcal methodology begins with the creation of a taxon-character matrix. Each matrix is a 2-d array, with the taxa, the objects of interest, in the rows, and each characteristic in the columns. The taxa used in this study are the rings and satellites of the Jovian and Saturnian Systems. The orbital, physical and compositional properties of the rings and satellites are used as characteristics, see Section \ref{Characteristics}. For a given taxa, each corresponding characteristic is defined as a numerical state, usually a 0 or 1, though multiple, discrete states may be used. A 0 numerical state is used to indicate the original or `base' state. An \textit{outgroup}, or a taxa outside the area of interest, is used to dictate the 0 base state of a characteristic. For this study, we use the Sun as an outgroup. An unknown character state can be accounted for, with a question mark (?). This taxon-character matrix is created using the Mesquite software package \citep{Mesquite}.

A set of phylogenetic trees are subsequently created  from the Mesquite taxon-character matrix, using Tree analysis using New Technology (TNT) 1.5 \citep{Goloboff2008TNT, Golboff2016TNT15}, via the Zephyr Mesquite package \citep{MesquiteZephyr}. The trees are created on the concept of maximum parsimony \citep{Maddison1984outgroup}, that the tree with the shortest lengths, the smallest number of changes, is most likely to show the true relationships. TNT uses a method of indirect tree length estimation \citep{Goloboff1994Treelengths, Goloboff1996FastPasrimony}, in its heuristic search for trees with the smallest length. TNT starts the drift algorithm \citep{Goloboff1996FastPasrimony} search by generating 100 Wagner trees \citep{Farris1970MethodsComp}, with 10 drifting trees per replicate. These starting trees are then checked using a Tree bisection and reconnection (TBR) algorithm \citep{Goloboff1996FastPasrimony} to generate a block of \replaced{equality}{equally} parsimonious trees. Closely related taxa are grouped together in the tree. Ideally, all equally parsimonious trees would be stored, but this is computationally prohibitive. For this analysis, 10000 equally parsimonious trees are requested from TNT, to create the tree block. Once a tree block has been generated and imported into Mesquite \citep{Mesquite} for analysis, a 0.5 majority-rules consensus tree can be constructed using \replaced{the a well established}{a well-established} algorithm \citep{Margush1981MajorityRules}. This tree is generated as a consensus of the block, with a tree branch being preserved if it is present in the majority of the trees. The resulting branching taxonomic tree is then a hypothesis for the relations between taxa, the satellites and rings of the gas giants.  

We can assess how accurately a tree represents true relationships between taxa. The number of steps it takes to create a tree is call the \textit{tree length}. A smaller tree length \replaced{indicates}{implies} a more likely tree, as it is more parsimonious. Tree length estimation algorithms \citep{Farris1970MethodsComp} continue to be improved, and are fully explored in a modern context by \cite{Goloboff2015Parsimony}. Two other tree metrics, the consistency and retention indices, are a measure of \textit{homoplasy}, or the independent loss or gain of a characteristic \citep{Givnish1997consistency}. High amounts of homoplasy in a tree is \replaced{indicative}{suggestive} of random events, rather than the desired relationships between taxa \citep{Brandley2009Homoplasy}. Mathematically, homoplasy can be represented by the consistency index ($CI$) of a tree,  (equation (\ref{ConIndexEq}) \citep{Kluge1969Cladistics}) and is related to the minimum number of changes ($M$) and the number of changes on the tree actually observed ($S$). 

\begin{equation}
CI = M/S
\label{ConIndexEq}
\end{equation}

A tree with no \textit{homoplasy} would have a consistency index of 1. \added{One of the criticisms of the consistency index is that it shows a negative correlation with the number of taxa and characteristics \citep{Archie1989homoplasy, Naylor1995RetentionIndex}. In order to combat the issues with the consistency index, a new measure of homoplasy, the retention index, was created \citep{Farris1989retention}.} The retention index ($RI$) \citep{Farris1989retention} \deleted{, a second measure of homoplasy} introduces the maximum number of changes ($G$) required into equation (\ref{RetentionIndexEq}). 

\begin{equation}
RI = \frac{G - M}{G - S}
\label{RetentionIndexEq}
\end{equation}

As with the consistency index, a tree with a retention index of 1 indicates a perfectly reliable tree. Both of these metrics \replaced{give an indication of}{show} how confidently the tree represents the \replaced{true}{most plausible} relationships between taxa. Values closer to 1 of both the consistency and retention indices indicate that the tree represents the true relationships between taxa \citep{Sanderson1989VaiationHomoplasy}. For a detailed examination of the mathematics behind the algorithms and statistics used in cladistical analysis, we direct the interested reader to \cite{Gascuel2005MathEvolPhylogeny}.

A traditional form of multivariate hierarchical clustering is used in the detection of asteroid collisional families \citep{Zappala1990HierarchicalClustering1, Zappala1994HierarchicalClustering2}.
This method of clustering uses Gauss equations to find clusters in n parameter space, typically using semi-major axis, eccentricity and inclination \citep{Zappala1990HierarchicalClustering1}. Work has also been undertaken incorporating the known colors \citep{Parker2008AsteroidFamSDSS} and albedo \citep{Carruba2013AsteroidFamilies} of the asteroids \citep{Milani2014AsteroidFamilies} into the classical method, though this reduces the dataset significantly. The classical method of multivariate hierarchical clustering was used by \citep{Nesvorny2003IrrSatEvol} to identify the Jovian irregular satellite families. \citet{Turrini2008IrregularSatsSaturn} expanded the classical method into the Saturnian irregular satellites, and utilized the Gauss equations, solved for velocities, in a similar way to \cite{Nesvorny2003IrrSatEvol} to verify the families found, using semi-major axis ($a$), eccentricity ($e$) and inclination ($i$) of the satellites. The rational behind these calculations is that the dispersal velocities of the clusters would be similar to the escape velocities of the parent body. In this work we use the inverse Gauss equations, equations \ref{InvGauss1}, \ref{InvGauss2} and \ref{InvGauss3}, substituted into equation \ref{VelocityEq}, to test the dispersal velocities of the clusters found through cladistics. $\delta a$, $\delta e$ and $\delta i$ are the difference between the individual satellites and the reference object. $a_r$, $e_r$, $i_r$ and orbital frequency ($n_r$) are parameters of the reference object. In this case, the reference object is taken as the largest member of the cluster. The true anomaly ($f$) and perihelion argument ($w + f$) at the time of disruption are unknown. Only in special cases, for example, for young asteroid families \citep[e.g.]{Nesvorny2002AsteroidBreakup}, the values of ($f$) and ($w + f$) can be inferred from observations. In this work we adopt $f = 90 \degree $ and $(f+w) = 45 \degree $ respectively as reasonable assumptions. Previous works by \cite{Nesvorny2003IrrSatEvol} and \cite{Turrini2008IrregularSatsSaturn} using this method, do not \replaced{indicate}{specify} the true anomaly ($f$) and perihelion argument ($w + f$) used, nor the central reference point, making any comparisons between them and this work relative rather than absolute. The final $\delta V_d$ for the cluster is composed of the velocities in the direction of orbital motion ($\delta V_T$), the radial direction ($\delta V_R$) and perpendicular to the orbital plane ($\delta V_W$). 

\begin{equation}
\delta V_T = \frac{n_r a_r (1+e_r \cos f)}{\sqrt{1-e_r^2}} \cdot \left [ \frac{\delta a}{2 a_r} - \frac{e_r \delta e}{1-e_r^2} \right ]
\label{InvGauss1}
\end{equation}

\begin{equation}
\delta V_R =  \frac{n_r a_r}{(\sqrt{1-e_r^2})\sin f} \cdot \left [ \frac{\delta e_r (1 + e_r \cos f)^2}{1-e_r^2} - \frac{\delta a (e_r + e_r \cos^2 f + 2\cos f)}{2a_r}   \right ]
\label{InvGauss2}
\end{equation}

\begin{equation}
\delta V_W = \frac{\delta i \cdot n_r a_r }{\sqrt{1-e_r^2}} \cdot \frac{1+e_r \cos f}{\cos (w+f)}
\label{InvGauss3}
\end{equation}

\begin{equation}
\delta V_d = \sqrt{\delta V_T^2 + \delta V_R^2 +  \delta V_W^2}
\label{VelocityEq}
\end{equation}

Cladistics offers a fundamental advantage over this primarily dynamics based clustering, and that is the incorporation of unknown values. Classical multivariate hierarchical clustering \citep{Zappala1990HierarchicalClustering1} requires the use of a complete dataset, and as such a choice is required. The parameters are either restricted to only known dynamical elements, or the dataset is reduced to well studied objects. Cladistical analysis can incorporate objects with large amounts of unknown information, originally fossil organisms \citep{Cobbett2007FossilsCladistics}, without a reduction in the number of parameters.

\subsection{Characteristics}
\label{Characteristics}
We define 38 characteristics that can be broken into three broad categories: orbital, physical and compositional parameters. All numerical states are considered having equal weight. The discrete character sets are unordered. Any continuous characteristics are broken into bins, as cladistical analysis requires discrete characteristics. We developed a Python program to establish the binning of continuous characteristics. \added{The pandas Cut module \citep{Mckinney2010Pandas} is used to create the bins.} Each characteristic is binned independently of each other and for each of the Jovian and Saturnian systems. The aforementioned \replaced{python}{Python} program iterates the number of bins until \replaced{an $r^2$ score of $>0.99$ is reached for that characteristic set}{a linear regression model between binned and unbinned sets achieves a coefficient of determination ($r^2$) score of $>0.99$. This is calculated using the stats package in SciPy \citep{Jones2010SciPy}}. Thus each character set will have a different number of bins, $r^2$ score and delimiters. All characteristics are binned in a linear fashion, with the majority increasing in progression. The exception to the linear increase is the density character set, with a reversed profile. All of the continuous, binned characteristic sets are ordered, as used by \cite{FraixBurnet2006DwarfGalaxies}. A full list of the characteristics used, the $r^2$ score for each of the binned characteristics, along with the delimiters are listed in Appendix \ref{ListCharacters}.

The first broad category includes the five orbital characteristics (Appendix \ref{listOrbitChar}). This category is comprised of two discrete characteristics, presence in orbit around the gas giant, and prograde or retrograde orbit. The three remaining characteristics, semi-major axis (a), orbital inclination (i) and eccentricity (e), are continuous and require binning using the aforementioned \replaced{python}{Python} program. 

The second category used to construct the matrix consists of two continuous physical characteristics, density and visual geometric albedo (Appendix \ref{listPhysChar}). We chose to not include mass, or any properties related to mass, as characters in the analysis. The inclusion of these characteristics could hide any relationships between a massive object and any daughter objects, as the result of collisions. 

The third category describes the discrete compositional characteristics and details the presence or absence of 31 different chemical species (Appendix \ref{listCompChar}). In order to account for any positional bias, the fundamental state, solid, liquid, gas or plasma was not considered. In this anyalysis, we make no distinction between surface, bulk and trace compositions. This is to account for the potential of daughter objects to have their bulk composition comprised of surface material from the parent. The majority of compounds have absence as a base state (0), and presence as the derived (1). The exception are the first three molecules, elemental hydrogen (eH), hydrogen (H$_2$) and helium (He), all of which are found in the Sun. As the Sun is the designated outgroup, the base state (0) indicates the presence of these species. With the exception of elemental hydrogen, the remaining single element species are those found in compounds. The spectroscopy of an object often only reports on the presence of an ion, as opposed to a full chemical analysis. \deleted{As the full chemical composition of a body.} As more detailed analysis becomes available, characters may be added to the matrix. Several chemical species are used in this particular matrix which are either not present in any of the satellites or unknown. These are included for future comparisons with other orbital bodies.

\subsection{Matrices}
The Jovian taxon-character matrix holds 68 taxa consisting of: the Sun (outgroup), four inner satellites, the main ring, four Galilean satellites and 59 irregular satellites. Appendix \ref{JupiterMatrix} contains the matrix, along with the references used in its construction.

The Saturnian matrix, presented in Appendix \ref{SaturnMatrix}, is created with 76 taxa. These taxa are the Sun (outgroup), six main rings, nine inner small satellites, four minor rings, eight large icy satellites, four Trojan satellites, three Alkynoids and their associated rings, and the 38 irregular satellites. The references used in the construction of the Saturnian matrix are located in Appendix \ref{SaturnMatrix}. Both matricies use the same characteristics, as discussed in Section \ref{Characteristics}, and are available in machine readable format.

\section{Results}

In this section we present the resulting taxonomic trees from the analysis of the Jovian and Saturnian satellites. The taxonomic trees are used to form the systematic classification, of the Jovian (Table \ref{JupiterClassTable}) and Saturnian (Table \ref{SaturnClassTable}) satellite systems. Using inverse Gauss equations \citep{Zappala1990HierarchicalClustering1}, in a similar method to \cite{Nesvorny2003IrrSatEvol} and \cite{Turrini2008IrregularSatsSaturn}, we show in Tables \ref{JupiterClassTable} and \ref{SaturnClassTable}, dispersal velocities ($\delta V$) for each of the taxonomic groups where a single origin object is hypothesized, namely the irregular satellites. For these calculations we assume the largest representative of the cluster as the origin point. See section \ref{cladistics} for further discussion.

\subsection{Jovian Taxonomy}
\label{JupiterTax}
\begin{figure*}
\includegraphics[height=0.65\paperheight]{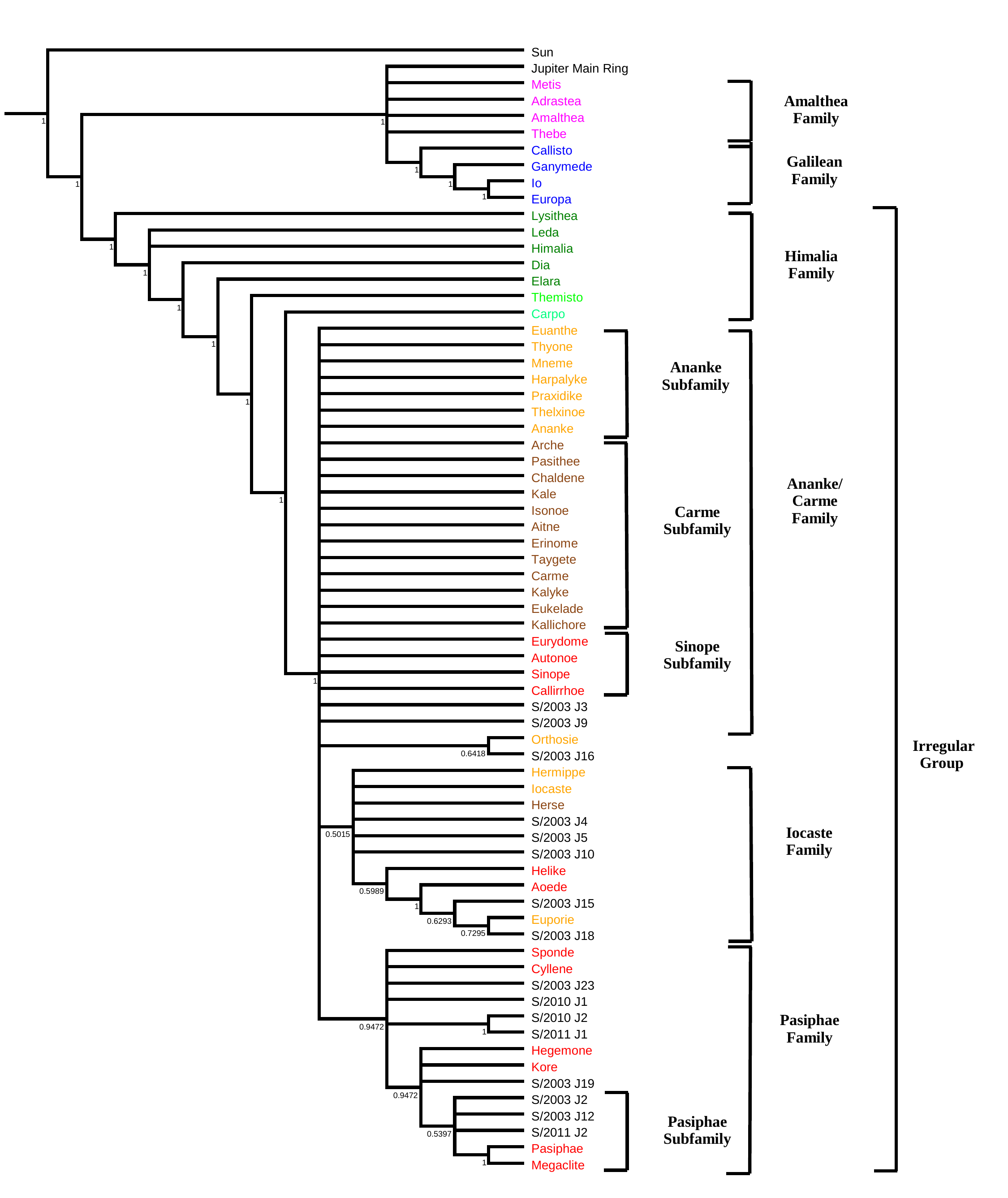} 
\caption{Majority consensus taxonomic tree of objects in the Jovian system. This tree has a tree length score of 128, with a consistency index of 0.46 and a retention index of 0.85. Numbers indicate frequency of the node in the 10000 most parsimonious tree block. \replaced{Colors are indicative of}{Colors represent terminology used in} traditional classification: \textcolor{Amalthea}{Amalthea inner regular family} ;      
 \textcolor{Galilians}{Galilean family};
  \textcolor{Themisto}{Themisto prograde irregular}  ;
   \textcolor{Himalia}{Himalia prograde irregular family};
  \textcolor{Carpo}{Carpo prograde irregular} ;     
  \textcolor{Ananke}{Ananke irregular family};
  \textcolor{Carme}{Carme irregular family}  ;           
  \textcolor{Pasiphae}{Pasiphae irregular group};
 Unnamed and unclassified. Proposed groups and families are \replaced{indicated}{shown} on the right.} 
\label{JupiterTree}
\end{figure*}

The results of the cladistical analysis of the individual Jovian satellites is shown in Figure \ref{JupiterTree}. This 0.5 majority-rules consensus tree has a tree length score of 128, with a consistency index of 0.46 and a retention index of 0.85. \added{The low value of the consistency index is possibly due to the mixed use of ordered, multi-state, continuous characteristics and bi-modal compositional characteristics \citep{Farris1990phenetics}.}  \replaced{These values indicate}{The high retention index suggests} that the consensus tree is robust and \replaced{indicative of the true}{demonstrates the most likely} relationships between the satellites. 

\begin{deluxetable}{p{3cm}p{3.8cm}ccccccp{2cm}cc}
\label{JupiterClassTable}

\rotate

\tabletypesize{\scriptsize}

\tablecaption{Jovian Satellite Systematic Classification}

\tablenum{1}

\tablehead{\colhead{Taxonomy} & \colhead{Members} & \colhead{Orbit} & \colhead{Semi-major Axis} & \colhead{Inclination} & \colhead{Eccentricity} & \colhead{Density} & \colhead{Albedo} & \colhead{Composition} & \colhead{Velocity ($\delta V$)} & \colhead{Ref.} \\ 
\colhead{} & \colhead{} & \colhead{} & \colhead{(km)} & \colhead{} & \colhead{} & \colhead{($kg~m^{-3}$)} & \colhead{} & \colhead{} & \colhead{($m~s^{-1}$)} & \colhead{} } 

\startdata
Amalthea family  & Thebe, Amalthea, Metis and Adrastea & Prograde & $< 3.0 \times 10^5$ & $< 0.02\degree$ & $< 2\degree$ & $< 900$ & $< 0.1$  & predominately water ice and silicates & $3570.4  \pm 491.8 $ & 1 \\
Galilean family  & Io, Ganymede, Europa, Callisto & Prograde & $4.0\times 10^5$ – $2.0\times 10^6$ & $< 0.5\degree$ & $< 0.01$ & $> 1800$ & $> 0.18$ & water ice and silicates dominate; presence of SO$_2$; other chemical species present & \nodata & 2 \\
Jovian Irregular Satellite group  &  &  &  &  &  &  &  &  &  &  \\
Himalia family  & Leda, Elara, Lyithea, Himalia and Themisto. & Prograde & $7.5 \times 10^6$ - $1.8 \times 10^6$  & $25\degree$ - $55\degree$ & $0.1$ - $0.3$ & \nodata & $< 0.1$ & silicate based & $623.8 \pm 750.3$ & 3,4 \\
Ananke/Carme Family & S/2003 J3, S/2003 J9, Ananke subfamily, Carme subfamily and Sinope subfamily. & Retrograde & $1.88 \times 10^7$ - $2.5\times 10^7$  & $143\degree$ - $166\degree$ & $0.2$ - $0.4$ & \nodata & $< 0.07$ & \nodata & $457.2 \pm 445.7$ & 3,4 \\
Ananke Subfamily & Euanthe, Thyone, Mneme, Harpalyke, Praxidike, Thelxinoe and Ananke. & Retrograde & $2.0 \times 10^7$ - $2.15 \times 10^7$  & $145\degree$ - $152\degree$ & $0.2$ - $0.25$ & \nodata & $< 0.07$ & \nodata & $61.0 \pm 45.6$ & 3,4 \\
Carme Subfamily & Arche, Pasithee, Chaldene, Isonoe, Kale, Aitne, Erinome, Taygete, Carme, Kalyke, Eukelade and Kallichore. & Retrograde & $2.2 \times 10^7$ - $2.4 \times 10^7$  & $164\degree$ - $166\degree$ & $0.24$ - $0.27$ & \nodata & $< 0.07$ & \nodata & $36.1 \pm 13.1$ & 3,4 \\
Sinope Subfamily & Eurydome, Autonoe, Sinope and Callirrhoe. & Retrograde & $2.2 \times 10^7$ - $2.42 \times 10^7$  & $147\degree$ - $159\degree$ & $0.27$ - $0.35$ & \nodata & $< 0.06$ & \nodata & $323.9 \pm 97.3$ &  \\
Iocaste Family & Euporie, S/2003 J18, Hermippe, Helike, Iocaste, S/2003 J15, Herse, S/2003 J4, Aoede, S/2003 J5 and S/2003 J10 & Retrograde & $1.9 \times 10^7$ - $2.5 \times 10^7$  & $140\degree$ - $165\degree$ & $0.1$ - $0.45$ & \nodata & $< 0.05$ & \nodata & $510.2 \pm 303.3$ &  \\
Pasiphae Family & S/2003 J12, S/2011 J1, S/2010 J2, S/2003 J19, S/2010 J1, S/2011 J2, Sponde, Pasiphae, Megaclite, Hegemone, S/2003 J23, Cyllene, Kore and S/2003 J2. & Retrograde & $1.9 \times 10^7$ - $2.9 \times 10^7$  & $145\degree$ - $164\degree$ & $0.30$ - $0.421$ & \nodata & $< 0.1$ & \nodata & $412.3 \pm 224.5$ & 3,4 \\
\enddata

\tablerefs{(1) \citet{Barnard1892Amalthea};
(2) \citet{Galileo1610SidereusNuncius};
(3) \citet{Nesvorny2003IrrSatEvol};
(4) \citet{Sheppard2003IrrSatNature}.}

\end{deluxetable}

As can be seen in the Jovian taxonomic tree in Figure \ref{JupiterTree}, the satellites cluster in to clades resembling the taxonomy proposed by \citet{Nesvorny2003IrrSatEvol} and \citet{Sheppard2003IrrSatNature}. The irregular satellites are a separate cluster to the prograde regular satellites. 

We maintain the closest family to Jupiter, the Amalthea family, as a valid taxonomic cluster. The dispersal velocity is very large and may \replaced{indicative}{suggest} that the Amalthea family did not form from a single object. This family, along with Jupiter's main ring, \replaced{are}{is} associated with the well known Galilean family. 

In the analysis, we maintain the 'irregular' satellite group. The Himalia family clusters with the retrograde satellites, separate to the other prograde satellites.  The Himalia family has relatively low inclinations, in comparison with the Jovian retrograde satellites and their high eccentricity could be explained by disruptions \citep{Christou2005HimaliaScattering}. The small satellites Themisto and Carpo cluster together with the other prograde satellites in the Himalia family. We propose that Themisto and Carpo be included in the Himalia family, as they are the sole members of the groups proposed by \citet{Sheppard2003IrrSatNature}, and show similar orbital characteristics. The large mean dispersal velocity calculated for the Himalia family (see Table \ref{JupiterClassTable}) was also noticed by \citet{Nesvorny2003IrrSatEvol} for the Prograde satellites. The large mean dispersal velocity due to the dispersal velocities of Themisto and Carpo. Without including these members, the mean dispersal velocity for the classical Himalia family is $154.6 \pm 72.5 m/s$, close to the escape velocity of Himalia, $121.14 m/s$. This dispersal velocity of the classical Himalia family, was explained via gravitational scattering from Himalia by \citet{Christou2005HimaliaScattering}. Disruption and scattering could also be used to explain the large dispersal velocities of Themisto and Carpo, though further modeling is required. 

The term `irregular' is maintained through the retrograde family for consistency with the literature \citep{Nesvorny2003IrrSatEvol, Sheppard2003IrrSatNature, Nesvorny2004IrrSatFamilyOrigin,  Beauge2007IrregSatRsonance, Jewitt2007IrregularSats}. The retrograde irregular satellites are separate, but related cluster to the Himalia, prograde irregulars. The broad classifications introduced by \citet{Sheppard2003IrrSatNature} and \citet{Nesvorny2003IrrSatEvol} are preserved, though the Ananke/Carme family is unresolved and may be split into subfamilies. Separating out the traditional families \citep{Nesvorny2003IrrSatEvol, Sheppard2003IrrSatNature}, see colors in figure \ref{JupiterTree}, give smaller dispersal velocities. The traditional Ananke (escape velocity (eV) $23.10 m/s$) family has a $\delta V$ of $61.0 \pm 45.6 m/s$, traditional Carme (eV $29.83 m/s$) has $36.2 \pm 13.1 m/s$, and a created Sinope (eV $27.62 m/s$) family has $323.9 \pm 97.3 m/s$. These are smaller than the $\delta V$ of our unresolved Ananke/Carme Family ($457.2 \pm 445.7 m/s$, see table \ref{JupiterClassTable}). \citet{Nesvorny2003IrrSatEvol} used similar small $\delta V$ values to establish the Ananke and Carme dynamical families. The dynamical situation could be explained through a more recent capture and breakup event for Ananke, Carme and Sinope, that disrupted the ancestral irregular satellites. The identified Iocaste and Pasiphae families also have large dispersal velocities, \replaced{indicative}{suggestive} of disruptions. Following the nomenclature of \citet{Sheppard2003IrrSatNature}, each of the families and subfamilies are represented by the name of the largest contained satellite. Satellites within families are related by their retrograde orbit, high inclinations and eccentricities. In addition to their linked orbital characteristics, the satellites of the retrograde irregular group all show a low albedo \citep{Beauge2007IrregSatRsonance}. 

The Ananke subfamily is tightly constrained in its orbital characteristics, with a small dispersal velocity. While the characteristics listed in Table \ref{JupiterClassTable} would preclude them from being included in the Pasiphae family, their clustering around a common semi-major axis, inclination and eccentricity, \replaced{indicated}{suggesting} that they are a distinct young dynamical family. The members we include in the Ananke family for this analysis are all historical members of the family \citep{Jewitt2007IrregularSats}. Some of the satellites that have been historically included in the Ananke family \citep{Jewitt2007IrregularSats} are moved to other families. We do not add any new satellites to this family. 

The orbital characteristics of the Carme subfamily are tightly constrained. Satellites in this family orbit further from Jupiter, with higher orbital inclinations, but similar eccentricities to the Ananke family. As with the Ananke family, it is the highly constrained orbital characteristics and low mean dispersal velocity, that justify the classification of this traditional family \citep{Jewitt2007IrregularSats}. According to the tree presented in Figure \ref{JupiterTree}, there is a continuum between the Ananke and Carme families. However, differences in orbital characteristics, broken down in Table \ref{JupiterClassTable}, distinguish both of these families from each other.

A new cluster, the Iocaste family, is defined as shown in Figure \ref{JupiterTree} and Table \ref{JupiterClassTable}. The semi-major axis of this family spans most of the orbital space where irregular satellites have been discovered. The lower eccentricities and albedo are used to separate this family from the Pasiphae family. As with the Passiphae family, the Iocaste family has a high mean dispersal velocity ($510.2 \pm 303.3$ compared with a escape velocity of $3.16 m/s$), \replaced{indicative}{suggestive} of disruptions taking place at some point since the break-up of the original object \citep{Christou2005HimaliaScattering}. Iocaste, being the largest member of this family, is proposed as the representative object. Also included are several members that have been previously included in other families \citep{Jewitt2007IrregularSats}, along with new unnamed satellites. For full details on included satellites and the descriptive properties of the family, see Table \ref{JupiterClassTable}.

The Pasiphae family show a broad range of orbital characteristics that, along with the large dispersal velocity ($412.3 \pm 224.5$ compared with an escape velocity of $47.16 m/s$), are \replaced{indicative}{suggestive} of disruptions during the family's life-time \citep{Christou2005HimaliaScattering}. The Pasiphae family has a broad range semi-major axies and inclinations, with the Pasiphae family orbiting further from Jupiter and having larger eccentricities on average than the new Iocaste family, see table \ref{JupiterClassTable}. A Pasiphae subfamily, see figure \ref{JupiterTree}, with a $\delta V$ of $230.1 \pm 174.3 m/s$, can be identified. This may \replaced{indicate}{imply} a secondary, more recent break-up from Pasiphae. In addition, many of the unnamed satellites from recent observations  \citep{Gladman2003JupIAU1, Gladman2003JupIAU2, Sheppard2003JupIAU1, Sheppard2003JupIAU2, Sheppard2003JupIAU3, Sheppard2003IAUJupSat,
Sheppard2004JupIAU1,Sheppard2004JupIAU2,
Jacobson2011NewJupSats, Sheppard2012JupIAU} are associated with this family, see Table \ref{JupiterClassTable} and Figure \ref{JupiterTree} for a complete list.

\subsection{Saturnian Taxonomy}
\label{SaturnTax}
\begin{figure*}

\includegraphics[height=0.65\paperheight]{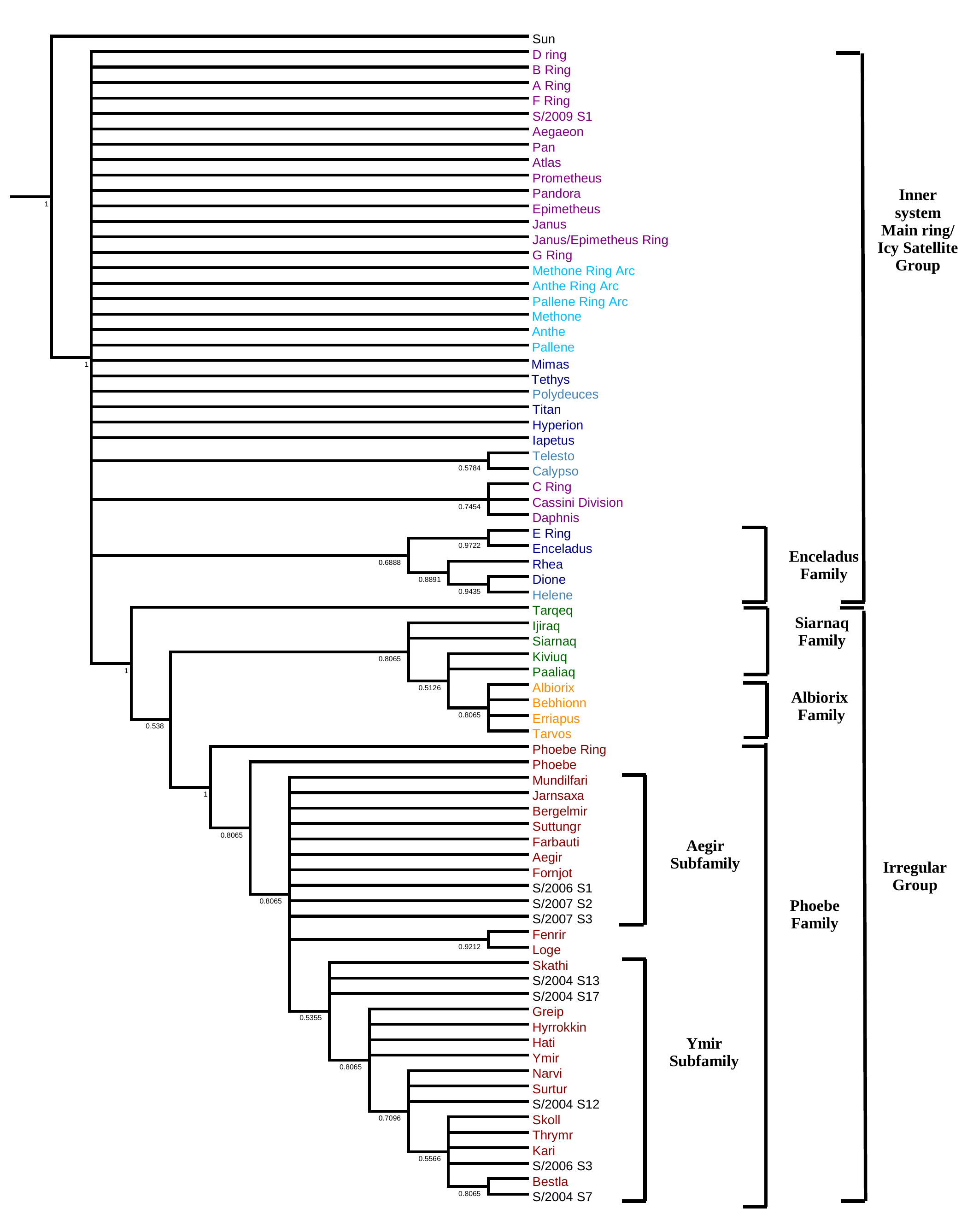} 
\caption{Majority Consensus taxonomic tree of objects in the Saturnian system. The tree has a consistency index of 0.30 and a retention index of 0.81. Numbers indicate frequency of the node in the 10000 most parsimonious tree block. \replaced{Colors are indicative of}{Colors represent terminology used in} classical classification: \textcolor{MainRing}{Main ring group, with associated shepherd satellites} ;      
 \textcolor{IcySats}{Mid-sized Icy satellites and Titan};
  \textcolor{Trojans}{Trojan satellites}  ;
   \textcolor{Alkanoids}{Alkanoids and associated rings};
  \textcolor{Inuit}{`Inuit' prograde irregular family} ;     
  \textcolor{Gallic}{`Gallic' prograde irregular family};
  \textcolor{Norse}{`Norse' retrograde irregular family}  ; 
  Unnamed and unclassified.           
  Proposed groups and families are \replaced{indicated}{shown} to the right. }
\label{SaturnTree}
\end{figure*}

Cladistical analysis of the Saturnian \replaced{System}{system} yields the 0.5 majority-rules consensus tree, Figure \ref{SaturnTree}, constructed from the 10000 parsimonious trees, with a tree length score of 186. The tree has a consistency index of 0.30 and a retention index of 0.81. The consistency index of the Saturnian tree is lower than that of the Jovian tree, though this could be due to the number of taxa used \citep{Sanderson1989VaiationHomoplasy}. \added{As with the Jovian tree, this low consistency index could be due to the mixed character states. This effect is to be explored further in a future paper.} The high retention index indicates that the tree \replaced{is indicative}{is suggestive} of the true relationships \citep{Farris1989retention}.

\begin{deluxetable}{p{3cm}p{3.8cm}ccccccp{2cm}cc}

\rotate

\tabletypesize{\scriptsize}

\tablecaption{Saturnian Satellite Systematic Classification}
\label{SaturnClassTable}

\tablenum{2}

\tablehead{\colhead{Taxonomy} & \colhead{Members} & \colhead{Orbit} & \colhead{Semi-major Axis} & \colhead{Inclination} & \colhead{Eccentricity} & \colhead{Density} & \colhead{Albedo} & \colhead{Composition} & \colhead{Velocity ($\delta V$)} & \colhead{Ref.} \\ 
\colhead{} & \colhead{} & \colhead{} & \colhead{(km)} & \colhead{} & \colhead{} & \colhead{($kg~m^{-3}$)} & \colhead{} & \colhead{} & \colhead{($m~s^{-1}$)} & \colhead{} } 

\startdata
Saturnian Inner system Group, Main ring and Icy satellites & Atlas, Janus, Epimetheus, Prometheus, Janus/Epimetheus ring, G ring, D ring, Pan, Aegaeon, S/2009 S1, F ring, B ring, Cassini Division, C ring, Daphnis and A ring. Possible members: Telesto, Calypso, Methone ring arc, Anthe ring arc, Pallene ring arc, Methone, Anthe, Pallene,  Polydeuces  Mimas, Tethys,  Enceladus family, Hyperion, Titan and Iapetus ; see Section \ref{SaturnTax} for discussion. & Prograde & $< 4.0 \times 10^6$ & $< 15\degree$ & $< 0.03$ & $550$ – $1900$ & $0.1$ - $1$ & Composition of water ice with silicates and presence of CO$_2$. Other chemical species may be present  & \nodata & 1, 2 \\
Enceladus Family & E ring, Enceladus, Rhea, Dione and Helene. & Prograde & $1.8 \times 10^5$ - $5.3 \times 10^5$ & $< 0.5\degree$ & $0$ & $1200$ – $1700$ & $> 0.7$ & Complex composition, predominately water ice and silicates, with Hydrocarbons and CO$_2$ present & \nodata &  \\
Saturnian Irregular Satellite group  &  &  &  &  &  &  &  &  &  &  \\
Albiorix family  & Bebhionn, Erriapus, Albiorix and Tarvos & Prograde & $1.6 \times 10^7$ - $1.8 \times 10^7$ & $30\degree$ - $40\degree$ & $0.4$ - $0.6$ & \nodata & $< 0.1$ & \nodata & $80.9 \pm 1.6 $ & 3,4,5 \\
Siarnaq family  & Tarqeq, Kiviuq, Ijiraq, Paaliaq and Siarnaq & Prograde & $1.1 \times 10^7$ - $1.9 \times 10^7$ & $40\degree$ - $50\degree$ & $0.1$ – $0.4$ & \nodata & $< 0.1$ & \nodata & $266.8 \pm 60.0$ & 3,4,5 \\
Phoebe family  & Phoebe Ring, Phoebe,  Fenrir, Loge, Aegir subfamily, and Ymir subfamily. & Retrograde & $1.1 \times 10^7$ - $2.51 \times 10^7$ & $> 145\degree$ & $> 0.1$ & \nodata & $< 0.1$ & \nodata & $763.3 \pm 259.0 $ & 3,4,5 \\
Aegir subfamily & S/2007 S2, Mundilfari, Jarnsaxa, S/2006 S1, Bergelmir, Suttungr, Farbauti, S/2007 S3, Aegir and Fornjot. & Retrograde & $1.6 \times 10^7$ - $2.51 \times 10^7$ & $> 150\degree$ & $0.1$ – $0.25$ & \nodata & \nodata & \nodata & $295.1 \pm 125.0$ & 5 \\
Ymir subfamily & Skathi, Skoll, Greip, Hyrrokkin, S/2004 S13, S/2004 S17, Narvi, S/2004 S12, S/2004 S07, Hati, Bestla, Thrymr, S/2006 S3, Kari, Surtur and Ymir & Retrograde & $1.55 \times 10^7$ - $2.30 \times 10^7$  & $> 145\degree$ & $0.25$ - $0.6$ & \nodata & $< 0.1$ & \nodata & $497.5 \pm 247.7$ & 5 \\
\enddata


\tablerefs{(1) \citet{Huygens1659systema};
(2) \citet{Cassini1673Sat2Sats,Cassini1686Sat2Sats}
(3) \citet{Nesvorny2003IrrSatEvol};
(4) \citet{Sheppard2003IrrSatNature};
(5) \citet{Turrini2008IrregularSatsSaturn}.}

\end{deluxetable}

The tree shown in Figure \ref{SaturnTree} highlights the diversity of structures found in the orbit of Saturn. Satellites cluster into two main grouping around Saturn, the Inner group, comprised of rings and icy satellites, and the Irregular satellite group, see Table \ref{SaturnClassTable} for members and diagnostic properties of each clade. While the traditional classification nomenclature \citep{Nesvorny2003IrrSatEvol, Sheppard2003IrrSatNature, Jewitt2007IrregularSats} is broadly conserved, several \replaced{anomalies}{discrepancies} require discussion. Table \ref{SaturnClassTable} shows our new taxonomy, along with included members of the families and their descriptive properties.

The Main ring and Icy satellite group form an unresolved, inner system group. This group includes the Saturnian ring system, the Alkynoids and their associated ring arcs, as well as the larger Icy satellites and their Trojans. We have confirmed the recently discovered S/2009 S1 \citep{Spitale2012s2009s1} is part of this group due to its orbital characteristics. Within this large group, there is the resolved Enceladus family.

Our results suggest the traditionally classified Alkyonides, Methone, Anthe and Pallene, along with their associated rings, are not clustered with the the Enceladus family, as would be expected by their orbital location, between Mimas and Enceladus, within the E-ring. Due to their bulk water ice composition, the Alkynoides associate with the Main ring objects, see Figure \ref{SaturnTree}. The low density and mid-range albedo of Pallene and Methone \citep{Hedman2009SatRingArcs} \replaced{indicates}{suggests} that the association with the Main ring group is genuine. The dynamic resonances of both Methone and Anthe \citep{Callegari2010SmallSaturnSatsDynamics} \replaced{are also indicative of these objects being}{implies that these objects where} captured, rather than forming in-situ. As there is very little known about the composition of these objects, beyond their bulk water ice composition \citep{Hedman2009SatRingArcs}, further study and dynamical modeling of the capture process is required to resolve their true origins.
 
Like the Alkynoids, the Trojan satellites of Tethys, Calypso and Telesto, also form an association with the main rings. The reason for this could be that Calypso and Telesto, like the Alkynoids, are also possible captured main ring objects. The capture dynamics could be similar to that of the Jovian Trojan asteroids \citep{Morbidelli2005TrojanCapture, Lykawka2010TrojanCaputer, Nesvorny2013TrojanCaptureJJ}. Both of the Tethys Trojans \citep{Buratti2010SatInnerSat} and main ring objects, are chiefly comprised of water ice, \replaced{indicative of}{implying} a common origin. The bulk composition of Tethys is also prominently water ice \citep{Buratti2010SatInnerSat}, with a very small fraction of silicates. Trojans may instead have formed from the same material as Tethys itself, either during accretion \citep{Charnoz2011SaturnSatAccretion} or in the same orbit from a large debris disk \citep{Canup2010SaturnSatOrigin}. As Tethys is also in the unresolved Main ring and Satellite group, we can not differentiate between the two scenarios. Further compositional information about the Tethys Trojans could shed light on this issue. Polydeuces, a Trojan of Dione, also forms an association with the Main ring group in our analysis. This could be due to overemphasis on orbital and physical characteristics, since the bulk composition of Polydeauces is unknown \citep{Thomas2013InnerSatSatu}. Helene, the more well studied Trojan of Dione \citep{Thomas2013InnerSatSatu}, is well within the Enceladus Family. Helene and Dione are closely associated in our analysis, \replaced{indicating}{implying} that Helene is a daughter object of Dione.

The outer icy satellites, Titan, Hyperion and Iapetus, do not form a single cluster, and are therefore not considered a valid taxonomic group. They are associated with the Main ring and Icy Satellite group. The Enceladus family is formed by the known association between the E-ring, Enceladus and Icy Satellites \citep{Verbiscer2007Enceladus}, which is mainly due to the detection of volatile chemicals, such as NH$_3$, CH$_4$ and other hydrocarbons. Plumes from Enceleadus containing these chemicals \citep{Porco2006EnceladusPlume}, thought to be representative  of the subcrust ocean \citep{Porco2006EnceladusPlume}, \deleted{and} are the source of the E ring \citep{Sphan2006EnceladusEring}. Titan itself also has an abundance of these volatiles \citep{Hirtzig2013VIMSTitan}, \replaced{indicating}{implying} a possible association between the Icy satellites of Saturn that remains unresolved in our analysis.
Material from the outer satellites, particularly Pheobe and its associated ring \citep{Tosi2010IapetusDark, Tamayo2011IapetusDust} is thought to play a role in the observed hemispherical dichotomy on Iapetus \citep{Tosi2010IapetusDark}. In Figure \ref{SaturnTree}, Iapetus is unresolved in the Main ring and Icy Satellite group.

The irregular satellites form a major cluster with each other separate from the inner Saturnian system, and are therefore collected under the Irregular satellite group. Along with their high inclinations, eccentricities and semi-major axes, the Irregular satellite group is characterized by a dark albedo, comparative to the other objects in the Saturnian system. We follow the naming convention introduced with the Jovian satellites, Section \ref{JupiterTax},  where each irregular satellite family is represented by the largest member \citep{Jewitt2007IrregularSats}. We therefore rename the classical Inuit group \citep{Blunck2010SaturnSats} to the Siarnaq family and the Gallic group \citep{Blunck2010SaturnSats} to the Albiorix family. Though this does change the formal name of the clusters, we encourage the discoverers of the unnamed satellites \citep{Gladman200112Sat,Sheppard2003IAUJupSat,Jewitt2005IAUCSat,Sheppard2006SatIAUC,Sheppard2007SatIAUC} and any future discoveries that are placed in these groups, to follow IAU convention and use names from Inuit and Gallic mythology for satellites in the Siarnaq and Albiorix families respectively. 
As in \cite{Turrini2008IrregularSatsSaturn}, the Albiorix family is distinct and has a low mean dispersal velocity ($\delta V$). The Siarnaq family has a higher $\delta V$, again \replaced{indicative}{suggestive} of disruptions \citep{Christou2005HimaliaScattering}. The mean $\delta V$ of all prograde satellites is $364.8 \pm 114.9 m/s$, only slightly higher than that of the Siarnaq family \citep{Turrini2008IrregularSatsSaturn}. This could \replaced{be indicative of}{imply} a disruption scenario, with a more recent capture of the Albiorix family parent body disrupting the older Siarnaq family. Our cladistical analysis supports this scenario, as the Siarnaq family shows a more branching structure than the Albiorix family. Further compositional information about these bodies, as well as dynamical modeling, could resolve this complex situation.

\replaced{Our study shows that the classically named retrograde Norse group \citep{Blunck2010SaturnSats} is determined to be the Phoebe family, which can be split into at least two subfamilies.}{In our analysis, we separate out the retrograde irregular satellites, including Phoebe, from the prograde irregular satellites. In previous taxonomy, this group has been classified as the 'Norse' group \citep{Blunck2010SaturnSats}. In our revised nomenclature, this group should be termed the Phoebe family. We further separate out two clades, distinct from Phoebe and its associated ring.} The first clade, the unresolved Aegir subfamily\added{ (previously identified as the S/2004 S10 group in \cite{Turrini2008IrregularSatsSaturn})}, is characterized as having on average, orbits further from Saturn, with low eccentricities and higher inclinations. The second clade is the Ymir subfamily and is categorized, on average, by being closer to Saturn, but with high eccentricities. This subfamily shows a branching structure and may be further split \citep{Grav2007IrregSatCol}. \added{This family was also identified by \cite{Turrini2008IrregularSatsSaturn}.} We identify an association between Fenrir and Loge, with a low dispersal velocity ($\delta V = 114.4 m/s$), \replaced{indicative}{suggestive} of a recent breakup. The high dispersal velocity ($\delta V$) of the Phoebe family is due to the selection of Phoebe as a reference point. If Phoebe and the \replaced{associate}{associated} ring are removed from the \replaced{subfamily}{family}, \added{and Ymir (with an escape velocity of $8.56 m/s$) selected as the reference object,} the $\delta V$ is halved from $763.3 \pm 259.0 m/s$ to $439.9 \pm 215.1 m/s$. The satellite with the lowest $\delta V$ to Phoebe is S/2007 S2, with $\delta V = 248.0 m/s$, still \added{significantly}larger than the escape velocity of Phoebe ($100.8 m/s$). \deleted{If Phoebe is removed from the family, and Ymir (escape velocity $8.56 m/s$) selected as the reference object, the mean $\delta V$ of the cluster is $439.9 \pm 215.1 m/s$, lower than the $\delta V$ of the Ymir subfamily.} \added{\cite{Turrini2008IrregularSatsSaturn} also found a dynamical separation between Phoebe and the other retrograde satellites.} This is supportive of the narrative \replaced{for a different origin of Phoebe and}{that Phoebe has a different origin to} the other retrograde irregular satellites of Saturn \citep{Turrini2008IrregularSatsSaturn}. The high $\delta V$ among all the subfamilies shows \added{that} a complex dynamical situation is present in the Saturnian irregular satellites. Phoebe has been shown to clear its orbital parameter space \citep{Turrini2008IrregularSatsSaturn}, which could have had a major disruptive effect on those remaining satellites \citep{Turrini2008IrregularSatsSaturn}. \added{The similarities between our analysis and that of \cite{Turrini2008IrregularSatsSaturn} further validates cladistics as a method suitable for applications in Solar system astronomy.} The addition of \replaced{detail}{detailed} compositional information from the other irregular satellites to an updated cladistical analysis could solve some of the \added{minor} discrepancies found between this analysis and that of \cite{Turrini2008IrregularSatsSaturn}.

We assign the currently unnamed irregular satellites to each of the subfamilies. S/2006 S1, S/2007 S2 and S/2007 S3 are part of the Aegir subfamily. We include S/2004 S13, S/2004 S17, S/2004 S12, S/2006 S3 and S/2007 S7 in the Ymir subfamily. See Table \ref{SaturnClassTable} for a full list of members in each subfamily. As with the Albiorix and Siarnaq families, we encourage discoverers of new satellites that fall within the Phoebe family to follow the Norse mythological naming convention as set by the IAU.

\section{Discussion}
\label{Discussion}

In this study we have shown, using the Jovian and Saturnian satellite systems, that cladistics can be used in a planetary science context. We have ensured that the technique is objective, by statistically creating bins for characteristics that are continuous in nature, see Section \ref{Characteristics}. By thus ensuring the objectivity of our analysis, we increase the confidence that cladistics is a valid technique that can be applied in the planetary sciences. Our results largely support the traditional classifications used in both the Jovian and Saturnian systems. However, the power of cladistics is shown in the ease of classifying new satellites as well as identifying substructures within larger clusters. Cladistics also offers a method of analysis where limited information is available. In our study we have examined well studied satellites, as well as those where only dynamical information available. In traditional methods of analysis, either only dynamical information is considered, or the dataset is truncated in favor of more well studied bodies. Cladistics offers a method that can incorporate as much information about an object as is available, while accounting for any unknown characteristics. As more detailed information becomes available, either of known or newly discovered satellites, cladistics offers a systematic method for inclusion or revision of the classification system. 

The relationships that we noted between the satellites suggest common formation scenarios within the clusters. The prograde, inner families of Jupiter are the products of accretion from a circumplanetary disk \citep{Canup2002GalSatAcc}. The association of the Amalthea and Galilean families, along with the Main ring of Jupiter, in our analysis supports this hypothesis. Clustering of the Himalia family with other 'irregular' satellites, \replaced{indicates}{implying} a capture scenario. The prograde nature of the Himalia family is possibly explained via a nebula drag capture mechanism \citep{Cuk2004HimaliaGasDrag}. Further modeling of the Himalia family is required to ascertain their true origins, particularly in light of the Jovian pebble formation hypothesis that may not include an extended nebula \citep{Levison2015GasGiantsPebbles}. 

With the proposal that Sinope forms its own subfamily, each Jovian irregular satellite subfamilies contain only a single large satellite. This strengthens the hypothesis that each of the families represents a capture event and subsequent breakup \citep{Nesvorny2007IrrSatCap} of an object external to the Jovian system. Two of the subfamiles, the Pasiphae and Sinope subfamiles, show a broad range of orbital characteristics and larger dispersal velocities. The other two, the Ananke and Carme subfamiles, show much more constrained characteristics and smaller dispersal velocities. This dichotomy between the two types of subfamiles, broad versus constrained, could \replaced{indicate}{imply} at least two capture events, with the earlier Pasiphae and Sinope families being disrupted by later Ananke and Carme captures. The Iocaste family does not contain a large progenitor satellite, but has high dispersal velocities. This is \replaced{indicative}{suggestive} of a possible ejection scenario. An alternative hypothesis is that the capture events happen simultaneously, but there were multiple disruption events. Both scenarios are supported by the dichotomy in dispersal velocities. Future analysis and simulations into the origins of the irregular satellites could help determine which theory is more probably correct. 

As with the Jovian satellites, there are multiple origins for the origin of the Saturnian rings and satellites. The results from our analysis support a growing body of work showing the complexity of formation scenarios in the Saturnian system. The rings themselves possibly formed after the breakup of an inner icy satellite \citep{Canup2010SaturnSatOrigin}. 

The unresolved nature of the inner Saturnian system shows a complexity of formation scenarios. The main ring satellites, along with the Alkyonides and Tethys Trojans possibly formed via accretion from the current ring system \citep{Charnoz2010SaturnMooletsfromMainRings}. The Alkynoides and Tethys Trojans are then secondarily captured in their current orbits. The major icy satellites, those in the E-ring and outer satellites, probably formed in an accretion scenario, with delivery of the silicate from the outer system \citep{Salmon2017SaturnMidAccretion}. Titan could be secondarily derived from multiple subsatellites that formed in the same disk \citep{Asphaug2013SatMerger}. The volatiles are delivered from comets, with at least one, Phoebe, being captured in orbit. The size of Phoebe is not traditionally associated with comet nuclei, but atleast one comet, C/2002 VQ94 with a similar ~100km diameter has been observed \citep{Korsun2014c2002vq94100kmComet}. The irregular satellite families and subfamiles form from collisional breakup events \citep{Nesvorny2004IrrSatFamilyOrigin} resulting from the captured comet nuclei. The large dispersal velocities of the subfamilies \replaced{indicate}{imply} that this capture and disruption process is complex and requires detailed modeling.

We have shown that cladistics can be used in the classification of the Jovian and Saturnian satellite systems. Consequently, several related studies may be attempted in the future. Uranus and Neptune have similarly complex satellite systems to those of Jupiter and Saturn \citep{Jewitt2007IrregularSats}. These satellite systems could also be classified using cladistics, particularly the irregular satellites. Such a study is hampered by a lack of completeness in the irregular satellite dataset \citep{Sheppard2005UransIrr, Sheppard2006NeptuneIrr}, but may become practical as observational technology improves and the hypothesized small irregular satellites are discovered. Cladistics could be used to further investigate the origins of the irregular satellites of Saturn and Jupiter. As the irregular satellites are thought to be captured bodies \replaced{\citep{Nesvorny2007IrrSatCap}}{\citep[e.g.][]{Nesvorny2007IrrSatCap}}, the question becomes from \replaced{what}{which} small body population they originated. Comparisons between the well studied irregular satellites and other \replaced{solar}{Solar} system bodies could help constrain the origins of these satellites.

\section{Conclusions}
\label{Conclusion}

We have shown that the new application of cladistics on the Jovian and Saturnian satellite systems is valid for investigating the relationships between orbital bodies. In the Jovian system, the traditional classification categories \citep{Nesvorny2003IrrSatEvol,Sheppard2003IrrSatNature,Jewitt2007IrregularSats} are preserved. We support the hypothesis put forward by \cite{Nesvorny2007IrrSatCap} that each Jovian irregular satellite family can be represented by the largest member, and that each family is the remnants of a dynamical capture event, and subsequent breakup. We can also assign recently discovered, as yet unnamed, satellites to each of their respective Jovian families. Cladistical analysis of the Saturnian system broadly preserves the traditional classifications \citep{Nesvorny2003IrrSatEvol, Sheppard2003IrrSatNature, Jewitt2007IrregularSats,Turrini2008IrregularSatsSaturn}, strengthening the validity of the cladistical method. In the Phoebe family of retrograde, irregular satellites, we assign two subfamilies\added{similar to those found by \citep{Turrini2008IrregularSatsSaturn}}. We rename the classical mythological designations for the Saturnian irregular satellites, to represent the largest member of the subfamily, in order to be consistent with the Jovian naming convention. Newly discovered, unnamed Saturnian satellites are easily assigned to various subfamiles. Through the application of the technique to the Jovian and Saturnian systems, we show that cladistics can be used as a valuable tool in a planetary science context, providing a systematic method for future classification.

\acknowledgments
This research was in part supported by the University of Southern Queensland's Strategic Research Initiative program. We wish to thank an anonymous reviewer for \replaced{their}{his/her} comments, particularly on Multivariate Hierarchical Clustering. The AAS Statistics Reviewer provided valuable feedback on the methodology. Dr. Guido Grimm assisted with the cladistical methodology and terminology used in this paper. Dr. Pablo Goloboff provided assistance with TNT, which is subsidized by the Willi Hennig Society, as well as additional comments on the methodology. We would like to thank \added{Dr.} Henry Throop for discussions regarding the Ring systems.

\software{
Mesquite 3.10 \citep{Mesquite}, 
Python 3.5, 
Spyder 2.3.8 \citep{Spyder238}, 
Anaconda Python distribution package 2.40 \citep{Anaconda240},
\added{pandas Python package \citep{Mckinney2010Pandas},
ScyPy Python package \citep{Jones2010SciPy},}
TexMaker 4.1.1, 
Tree analysis using New Technology (TNT) 1.5 \citep{Goloboff2008TNT, Golboff2016TNT15}.
Zephyr 1.1: Mesquite package \citep{MesquiteZephyr}.
}

\bibliographystyle{aasjournal} 
\bibliography{Holtetal2018CladisticsJupiterSaturn-ReSub2}

\begin{thebibliography}{}
\expandafter\ifx\csname natexlab\endcsname\relax\def\natexlab#1{#1}\fi
\providecommand{\url}[1]{\href{#1}{#1}}

\bibitem[{Archie(1989)}]{Archie1989homoplasy}
Archie, J.~W. 1989, SystZoo, 38, 253

\bibitem[{Aria \& Caron(2017)}]{Aria2017burgess}
Aria, C., \& Caron, J.-B. 2017, \nat, 545, 89

\bibitem[{{Asphaug} \& {Reufer}(2013)}]{Asphaug2013SatMerger}
{Asphaug}, E., \& {Reufer}, A. 2013, \icarus, 223, 544

\bibitem[{Bakker \& Galton(1974)}]{Bakker1974dinosaur}
Bakker, R.~T., \& Galton, P.~M. 1974, \nat, 248, 168

\bibitem[{{Barnard}(1892)}]{Barnard1892Amalthea}
{Barnard}, E.~E. 1892, \aj, 12, 81

\bibitem[{{Baum} {et~al.}(1981){Baum}, {Kreidl}, {Westphal}, {Danielson},
  {Seidelmann}, {Pascu}, \& {Currie}}]{Baum1981SatEring}
{Baum}, W.~A., {Kreidl}, T., {Westphal}, J.~A., {et~al.} 1981, \icarus, 47, 84

\bibitem[{{Beaug{\'e}} \& {Nesvorn{\'y}}(2007)}]{Beauge2007IrregSatRsonance}
{Beaug{\'e}}, C., \& {Nesvorn{\'y}}, D. 2007, \aj, 133, 2537

\bibitem[{Blunck(2010)}]{Blunck2010SaturnSats}
Blunck, J. 2010, The Satellites of Saturn (Berlin, Heidelberg: Springer Berlin
  Heidelberg), 53--90

\bibitem[{Brandley {et~al.}(2009)Brandley, Warren, Leaché, \&
  McGuire}]{Brandley2009Homoplasy}
Brandley, M.~C., Warren, D.~L., Leaché, A.~D., \& McGuire, J.~A. 2009,
  SystBio, 58, 184

\bibitem[{{Brooks} {et~al.}(2004){Brooks}, {Esposito}, {Showalter}, \&
  {Throop}}]{Brooks2004JupiterRing}
{Brooks}, S.~M., {Esposito}, L.~W., {Showalter}, M.~R., \& {Throop}, H.~B.
  2004, \icarus, 170, 35

\bibitem[{{Brown}(2014)}]{Brown2014Rayleigh}
{Brown}, A.~J. 2014, \icarus, 239, 85

\bibitem[{{Brown} {et~al.}(2003){Brown}, {Baines}, {Bellucci}, {Bibring},
  {Buratti}, {Capaccioni}, {Cerroni}, {Clark}, {Coradini}, {Cruikshank},
  {Drossart}, {Formisano}, {Jaumann}, {Langevin}, {Matson}, {McCord},
  {Mennella}, {Nelson}, {Nicholson}, {Sicardy}, {Sotin}, {Amici},
  {Chamberlain}, {Filacchione}, {Hansen}, {Hibbitts}, \&
  {Showalter}}]{Brown2003CassiniJupiter}
{Brown}, R.~H., {Baines}, K.~H., {Bellucci}, G., {et~al.} 2003, \icarus, 164,
  461

\bibitem[{{Buratti} {et~al.}(2010){Buratti}, {Bauer}, {Hicks}, {Mosher},
  {Filacchione}, {Momary}, {Baines}, {Brown}, {Clark}, \&
  {Nicholson}}]{Buratti2010SatInnerSat}
{Buratti}, B.~J., {Bauer}, J.~M., {Hicks}, M.~D., {et~al.} 2010, \icarus, 206,
  524

\bibitem[{{Burns} {et~al.}(1999){Burns}, {Showalter}, {Hamilton}, {Nicholson},
  {de Pater}, {Ockert-Bell}, \& {Thomas}}]{Burns1999JupiterRingForm}
{Burns}, J.~A., {Showalter}, M.~R., {Hamilton}, D.~P., {et~al.} 1999, Science,
  284, 1146

\bibitem[{{Callegari} \&
  {Yokoyama}(2010)}]{Callegari2010SmallSaturnSatsDynamics}
{Callegari}, N., \& {Yokoyama}, T. 2010, in IAU Symposium, Vol. 263, Icy Bodies
  of the Solar System, ed. J.~A. {Fernandez}, D.~{Lazzaro}, D.~{Prialnik}, \&
  R.~{Schulz} (Cambridge University Press, Cambridge, UK), 161--166

\bibitem[{{Canup}(2010)}]{Canup2010SaturnSatOrigin}
{Canup}, R.~M. 2010, \nat, 468, 943

\bibitem[{{Canup} \& {Ward}(2002)}]{Canup2002GalSatAcc}
{Canup}, R.~M., \& {Ward}, W.~R. 2002, \aj, 124, 3404

\bibitem[{{Cardone} \& {Fraix-Burnet}(2013)}]{Cardone2013GRBClads}
{Cardone}, V.~F., \& {Fraix-Burnet}, D. 2013, \mnras, 434, 1930

\bibitem[{{Carruba} {et~al.}(2013){Carruba}, {Domingos}, {Nesvorn{\'y}},
  {Roig}, {Huaman}, \& {Souami}}]{Carruba2013AsteroidFamilies}
{Carruba}, V., {Domingos}, R.~C., {Nesvorn{\'y}}, D., {et~al.} 2013, \mnras,
  433, 2075

\bibitem[{{Cassini}(1673)}]{Cassini1673Sat2Sats}
{Cassini}, G.~D. 1673, RSPT, 8, 5178

\bibitem[{{Cassini}(1686)}]{Cassini1686Sat2Sats}
---. 1686, RSPT, 16, 79

\bibitem[{Chamberlain \& Wood(1987)}]{Chamberlain1987EarlyHominid}
Chamberlain, A., \& Wood, B.~A. 1987, JHuEv, 16, 119

\bibitem[{{Chamberlain} \& {Brown}(2004)}]{Chamberlain2004Himalia}
{Chamberlain}, M.~A., \& {Brown}, R.~H. 2004, \icarus, 172, 163

\bibitem[{{Charnoz} {et~al.}(2010){Charnoz}, {Salmon}, \&
  {Crida}}]{Charnoz2010SaturnMooletsfromMainRings}
{Charnoz}, S., {Salmon}, J., \& {Crida}, A. 2010, \nat, 465, 752

\bibitem[{{Charnoz} {et~al.}(2011){Charnoz}, {Crida}, {Castillo-Rogez},
  {Lainey}, {Dones}, {Karatekin}, {Tobie}, {Mathis}, {Le Poncin-Lafitte}, \&
  {Salmon}}]{Charnoz2011SaturnSatAccretion}
{Charnoz}, S., {Crida}, A., {Castillo-Rogez}, J.~C., {et~al.} 2011, \icarus,
  216, 535

\bibitem[{{Christou}(2005)}]{Christou2005HimaliaScattering}
{Christou}, A.~A. 2005, \icarus, 174, 215

\bibitem[{{Clark} {et~al.}(2005){Clark}, {Brown}, {Jaumann}, {Cruikshank},
  {Nelson}, {Buratti}, {McCord}, {Lunine}, {Baines}, {Bellucci}, {Bibring},
  {Capaccioni}, {Cerroni}, {Coradini}, {Formisano}, {Langevin}, {Matson},
  {Mennella}, {Nicholson}, {Sicardy}, {Sotin}, {Hoefen}, {Curchin}, {Hansen},
  {Hibbitts}, \& {Matz}}]{Clark2005Phoebe}
{Clark}, R.~N., {Brown}, R.~H., {Jaumann}, R., {et~al.} 2005, \nat, 435, 66

\bibitem[{{Clark} {et~al.}(2012){Clark}, {Cruikshank}, {Jaumann}, {Brown},
  {Stephan}, {Dalle Ore}, {Eric Livo}, {Pearson}, {Curchin}, {Hoefen},
  {Buratti}, {Filacchione}, {Baines}, \& {Nicholson}}]{Clark2012VIMSIapetus}
{Clark}, R.~N., {Cruikshank}, D.~P., {Jaumann}, R., {et~al.} 2012, \icarus,
  218, 831

\bibitem[{Cobbett {et~al.}(2007)Cobbett, Wilkinson, Wills, \&
  Sullivan}]{Cobbett2007FossilsCladistics}
Cobbett, A., Wilkinson, M., Wills, M.~A., \& Sullivan, J. 2007, SystBio, 56,
  753

\bibitem[{{Colombo} \& {Franklin}(1971)}]{Colombo1971JupSatsForm}
{Colombo}, G., \& {Franklin}, F.~A. 1971, \icarus, 15, 186

\bibitem[{{Continuum Analytics}(2016)}]{Anaconda240}
{Continuum Analytics}. 2016, Anaconda Software Distribution. Version 2.4.0,
  https://continuum.io, ,

\bibitem[{{Cooper} {et~al.}(2006){Cooper}, {Murray}, {Porco}, \&
  {Spitale}}]{Cooper2006CassiniAmaltheaThebe}
{Cooper}, N.~J., {Murray}, C.~D., {Porco}, C.~C., \& {Spitale}, J.~N. 2006,
  \icarus, 181, 223

\bibitem[{{{\'C}uk} \& {Burns}(2004)}]{Cuk2004HimaliaGasDrag}
{{\'C}uk}, M., \& {Burns}, J.~A. 2004, \icarus, 167, 369

\bibitem[{{Cuzzi} {et~al.}(2014){Cuzzi}, {Whizin}, {Hogan}, {Dobrovolskis},
  {Dones}, {Showalter}, {Colwell}, \& {Scargle}}]{Cuzzi2014FringPromethius}
{Cuzzi}, J.~N., {Whizin}, A.~D., {Hogan}, R.~C., {et~al.} 2014, \icarus, 232,
  157

\bibitem[{{Dalton}(2010)}]{Dalton2010IcyMoonSpec}
{Dalton}, J.~B. 2010, \ssr, 153, 219

\bibitem[{{Dalton} {et~al.}(2010){Dalton}, {Cruikshank}, {Stephan}, {McCord},
  {Coustenis}, {Carlson}, \& {Coradini}}]{Dalton2010IcySatComp}
{Dalton}, J.~B., {Cruikshank}, D.~P., {Stephan}, K., {et~al.} 2010, \ssr, 153,
  113

\bibitem[{Darwin(1859)}]{Darwin1859Origin}
Darwin, C. 1859, {On the Origin of the Species by Natural Selection} (London,
  UK: Murray)

\bibitem[{{Deienno} {et~al.}(2014){Deienno}, {Nesvorn{\'y}},
  {Vokrouhlick{\'y}}, \& {Yokoyama}}]{Deienno2014OrbitGalileanSat}
{Deienno}, R., {Nesvorn{\'y}}, D., {Vokrouhlick{\'y}}, D., \& {Yokoyama}, T.
  2014, \aj, 148, 25

\bibitem[{{El Moutamid} {et~al.}(2016){El Moutamid}, {Nicholson}, {French},
  {Tiscareno}, {Murray}, {Evans}, {French}, {Hedman}, \&
  {Burns}}]{ElMoutamid2016JansSwapAring}
{El Moutamid}, M., {Nicholson}, P.~D., {French}, R.~G., {et~al.} 2016, \icarus,
  279, 125

\bibitem[{{Emelyanov}(2005)}]{Emelyanov2005HimaliaMass}
{Emelyanov}, N.~V. 2005, \aap, 438, L33

\bibitem[{Farris(1970)}]{Farris1970MethodsComp}
Farris, J.~S. 1970, SystBio, 19, 83

\bibitem[{Farris(1989)}]{Farris1989retention}
---. 1989, Cladistics, 5, 417

\bibitem[{Farris(1990)}]{Farris1990phenetics}
---. 1990, Cladistics, 6, 91

\bibitem[{{Feibelman}(1967)}]{Feibelman1967SatEring}
{Feibelman}, W.~A. 1967, \nat, 214, 793

\bibitem[{{Filacchione} {et~al.}(2007){Filacchione}, {Capaccioni}, {McCord},
  {Coradini}, {Cerroni}, {Bellucci}, {Tosi}, {D'Aversa}, {Formisano}, {Brown},
  {Baines}, {Bibring}, {Buratti}, {Clark}, {Combes}, {Cruikshank}, {Drossart},
  {Jaumann}, {Langevin}, {Matson}, {Mennella}, {Nelson}, {Nicholson},
  {Sicardy}, {Sotin}, {Hansen}, {Hibbitts}, {Showalter}, \&
  {Newman}}]{Filacchione2007VIMS1}
{Filacchione}, G., {Capaccioni}, F., {McCord}, T.~B., {et~al.} 2007, \icarus,
  186, 259

\bibitem[{{Filacchione} {et~al.}(2010){Filacchione}, {Capaccioni}, {Clark},
  {Cuzzi}, {Cruikshank}, {Coradini}, {Cerroni}, {Nicholson}, {McCord}, {Brown},
  {Buratti}, {Tosi}, {Nelson}, {Jaumann}, \& {Stephan}}]{Filacchione2010VIMS2}
{Filacchione}, G., {Capaccioni}, F., {Clark}, R.~N., {et~al.} 2010, \icarus,
  206, 507

\bibitem[{{Filacchione} {et~al.}(2012){Filacchione}, {Capaccioni},
  {Ciarniello}, {Clark}, {Cuzzi}, {Nicholson}, {Cruikshank}, {Hedman},
  {Buratti}, {Lunine}, {Soderblom}, {Tosi}, {Cerroni}, {Brown}, {McCord},
  {Jaumann}, {Stephan}, {Baines}, \& {Flamini}}]{Filacchione2012VIMS3}
{Filacchione}, G., {Capaccioni}, F., {Ciarniello}, M., {et~al.} 2012, \icarus,
  220, 1064

\bibitem[{{Filacchione} {et~al.}(2014){Filacchione}, {Ciarniello},
  {Capaccioni}, {Clark}, {Nicholson}, {Hedman}, {Cuzzi}, {Cruikshank}, {Dalle
  Ore}, {Brown}, {Cerroni}, {Altobelli}, \&
  {Spilker}}]{Filacchione2014VIMSrings}
{Filacchione}, G., {Ciarniello}, M., {Capaccioni}, F., {et~al.} 2014, \icarus,
  241, 45

\bibitem[{{Filacchione} {et~al.}(2016){Filacchione}, {D'Aversa}, {Capaccioni},
  {Clark}, {Cruikshank}, {Ciarniello}, {Cerroni}, {Bellucci}, {Brown},
  {Buratti}, {Nicholson}, {Jaumann}, {McCord}, {Sotin}, {Stephan}, \& {Dalle
  Ore}}]{Filacchione2016VIMS4}
{Filacchione}, G., {D'Aversa}, E., {Capaccioni}, F., {et~al.} 2016, \icarus,
  271, 292

\bibitem[{{Fraix-Burnet} {et~al.}(2012){Fraix-Burnet}, {Chattopadhyay},
  {Chattopadhyay}, {Davoust}, \& {Thuillard}}]{FraixBurnet2012SixPermGal}
{Fraix-Burnet}, D., {Chattopadhyay}, T., {Chattopadhyay}, A.~K., {Davoust}, E.,
  \& {Thuillard}, M. 2012, \aap, 545, A80

\bibitem[{{Fraix-Burnet} {et~al.}(2006){Fraix-Burnet}, {Choler}, \&
  {Douzery}}]{FraixBurnet2006DwarfGalaxies}
{Fraix-Burnet}, D., {Choler}, P., \& {Douzery}, E.~J.~P. 2006, \aap, 455, 845

\bibitem[{{Fraix-Burnet} \& {Davoust}(2015)}]{FraixBurnet2015StarClads}
{Fraix-Burnet}, D., \& {Davoust}, E. 2015, \mnras, 450, 3431

\bibitem[{{Fraix-Burnet} {et~al.}(2009){Fraix-Burnet}, {Davoust}, \&
  {Charbonnel}}]{FraixBurnet2009GlobularClusters}
{Fraix-Burnet}, D., {Davoust}, E., \& {Charbonnel}, C. 2009, \mnras, 398, 1706

\bibitem[{{Fraix-Burnet} {et~al.}(2010){Fraix-Burnet}, {Dugu{\'e}},
  {Chattopadhyay}, {Chattopadhyay}, \& {Davoust}}]{FraixBurnet2010EarlyGalx}
{Fraix-Burnet}, D., {Dugu{\'e}}, M., {Chattopadhyay}, T., {Chattopadhyay},
  A.~K., \& {Davoust}, E. 2010, \mnras, 407, 2207

\bibitem[{{Fraix-Burnet} {et~al.}(2015){Fraix-Burnet}, {Thuillard}, \&
  {Chattopadhyay}}]{FraixBurnet2015GalClad}
{Fraix-Burnet}, D., {Thuillard}, M., \& {Chattopadhyay}, A.~K. 2015, FrASS, 2,
  3

\bibitem[{Galilei(1610)}]{Galileo1610SidereusNuncius}
Galilei, G. 1610, Sidereus nuncius magna, longeque admirabilia spectacula
  pandens (Tommaso Baglioni, Venice)

\bibitem[{Gascuel(2005)}]{Gascuel2005MathEvolPhylogeny}
Gascuel, O. 2005, Mathematics of Evolution and Phylogeny. (OUP Oxford)

\bibitem[{{Giese} {et~al.}(2006){Giese}, {Neukum}, {Roatsch}, {Denk}, \&
  {Porco}}]{Giese2006PhoebeTopo}
{Giese}, B., {Neukum}, G., {Roatsch}, T., {Denk}, T., \& {Porco}, C.~C. 2006,
  \planss, 54, 1156

\bibitem[{{Gillon} {et~al.}(2016){Gillon}, {Jehin}, {Lederer}, {Delrez}, {de
  Wit}, {Burdanov}, {Van Grootel}, {Burgasser}, {Triaud}, {Opitom}, {Demory},
  {Sahu}, {Bardalez Gagliuffi}, {Magain}, \& {Queloz}}]{Gillon2016Trapist1}
{Gillon}, M., {Jehin}, E., {Lederer}, S.~M., {et~al.} 2016, \nat, 533, 221

\bibitem[{Givnish \& Sytsma(1997)}]{Givnish1997consistency}
Givnish, T., \& Sytsma, K. 1997, MolPhylogentEv, 7, 320

\bibitem[{{Gladman} {et~al.}(2003{\natexlab{a}}){Gladman}, {Sheppard}, \&
  {Marsden}}]{Gladman2003JupIAU1}
{Gladman}, B., {Sheppard}, S.~S., \& {Marsden}, B.~G. 2003{\natexlab{a}},
  \iaucirc, 8125

\bibitem[{{Gladman} {et~al.}(2003{\natexlab{b}}){Gladman}, {Sheppard}, \&
  {Marsden}}]{Gladman2003JupIAU2}
---. 2003{\natexlab{b}}, \iaucirc, 8138

\bibitem[{{Gladman} {et~al.}(2001){Gladman}, {Kavelaars}, {Holman},
  {Nicholson}, {Burns}, {Hergenrother}, {Petit}, {Marsden}, {Jacobson}, {Gray},
  \& {Grav}}]{Gladman200112Sat}
{Gladman}, B., {Kavelaars}, J.~J., {Holman}, M., {et~al.} 2001, \nat, 412, 163

\bibitem[{Goloboff(1994)}]{Goloboff1994Treelengths}
Goloboff, P.~A. 1994, Cladistics, 9, 433

\bibitem[{Goloboff(1996)}]{Goloboff1996FastPasrimony}
---. 1996, Cladistics, 12, 199

\bibitem[{Goloboff(2015)}]{Goloboff2015Parsimony}
---. 2015, Cladistics, 31, 210

\bibitem[{Goloboff \& Catalano(2016)}]{Golboff2016TNT15}
Goloboff, P.~A., \& Catalano, S.~A. 2016, Cladistics, 32, 221

\bibitem[{Goloboff {et~al.}(2008)Goloboff, Farris, \& Nixon}]{Goloboff2008TNT}
Goloboff, P.~A., Farris, J.~S., \& Nixon, K.~C. 2008, Cladistics, 24, 774

\bibitem[{{Grav} \& {Bauer}(2007)}]{Grav2007IrregSatCol}
{Grav}, T., \& {Bauer}, J. 2007, \icarus, 191, 267

\bibitem[{{Grav} {et~al.}(2015){Grav}, {Bauer}, {Mainzer}, {Masiero}, {Nugent},
  {Cutri}, {Sonnett}, \& {Kramer}}]{Grav2015NEOWISEIrregulars}
{Grav}, T., {Bauer}, J.~M., {Mainzer}, A.~K., {et~al.} 2015, \apj, 809, 3

\bibitem[{{Grav} {et~al.}(2003){Grav}, {Holman}, {Gladman}, \&
  {Aksnes}}]{Grav2003IrregSatPhoto}
{Grav}, T., {Holman}, M.~J., {Gladman}, B.~J., \& {Aksnes}, K. 2003, \icarus,
  166, 33

\bibitem[{{Greenberg}(2010)}]{Greenberg2010IcyJovian}
{Greenberg}, R. 2010, RPPh, 73, 036801

\bibitem[{{Grundy} {et~al.}(2007){Grundy}, {Buratti}, {Cheng}, {Emery},
  {Lunsford}, {McKinnon}, {Moore}, {Newman}, {Olkin}, {Reuter}, {Schenk},
  {Spencer}, {Stern}, {Throop}, \& {Weaver}}]{Grundy2007NewHorizonsJupiterSats}
{Grundy}, W.~M., {Buratti}, B.~J., {Cheng}, A.~F., {et~al.} 2007, Science, 318,
  234

\bibitem[{Hamilton(2014)}]{Hamilton2014EvolSystem}
Hamilton, A. 2014, The Evolution of Phylogenetic Systematics., Species and
  Systematics No. volume 5 (University of California Press)

\bibitem[{{Hedman} {et~al.}(2012){Hedman}, {Burns}, {Hamilton}, \&
  {Showalter}}]{Hedman2012EringStruc}
{Hedman}, M.~M., {Burns}, J.~A., {Hamilton}, D.~P., \& {Showalter}, M.~R. 2012,
  \icarus, 217, 322

\bibitem[{{Hedman} {et~al.}(2010){Hedman}, {Cooper}, {Murray}, {Beurle},
  {Evans}, {Tiscareno}, \& {Burns}}]{Hedman2010Aegaeon}
{Hedman}, M.~M., {Cooper}, N.~J., {Murray}, C.~D., {et~al.} 2010, \icarus, 207,
  433

\bibitem[{{Hedman} {et~al.}(2009){Hedman}, {Murray}, {Cooper}, {Tiscareno},
  {Beurle}, {Evans}, \& {Burns}}]{Hedman2009SatRingArcs}
{Hedman}, M.~M., {Murray}, C.~D., {Cooper}, N.~J., {et~al.} 2009, \icarus, 199,
  378

\bibitem[{{Hedman} {et~al.}(2007{\natexlab{a}}){Hedman}, {Burns}, {Tiscareno},
  {Porco}, {Jones}, {Roussos}, {Krupp}, {Paranicas}, \&
  {Kempf}}]{Hedman2007Gring}
{Hedman}, M.~M., {Burns}, J.~A., {Tiscareno}, M.~S., {et~al.}
  2007{\natexlab{a}}, Science, 317, 653

\bibitem[{{Hedman} {et~al.}(2007{\natexlab{b}}){Hedman}, {Burns}, {Showalter},
  {Porco}, {Nicholson}, {Bosh}, {Tiscareno}, {Brown}, {Buratti}, {Baines}, \&
  {Clark}}]{Hedman2007Dring}
{Hedman}, M.~M., {Burns}, J.~A., {Showalter}, M.~R., {et~al.}
  2007{\natexlab{b}}, \icarus, 188, 89

\bibitem[{{Hemingway} {et~al.}(2013){Hemingway}, {Nimmo}, {Zebker}, \&
  {Iess}}]{Hemingway2013TitanIceshell}
{Hemingway}, D., {Nimmo}, F., {Zebker}, H., \& {Iess}, L. 2013, \nat, 500, 550

\bibitem[{Hennig(1965)}]{Hennig1965PhylogeneticSystem}
Hennig, W. 1965, AREntomol, 10, 97

\bibitem[{{Heppenheimer} \& {Porco}(1977)}]{Heppenheimer1977Capture}
{Heppenheimer}, T.~A., \& {Porco}, C. 1977, \icarus, 30, 385

\bibitem[{{Hillier} {et~al.}(2007){Hillier}, {Green}, {McBride}, {Schwanethal},
  {Postberg}, {Srama}, {Kempf}, {Moragas-Klostermeyer}, {McDonnell}, \&
  {Gr{\"u}n}}]{Hillier2007EringComp}
{Hillier}, J.~K., {Green}, S.~F., {McBride}, N., {et~al.} 2007, \mnras, 377,
  1588

\bibitem[{{Hirtzig} {et~al.}(2013){Hirtzig}, {B{\'e}zard}, {Lellouch},
  {Coustenis}, {de Bergh}, {Drossart}, {Campargue}, {Boudon}, {Tyuterev},
  {Rannou}, {Cours}, {Kassi}, {Nikitin}, {Mondelain}, {Rodriguez}, \& {Le
  Mou{\'e}lic}}]{Hirtzig2013VIMSTitan}
{Hirtzig}, M., {B{\'e}zard}, B., {Lellouch}, E., {et~al.} 2013, \icarus, 226,
  470

\bibitem[{{Holt} {et~al.}(2016){Holt}, {Brown}, \&
  {Nesvorny}}]{Holt2016JovSatCald}
{Holt}, T.~R., {Brown}, A.~J., \& {Nesvorny}, D. 2016, in Lunar and Planetary
  Science Conference (The Woodlands, TX: Lunar and Planetary Institute),
  Vol.~47, Lunar and Planetary Science Conference (The Woodlands, TX: Lunar and
  Planetary Institute), 2676

\bibitem[{Hug {et~al.}(2016)Hug, Baker, Anantharaman, Brown, Probst, Castelle,
  Butterfield, Hernsdorf, Amano, Ise, {et~al.}}]{Hug2016TreeLife}
Hug, L.~A., Baker, B.~J., Anantharaman, K., {et~al.} 2016, NaMic, 1, 16048

\bibitem[{{Hussmann} {et~al.}(2006){Hussmann}, {Sohl}, \&
  {Spohn}}]{Haussmann2006SatTNO}
{Hussmann}, H., {Sohl}, F., \& {Spohn}, T. 2006, \icarus, 185, 258

\bibitem[{Huygens(1659)}]{Huygens1659systema}
Huygens, C. 1659, Systema saturnium (Ex typographia Adriani Vlacq)

\bibitem[{{Jacobson} {et~al.}(2011){Jacobson}, {Brozovic}, {Gladman},
  {Alexandersen}, {Veillet}, \& {Williams}}]{Jacobson2011NewJupSats}
{Jacobson}, R., {Brozovic}, M., {Gladman}, B., {et~al.} 2011, \iaucirc, 9222

\bibitem[{{Jewitt} \& {Haghighipour}(2007)}]{Jewitt2007IrregularSats}
{Jewitt}, D., \& {Haghighipour}, N. 2007, \araa, 45, 261

\bibitem[{{Jewitt} {et~al.}(2005){Jewitt}, {Sheppard}, {Kleyna}, \&
  {Marsden}}]{Jewitt2005IAUCSat}
{Jewitt}, D., {Sheppard}, S., {Kleyna}, J., \& {Marsden}, B.~G. 2005, \iaucirc,
  8523

\bibitem[{{Jewitt} {et~al.}(1979){Jewitt}, {Danielson}, \&
  {Synnott}}]{Jewitt1979AdrasteaDiscov}
{Jewitt}, D.~C., {Danielson}, G.~E., \& {Synnott}, S.~P. 1979, Science, 206,
  951

\bibitem[{{Jofr{\'e}} {et~al.}(2017){Jofr{\'e}}, {Das}, {Bertranpetit}, \&
  {Foley}}]{Jofre2017StarsClads}
{Jofr{\'e}}, P., {Das}, P., {Bertranpetit}, J., \& {Foley}, R. 2017, \mnras,
  467, 1140

\bibitem[{{Johnson} \& {Lunine}(2005)}]{Johnson2005PhoebeKuiper}
{Johnson}, T.~V., \& {Lunine}, J.~I. 2005, \nat, 435, 69

\bibitem[{Jones {et~al.}(2001)Jones, Oliphant, Peterson,
  {et~al.}}]{Jones2010SciPy}
Jones, E., Oliphant, T., Peterson, P., {et~al.} 2001, {SciPy}: Open source
  scientific tools for {Python}, , .
\newblock \url{"http://www.scipy.org/"}

\bibitem[{{Karkoschka}(1994)}]{Karkoschka1994TitanESO}
{Karkoschka}, E. 1994, \icarus, 111, 174

\bibitem[{Kluge \& Farris(1969)}]{Kluge1969Cladistics}
Kluge, A.~G., \& Farris, J.~S. 1969, SystZoo, 18, 1

\bibitem[{{Korsun} {et~al.}(2014){Korsun}, {Rousselot}, {Kulyk}, {Afanasiev},
  \& {Ivanova}}]{Korsun2014c2002vq94100kmComet}
{Korsun}, P.~P., {Rousselot}, P., {Kulyk}, I.~V., {Afanasiev}, V.~L., \&
  {Ivanova}, O.~V. 2014, \icarus, 232, 88

\bibitem[{{Kowal} {et~al.}(1975{\natexlab{a}}){Kowal}, {Roemer}, {Daniel},
  {McCarthy}, {Aksnes}, \& {Marsden}}]{Kowal1975Themisto}
{Kowal}, C., {Roemer}, E., {Daniel}, M.~A., {et~al.} 1975{\natexlab{a}},
  \iaucirc, 2855

\bibitem[{{Kowal} {et~al.}(1975{\natexlab{b}}){Kowal}, {Aksnes}, {Marsden}, \&
  {Roemer}}]{Kowal1975Leda}
{Kowal}, C.~T., {Aksnes}, K., {Marsden}, B.~G., \& {Roemer}, E.
  1975{\natexlab{b}}, \aj, 80, 460

\bibitem[{{Kr{\"u}ger} {et~al.}(2009){Kr{\"u}ger}, {Hamilton}, {Moissl}, \&
  {Gr{\"u}n}}]{Kruger2009JupiterRingIntitu}
{Kr{\"u}ger}, H., {Hamilton}, D.~P., {Moissl}, R., \& {Gr{\"u}n}, E. 2009,
  \icarus, 203, 198

\bibitem[{{Kuiper}(1944)}]{Kuiper1944TitanAtmos}
{Kuiper}, G.~P. 1944, \apj, 100, 378

\bibitem[{{Lebreton} {et~al.}(2005){Lebreton}, {Witasse}, {Sollazzo},
  {Blancquaert}, {Couzin}, {Schipper}, {Jones}, {Matson}, {Gurvits},
  {Atkinson}, {Kazeminejad}, \&
  {P{\'e}rez-Ay{\'u}car}}]{Lebreton2005HuygensTiten}
{Lebreton}, J.-P., {Witasse}, O., {Sollazzo}, C., {et~al.} 2005, \nat, 438, 758

\bibitem[{{Levison} {et~al.}(2015){Levison}, {Kretke}, \&
  {Duncan}}]{Levison2015GasGiantsPebbles}
{Levison}, H.~F., {Kretke}, K.~A., \& {Duncan}, M.~J. 2015, \nat, 524, 322

\bibitem[{{Lissauer}(1987)}]{Lissauer1987PlanetAccretion}
{Lissauer}, J.~J. 1987, \icarus, 69, 249

\bibitem[{{Lodders}(2003)}]{Lodders2003Abundances}
{Lodders}, K. 2003, \apj, 591, 1220

\bibitem[{{Lykawka} \& {Horner}(2010)}]{Lykawka2010TrojanCaputer}
{Lykawka}, P.~S., \& {Horner}, J. 2010, \mnras, 405, 1375

\bibitem[{Maddison {et~al.}(1984)Maddison, Donoghue, \&
  Maddison}]{Maddison1984outgroup}
Maddison, W.~P., Donoghue, M.~J., \& Maddison, D.~R. 1984, SystBio, 33, 83

\bibitem[{{Maddison} \& {Maddison}(2015)}]{MesquiteZephyr}
{Maddison}, W.~P., \& {Maddison}, D.~R. 2015, Zephyr: a Mesquite package for
  interacting with external phylogeny inference programs. Version 1.1,
  https://mesquitezephyr.wikispaces.com, ,

\bibitem[{{Maddison} \& {Maddison}(2017)}]{Mesquite}
---. 2017, Mesquite: a modular system for evolutionary analysis. Version 3.20,
  http://mesquiteproject.org, ,

\bibitem[{Margush \& McMorris(1981)}]{Margush1981MajorityRules}
Margush, T., \& McMorris, F.~R. 1981, BuMaBio, 43, 239

\bibitem[{{Matson} {et~al.}(2009){Matson}, {Castillo-Rogez}, {Schubert},
  {Sotin}, \& {McKinnon}}]{Matson2009SaturnSat}
{Matson}, D.~L., {Castillo-Rogez}, J.~C., {Schubert}, G., {Sotin}, C., \&
  {McKinnon}, W.~B. 2009, {The Thermal Evolution and Internal Structure of
  Saturn's Mid-Sized Icy Satellites}, ed. M.~K. {Dougherty}, L.~W. {Esposito},
  \& S.~M. {Krimigis} (Springer Science \& Business Media, Netherlands), 577

\bibitem[{McKinney(2010)}]{Mckinney2010Pandas}
McKinney, W. 2010, in Proceedings of the 9th Python in Science Conference
  (Austin, TX: SciPy), Vol. 445, Proceedings of the 9th Python in Science
  Conference (Austin, TX: SciPy), 51--56

\bibitem[{{Melotte} \& {Perrine}(1908)}]{Melotte1908Pasiphae}
{Melotte}, J., \& {Perrine}, C.~D. 1908, \pasp, 20, 184

\bibitem[{{Milani} {et~al.}(2014){Milani}, {Cellino}, {Kne{\v z}evi{\'c}},
  {Novakovi{\'c}}, {Spoto}, \& {Paolicchi}}]{Milani2014AsteroidFamilies}
{Milani}, A., {Cellino}, A., {Kne{\v z}evi{\'c}}, Z., {et~al.} 2014, \icarus,
  239, 46

\bibitem[{Mitchell(1901)}]{Mitchell1901BridsClads}
Mitchell, P.~C. 1901, TLinnSL, 8, 173

\bibitem[{{Morbidelli} {et~al.}(2005){Morbidelli}, {Levison}, {Tsiganis}, \&
  {Gomes}}]{Morbidelli2005TrojanCapture}
{Morbidelli}, A., {Levison}, H.~F., {Tsiganis}, K., \& {Gomes}, R. 2005, \nat,
  435, 462

\bibitem[{Naylor \& Kraus(1995)}]{Naylor1995RetentionIndex}
Naylor, G., \& Kraus, F. 1995, SystBio, 44, 559

\bibitem[{{Nesvorn{\'y}} {et~al.}(2003){Nesvorn{\'y}}, {Alvarellos}, {Dones},
  \& {Levison}}]{Nesvorny2003IrrSatEvol}
{Nesvorn{\'y}}, D., {Alvarellos}, J.~L.~A., {Dones}, L., \& {Levison}, H.~F.
  2003, \aj, 126, 398

\bibitem[{{Nesvorn{\'y}} {et~al.}(2004){Nesvorn{\'y}}, {Beaug{\'e}}, \&
  {Dones}}]{Nesvorny2004IrrSatFamilyOrigin}
{Nesvorn{\'y}}, D., {Beaug{\'e}}, C., \& {Dones}, L. 2004, \aj, 127, 1768

\bibitem[{{Nesvorn{\'y}} {et~al.}(2002){Nesvorn{\'y}}, {Bottke}, {Dones}, \&
  {Levison}}]{Nesvorny2002AsteroidBreakup}
{Nesvorn{\'y}}, D., {Bottke}, Jr., W.~F., {Dones}, L., \& {Levison}, H.~F.
  2002, \nat, 417, 720

\bibitem[{{Nesvorn{\'y}} \& {Morbidelli}(2012)}]{Nesvorny2012JumpingJupiter}
{Nesvorn{\'y}}, D., \& {Morbidelli}, A. 2012, \aj, 144, 117

\bibitem[{{Nesvorn{\'y}} {et~al.}(2014){Nesvorn{\'y}}, {Vokrouhlick{\'y}}, \&
  {Deienno}}]{Nesvorny2014IrrCapture}
{Nesvorn{\'y}}, D., {Vokrouhlick{\'y}}, D., \& {Deienno}, R. 2014, \apj, 784,
  22

\bibitem[{{Nesvorn{\'y}} {et~al.}(2007){Nesvorn{\'y}}, {Vokrouhlick{\'y}}, \&
  {Morbidelli}}]{Nesvorny2007IrrSatCap}
{Nesvorn{\'y}}, D., {Vokrouhlick{\'y}}, D., \& {Morbidelli}, A. 2007, \aj, 133,
  1962

\bibitem[{{Nesvorn{\'y}} {et~al.}(2013){Nesvorn{\'y}}, {Vokrouhlick{\'y}}, \&
  {Morbidelli}}]{Nesvorny2013TrojanCaptureJJ}
---. 2013, \apj, 768, 45

\bibitem[{{Nicholson} {et~al.}(1992){Nicholson}, {Hamilton}, {Matthews}, \&
  {Yoder}}]{Nicholson1992CoorbitalSaturn}
{Nicholson}, P.~D., {Hamilton}, D.~P., {Matthews}, K., \& {Yoder}, C.~F. 1992,
  \icarus, 100, 464

\bibitem[{{Nicholson} {et~al.}(2008){Nicholson}, {Hedman}, {Clark},
  {Showalter}, {Cruikshank}, {Cuzzi}, {Filacchione}, {Capaccioni}, {Cerroni},
  {Hansen}, {Sicardy}, {Drossart}, {Brown}, {Buratti}, {Baines}, \&
  {Coradini}}]{Nicholson2008VIMSRings}
{Nicholson}, P.~D., {Hedman}, M.~M., {Clark}, R.~N., {et~al.} 2008, \icarus,
  193, 182

\bibitem[{{Nicholson}(1914)}]{Nicholson1914Sinope}
{Nicholson}, S.~B. 1914, \pasp, 26, 197

\bibitem[{{Nicholson}(1938)}]{Nicholson1938LysitheaCarme}
---. 1938, \pasp, 50, 292

\bibitem[{{Nicholson}(1951)}]{Nicholson1951Ananke}
---. 1951, \pasp, 63, 297

\bibitem[{{Niemann} {et~al.}(2005){Niemann}, {Atreya}, {Bauer}, {Carignan},
  {Demick}, {Frost}, {Gautier}, {Haberman}, {Harpold}, {Hunten}, {Israel},
  {Lunine}, {Kasprzak}, {Owen}, {Paulkovich}, {Raulin}, {Raaen}, \&
  {Way}}]{Niemann2005TitanAtmos}
{Niemann}, H.~B., {Atreya}, S.~K., {Bauer}, S.~J., {et~al.} 2005, \nat, 438,
  779

\bibitem[{{Ockert-Bell} {et~al.}(1999){Ockert-Bell}, {Burns}, {Daubar},
  {Thomas}, {Veverka}, {Belton}, \& {Klaasen}}]{OckertBell1999JupiterRing}
{Ockert-Bell}, M.~E., {Burns}, J.~A., {Daubar}, I.~J., {et~al.} 1999, \icarus,
  138, 188

\bibitem[{Olsen {et~al.}(1994)Olsen, Woese, \& Overbeek}]{Olsen1994winds}
Olsen, G.~J., Woese, C.~R., \& Overbeek, R. 1994, JBact, 176, 1

\bibitem[{{Parker} {et~al.}(2008){Parker}, {Ivezi{\'c}}, {Juri{\'c}}, {Lupton},
  {Sekora}, \& {Kowalski}}]{Parker2008AsteroidFamSDSS}
{Parker}, A., {Ivezi{\'c}}, {\v Z}., {Juri{\'c}}, M., {et~al.} 2008, \icarus,
  198, 138

\bibitem[{Perrine(1905)}]{Perrine1905Himalia}
Perrine, C.~D. 1905, \pasp, 17, 62.
\newblock \url{http://stacks.iop.org/1538-3873/17/i=101/a=56}

\bibitem[{Perrine \& Aitken(1905)}]{Perrine1905Elara}
Perrine, C.~D., \& Aitken, R.~G. 1905, \pasp, 17, 62.
\newblock \url{http://stacks.iop.org/1538-3873/17/i=101/a=56}

\bibitem[{{Pickering}(1899)}]{Pickering1899Phoebe}
{Pickering}, E.~C. 1899, \apj, 9, 274

\bibitem[{{Pickering}(1905)}]{Pickering1905Phoebe}
{Pickering}, W.~H. 1905, AnHvaCOB, 53, 85

\bibitem[{{Pollack} {et~al.}(1979){Pollack}, {Burns}, \&
  {Tauber}}]{Pollack1979SatGasDrag}
{Pollack}, J.~B., {Burns}, J.~A., \& {Tauber}, M.~E. 1979, \icarus, 37, 587

\bibitem[{{Pollack} {et~al.}(1996){Pollack}, {Hubickyj}, {Bodenheimer},
  {Lissauer}, {Podolak}, \& {Greenzweig}}]{Pollack1996GiantPlanetAccretion}
{Pollack}, J.~B., {Hubickyj}, O., {Bodenheimer}, P., {et~al.} 1996, \icarus,
  124, 62

\bibitem[{{Porco} {et~al.}(2007){Porco}, {Thomas}, {Weiss}, \&
  {Richardson}}]{Porco2007SaturnSmallSats}
{Porco}, C.~C., {Thomas}, P.~C., {Weiss}, J.~W., \& {Richardson}, D.~C. 2007,
  Science, 318, 1602

\bibitem[{{Porco} {et~al.}(2005){Porco}, {Baker}, {Barbara}, {Beurle},
  {Brahic}, {Burns}, {Charnoz}, {Cooper}, {Dawson}, {Del Genio}, {Denk},
  {Dones}, {Dyudina}, {Evans}, {Giese}, {Grazier}, {Helfenstein}, {Ingersoll},
  {Jacobson}, {Johnson}, {McEwen}, {Murray}, {Neukum}, {Owen}, {Perry},
  {Roatsch}, {Spitale}, {Squyres}, {Thomas}, {Tiscareno}, {Turtle}, {Vasavada},
  {Veverka}, {Wagner}, \& {West}}]{Porco2005CassiniRingSat}
{Porco}, C.~C., {Baker}, E., {Barbara}, J., {et~al.} 2005, Science, 307, 1226

\bibitem[{{Porco} {et~al.}(2006){Porco}, {Helfenstein}, {Thomas}, {Ingersoll},
  {Wisdom}, {West}, {Neukum}, {Denk}, {Wagner}, {Roatsch}, {Kieffer}, {Turtle},
  {McEwen}, {Johnson}, {Rathbun}, {Veverka}, {Wilson}, {Perry}, {Spitale},
  {Brahic}, {Burns}, {Del Genio}, {Dones}, {Murray}, \&
  {Squyres}}]{Porco2006EnceladusPlume}
{Porco}, C.~C., {Helfenstein}, P., {Thomas}, P.~C., {et~al.} 2006, Science,
  311, 1393

\bibitem[{{Rettig} {et~al.}(2001){Rettig}, {Walsh}, \&
  {Consolmagno}}]{Rettig2001JovianIrrBVR}
{Rettig}, T.~W., {Walsh}, K., \& {Consolmagno}, G. 2001, \icarus, 154, 313

\bibitem[{{\v{R}}{\'\i}{\v{c}}an {et~al.}(2011){\v{R}}{\'\i}{\v{c}}an, Pialek,
  Almiron, \& Casciotta}]{Vrivcan2011two}
{\v{R}}{\'\i}{\v{c}}an, O., Pialek, L., Almiron, A., \& Casciotta, J. 2011,
  Zootaxa, 2982, 1.
\newblock \url{http://www.mapress.com/j/zt/article/view/11638}

\bibitem[{{Ross}(1905)}]{Ross1905Phoebe}
{Ross}, F.~E. 1905, AnHvaCOB, 53, 101

\bibitem[{Salisbury {et~al.}(2006)Salisbury, Molnar, Frey, \&
  Willis}]{Salisbury2006originCrocs}
Salisbury, S.~W., Molnar, R.~E., Frey, E., \& Willis, P.~M. 2006, RSPSB, 273,
  2439

\bibitem[{{Salmon} \& {Canup}(2017)}]{Salmon2017SaturnMidAccretion}
{Salmon}, J., \& {Canup}, R.~M. 2017, \apj, 836, 109

\bibitem[{Sanderson \& Donoghue(1989)}]{Sanderson1989VaiationHomoplasy}
Sanderson, M.~J., \& Donoghue, M.~J. 1989, Ev, 43, 1781

\bibitem[{{Scotti} {et~al.}(2000){Scotti}, {Spahr}, {McMillan}, {Larsen},
  {Montani}, {Gleason}, {Gehrels}, {Marsden}, \&
  {Williams}}]{Scotti2000Callirrhoe}
{Scotti}, J.~V., {Spahr}, T.~B., {McMillan}, R.~S., {et~al.} 2000, \iaucirc,
  7460

\bibitem[{{Sheppard} {et~al.}(2003{\natexlab{a}}){Sheppard}, {Gladman}, \&
  {Marsden}}]{Sheppard2003IAUJupSat}
{Sheppard}, S.~S., {Gladman}, B., \& {Marsden}, B.~G. 2003{\natexlab{a}},
  \iaucirc, 8116

\bibitem[{{Sheppard} {et~al.}(2004){Sheppard}, {Gladman}, \&
  {Marsden}}]{Sheppard2004JupIAU1}
---. 2004, \iaucirc, 8276

\bibitem[{{Sheppard} {et~al.}(2005){Sheppard}, {Jewitt}, \&
  {Kleyna}}]{Sheppard2005UransIrr}
{Sheppard}, S.~S., {Jewitt}, D., \& {Kleyna}, J. 2005, \aj, 129, 518

\bibitem[{{Sheppard} {et~al.}(2006{\natexlab{a}}){Sheppard}, {Jewitt}, \&
  {Kleyna}}]{Sheppard2006NeptuneIrr}
---. 2006{\natexlab{a}}, \aj, 132, 171

\bibitem[{{Sheppard} \& {Jewitt}(2003)}]{Sheppard2003IrrSatNature}
{Sheppard}, S.~S., \& {Jewitt}, D.~C. 2003, \nat, 423, 261

\bibitem[{{Sheppard} {et~al.}(2001){Sheppard}, {Jewitt}, {Fernandez},
  {Magnier}, {Marsden}, {Dahm}, \& {Evans}}]{Sheppard2001SatsJupDiscovery}
{Sheppard}, S.~S., {Jewitt}, D.~C., {Fernandez}, Y.~R., {et~al.} 2001,
  \iaucirc, 7555

\bibitem[{{Sheppard} {et~al.}(2003{\natexlab{b}}){Sheppard}, {Jewitt},
  {Kleyna}, {Fernandez}, {Hsieh}, \& {Marsden}}]{Sheppard2003JupIAU1}
{Sheppard}, S.~S., {Jewitt}, D.~C., {Kleyna}, J., {et~al.} 2003{\natexlab{b}},
  \iaucirc, 8087

\bibitem[{{Sheppard} {et~al.}(2006{\natexlab{b}}){Sheppard}, {Jewitt},
  {Kleyna}, \& {Marsden}}]{Sheppard2006SatIAUC}
{Sheppard}, S.~S., {Jewitt}, D.~C., {Kleyna}, J., \& {Marsden}, B.~G.
  2006{\natexlab{b}}, \iaucirc, 8727

\bibitem[{{Sheppard} {et~al.}(2007){Sheppard}, {Jewitt}, {Kleyna}, \&
  {Marsden}}]{Sheppard2007SatIAUC}
---. 2007, \iaucirc, 8836

\bibitem[{{Sheppard} {et~al.}(2002){Sheppard}, {Jewitt}, {Kleyna}, {Marsden},
  \& {Jacobson}}]{Sheppard2002JupSatDisc}
{Sheppard}, S.~S., {Jewitt}, D.~C., {Kleyna}, J., {Marsden}, B.~G., \&
  {Jacobson}, R. 2002, \iaucirc, 7900

\bibitem[{{Sheppard} \& {Marsden}(2003{\natexlab{a}})}]{Sheppard2003JupIAU2}
{Sheppard}, S.~S., \& {Marsden}, B.~G. 2003{\natexlab{a}}, \iaucirc, 8088

\bibitem[{{Sheppard} \& {Marsden}(2003{\natexlab{b}})}]{Sheppard2003JupIAU3}
---. 2003{\natexlab{b}}, \iaucirc, 8089

\bibitem[{{Sheppard} \& {Marsden}(2004)}]{Sheppard2004JupIAU2}
---. 2004, \iaucirc, 8281

\bibitem[{{Sheppard} \& {Williams}(2012)}]{Sheppard2012JupIAU}
{Sheppard}, S.~S., \& {Williams}, G.~V. 2012, \iaucirc, 9252

\bibitem[{{Sheppard} {et~al.}(2000){Sheppard}, {Jewitt}, {Fernandez},
  {Magnier}, {Marsden}, {Holman}, {Kowal}, {Roemer}, \&
  {Williams}}]{Sheppard2000ThemistoReDis}
{Sheppard}, S.~S., {Jewitt}, D.~C., {Fernandez}, Y., {et~al.} 2000, \iaucirc,
  7525

\bibitem[{{Showalter}(1991)}]{Showalter1991Pan}
{Showalter}, M.~R. 1991, \nat, 351, 709

\bibitem[{{Smith} {et~al.}(1979){Smith}, {Soderblom}, {Johnson}, {Ingersoll},
  {Collins}, {Shoemaker}, {Hunt}, {Masursky}, {Carr}, {Davies}, {Cook},
  {Boyce}, {Owen}, {Danielson}, {Sagan}, {Beebe}, {Veverka}, {McCauley},
  {Strom}, {Morrison}, {Briggs}, \& {Suomi}}]{Smith1979Jupiter}
{Smith}, B.~A., {Soderblom}, L.~A., {Johnson}, T.~V., {et~al.} 1979, Science,
  204, 951

\bibitem[{Smith {et~al.}(2017)Smith, Stockey, Rothwell, \&
  Little}]{Smith2017new}
Smith, S.~Y., Stockey, R.~A., Rothwell, G.~W., \& Little, S.~A. 2017,
  JSysPalaeo, 15, 69

\bibitem[{{Spahn} {et~al.}(2006){Spahn}, {Schmidt}, {Albers}, {H{\"o}rning},
  {Makuch}, {Sei{\ss}}, {Kempf}, {Srama}, {Dikarev}, {Helfert},
  {Moragas-Klostermeyer}, {Krivov}, {Srem{\v c}evi{\'c}}, {Tuzzolino},
  {Economou}, \& {Gr{\"u}n}}]{Sphan2006EnceladusEring}
{Spahn}, F., {Schmidt}, J., {Albers}, N., {et~al.} 2006, Science, 311, 1416

\bibitem[{{Sparks} {et~al.}(2016){Sparks}, {Hand}, {McGrath}, {Bergeron},
  {Cracraft}, \& {Deustua}}]{Sparks2016EuropaHST}
{Sparks}, W.~B., {Hand}, K.~P., {McGrath}, M.~A., {et~al.} 2016, \apj, 829, 121

\bibitem[{{Spencer} \& {Nimmo}(2013)}]{Spencer2013EnceladusRev}
{Spencer}, J.~R., \& {Nimmo}, F. 2013, AREPS, 41, 693

\bibitem[{{Spitale} \& {Tiscareno}(2012)}]{Spitale2012s2009s1}
{Spitale}, J.~N., \& {Tiscareno}, M. 2012, in AAS/Division for Planetary
  Sciences Meeting Abstracts (Reno, NV: American Astronomical Society),
  Vol.~44, AAS/Division for Planetary Sciences Meeting Abstracts (Reno, NV:
  American Astronomical Society), 414.04

\bibitem[{{Spyder Development Team}(2015)}]{Spyder238}
{Spyder Development Team}. 2015, Spyder: The Scientific PYthon Development
  EnviRonment. Version 2.3.8, https://pythonhosted.org/spyder/, ,

\bibitem[{Su{\'a}rez-D{\'\i}az \&
  Anaya-Mu{\~n}oz(2008)}]{Suarez2008HistoryPhylo}
Su{\'a}rez-D{\'\i}az, E., \& Anaya-Mu{\~n}oz, V.~H. 2008, SHPBioSci, 39, 451

\bibitem[{{Sun} {et~al.}(2017){Sun}, {Sei{\ss}}, {Hedman}, \&
  {Spahn}}]{Sun2017MethoneDust}
{Sun}, K.-L., {Sei{\ss}}, M., {Hedman}, M.~M., \& {Spahn}, F. 2017, \icarus,
  284, 206

\bibitem[{{Sykes} {et~al.}(2000){Sykes}, {Nelson}, {Cutri}, {Kirkpatrick},
  {Hurt}, \& {Skrutskie}}]{Sykes2000JupiterSatInfrared}
{Sykes}, M.~V., {Nelson}, B., {Cutri}, R.~M., {et~al.} 2000, \icarus, 143, 371

\bibitem[{{Synnott}(1980)}]{Synnott1980ThebeDiscov}
{Synnott}, S.~P. 1980, Science, 210, 786

\bibitem[{{Synnott}(1981)}]{Synnott1981MetisDiscov}
---. 1981, Science, 212, 1392

\bibitem[{{Takato} {et~al.}(2004){Takato}, {Bus}, {Terada}, {Pyo}, \&
  {Kobayashi}}]{Takato2004Amalthea}
{Takato}, N., {Bus}, S.~J., {Terada}, H., {Pyo}, T.-S., \& {Kobayashi}, N.
  2004, Science, 306, 2224

\bibitem[{{Tamayo} {et~al.}(2011){Tamayo}, {Burns}, {Hamilton}, \&
  {Hedman}}]{Tamayo2011IapetusDust}
{Tamayo}, D., {Burns}, J.~A., {Hamilton}, D.~P., \& {Hedman}, M.~M. 2011,
  \icarus, 215, 260

\bibitem[{{Tamayo} {et~al.}(2014){Tamayo}, {Hedman}, \&
  {Burns}}]{Tamayo2014PhoRingOptic}
{Tamayo}, D., {Hedman}, M.~M., \& {Burns}, J.~A. 2014, \icarus, 233, 1

\bibitem[{{Thomas}(2010)}]{Thomas2010SaturnSatsCassiniProps}
{Thomas}, P.~C. 2010, \icarus, 208, 395

\bibitem[{{Thomas} {et~al.}(2013){Thomas}, {Burns}, {Hedman}, {Helfenstein},
  {Morrison}, {Tiscareno}, \& {Veverka}}]{Thomas2013InnerSatSatu}
{Thomas}, P.~C., {Burns}, J.~A., {Hedman}, M., {et~al.} 2013, \icarus, 226, 999

\bibitem[{{Thomas} {et~al.}(1998){Thomas}, {Burns}, {Rossier}, {Simonelli},
  {Veverka}, {Chapman}, {Klaasen}, {Johnson}, {Belton}, \& {Galileo Solid State
  Imaging Team}}]{Thomas1998InnerSatJup}
{Thomas}, P.~C., {Burns}, J.~A., {Rossier}, L., {et~al.} 1998, \icarus, 135,
  360

\bibitem[{{Throop} {et~al.}(2004){Throop}, {Porco}, {West}, {Burns},
  {Showalter}, \& {Nicholson}}]{Throop2004JupiterRings}
{Throop}, H.~B., {Porco}, C.~C., {West}, R.~A., {et~al.} 2004, \icarus, 172, 59

\bibitem[{Tillyard(1926)}]{Tillyard1926insects}
Tillyard, R.~J. 1926, The Insects of Australia and New Zealand. (Angus \&
  Robertson, Ltd., Sydney, Australia)

\bibitem[{{Tiscareno} {et~al.}(2006){Tiscareno}, {Burns}, {Hedman}, {Porco},
  {Weiss}, {Dones}, {Richardson}, \&
  {Murray}}]{Tiscareno2006SaturnAringMoonlets}
{Tiscareno}, M.~S., {Burns}, J.~A., {Hedman}, M.~M., {et~al.} 2006, \nat, 440,
  648

\bibitem[{{Tosi} {et~al.}(2010){Tosi}, {Turrini}, {Coradini}, \&
  {Filacchione}}]{Tosi2010IapetusDark}
{Tosi}, F., {Turrini}, D., {Coradini}, A., \& {Filacchione}, G. 2010, \mnras,
  403, 1113

\bibitem[{{Treffenst{\"a}dt} {et~al.}(2015){Treffenst{\"a}dt}, {Mour{\~a}o}, \&
  {Winter}}]{Treffenstdt2015JanusEpiFormation}
{Treffenst{\"a}dt}, L.~L., {Mour{\~a}o}, D.~C., \& {Winter}, O.~C. 2015, \aap,
  583, A80

\bibitem[{{Tsiganis} {et~al.}(2005){Tsiganis}, {Gomes}, {Morbidelli}, \&
  {Levison}}]{Tsiganis2005NICEplanets}
{Tsiganis}, K., {Gomes}, R., {Morbidelli}, A., \& {Levison}, H.~F. 2005, \nat,
  435, 459

\bibitem[{{Turrini} {et~al.}(2008){Turrini}, {Marzari}, \&
  {Beust}}]{Turrini2008IrregularSatsSaturn}
{Turrini}, D., {Marzari}, F., \& {Beust}, H. 2008, \mnras, 391, 1029

\bibitem[{{Turrini} {et~al.}(2009){Turrini}, {Marzari}, \&
  {Tosi}}]{Turrini2009IrregularSatsSaturn}
{Turrini}, D., {Marzari}, F., \& {Tosi}, F. 2009, \mnras, 392, 455

\bibitem[{Van~Dung {et~al.}(1993)Van~Dung, Giao, Chinh, Tuoc, Arctander, \&
  MacKinnon}]{Van1993new}
Van~Dung, V., Giao, P.~M., Chinh, N.~N., {et~al.} 1993, \nat, 363, 443

\bibitem[{{Vasundhara} {et~al.}(2017){Vasundhara}, {Selvakumar}, \&
  {Anbazhagan}}]{Vasundhara2017GalileansGround}
{Vasundhara}, R., {Selvakumar}, G., \& {Anbazhagan}, P. 2017, \mnras, 468, 501

\bibitem[{{Verbiscer} {et~al.}(2007){Verbiscer}, {French}, {Showalter}, \&
  {Helfenstein}}]{Verbiscer2007Enceladus}
{Verbiscer}, A., {French}, R., {Showalter}, M., \& {Helfenstein}, P. 2007,
  Science, 315, 815

\bibitem[{{Verbiscer} {et~al.}(2009){Verbiscer}, {Skrutskie}, \&
  {Hamilton}}]{Verbiscer2009SatLarRing}
{Verbiscer}, A.~J., {Skrutskie}, M.~F., \& {Hamilton}, D.~P. 2009, \nat, 461,
  1098

\bibitem[{{Winter} {et~al.}(2016){Winter}, {Souza}, {Sfait}, {Giuliatti
  Winter}, {Mour{\~a}o}, \& {Foryta}}]{Winter2016JanusEpiRing}
{Winter}, O., {Souza}, A., {Sfait}, R., {et~al.} 2016, in AAS/Division for
  Planetary Sciences Meeting Abstracts (Pasadena, CA: American Astronomical
  Society), Vol.~48, AAS/Division for Planetary Sciences Meeting Abstracts
  (Pasadena, CA: American Astronomical Society), 203.03

\bibitem[{{Wong} {et~al.}(2006){Wong}, {de Pater}, {Showalter}, {Roe},
  {Macintosh}, \& {Verbanac}}]{Wong2006GBIRJupRingMoon}
{Wong}, M.~H., {de Pater}, I., {Showalter}, M.~R., {et~al.} 2006, \icarus, 185,
  403

\bibitem[{{Yoder} {et~al.}(1983){Yoder}, {Colombo}, {Synnott}, \&
  {Yoder}}]{Yoder1983SaturnCoorobiting}
{Yoder}, C.~F., {Colombo}, G., {Synnott}, S.~P., \& {Yoder}, K.~A. 1983,
  \icarus, 53, 431

\bibitem[{{Yoder} {et~al.}(1989){Yoder}, {Synnott}, \&
  {Salo}}]{Yoder1989JanusEpiMassOrbit}
{Yoder}, C.~F., {Synnott}, S.~P., \& {Salo}, H. 1989, \aj, 98, 1875

\bibitem[{{Zappala} {et~al.}(1990){Zappala}, {Cellino}, {Farinella}, \&
  {Knezevic}}]{Zappala1990HierarchicalClustering1}
{Zappala}, V., {Cellino}, A., {Farinella}, P., \& {Knezevic}, Z. 1990, \aj,
  100, 2030

\bibitem[{{Zappala} {et~al.}(1994){Zappala}, {Cellino}, {Farinella}, \&
  {Milani}}]{Zappala1994HierarchicalClustering2}
{Zappala}, V., {Cellino}, A., {Farinella}, P., \& {Milani}, A. 1994, \aj, 107,
  772

\bibitem[{{Zebker} {et~al.}(1985){Zebker}, {Marouf}, \&
  {Tyler}}]{Zebker1985SaturnRingParticle}
{Zebker}, H.~A., {Marouf}, E.~A., \& {Tyler}, G.~L. 1985, \icarus, 64, 531

\bibitem[{Zimmermann \& Schultz(1931)}]{Zimmermann1931arbeitsweise}
Zimmermann, W., \& Schultz, W. 1931, Arbeitsweise der botanischen Phylogenetik
  und anderer Gruppierungswissenschaften (Urban \& Schwarzenberg, Germany)

\end{thebibliography}

 \appendix
 
 \section{List of Characteristics}
 \label{ListCharacters}
 A list of characters used in the creation of the Jovian (Appendix \ref{JupiterMatrix}) and Saturnian (Appendix \ref{SaturnMatrix}) satellite matrices. See Section \ref{Characteristics} for full discussion.
 
 \subsection{Orbital Characteristics}  
 \label{listOrbitChar}
 \begin{enumerate}
\item In orbit around the gas giant (Orb): no (0) ; yes (1)
\item Revolution (Rev): Prograde revolution(0); Retrograde revolution (1)

\item Semi-major axis(a): 
\begin{itemize}
\item Jovian: $r^2$:0.990 Bin delimiters 0 km (0) ; 
$3.67892625 \times 10^6$ km (1) ; 
$7.2348525  \times 10^6$ km (2) ; 
$1.079077875 \times 10^7$ km (3) ; 
$1.4346705 \times 10^7$ km (4) ; 
$1.790263125 \times 10^7$ km (5) ;
$2.14585575 \times 10^7$ km (6) ;
$2.501448375 \times 10^7$ km (7) 
\item Saturnian: $r^2$:0.991 Bin delimiters: 0 km (0) ; 
$3.644200 \times 10^6$ km (1) ; 
$7.221500 \times 10^6$ km (2) ; 
$1.0798800 \times 10^7$ km (3) ; 
$1.4376100 \times 10^7$ km (4) ; 
$1.7953400 \times 10^7$ km (5) ;
$2.1530700 \times 10^7$ km (5)
\end{itemize}

\item Orbital inclination to the plane(i):
\begin{itemize}
\item Jovian: $r^2$:0.990 Bin delimiters: 0\degree (0) ; 
16.55\degree (1) ; 
33.1\degree (2) ;  
49.65\degree (3) ; 
66.2\degree (4) ; 
82.75\degree (5) ; 
99.3 \degree (6) ;
115.85 \degree (7) ;
132.4  \degree (8) ;
148.95 \degree (9) 
\item Saturnian: $r^2$:0.993 Bin delimiters: 0\degree (0) ; 
29.97\degree (1) ;  
59.93\degree (2) ;  
89.9\degree (3) ;  
119.87\degree (4) ;  
149.83\degree (5)
\end{itemize}

\item Orbital eccentricity(e): 
\begin{itemize}
\item Jovian: $r^2$:0.99 Bin delimiters:  0(0) ; 
0.036 (1) ; 
0.072 (2) ;  
0.108 (3) ; 
0.144 (4) ; 
0.18 (5) ; 
0.216 (6) ; 
0.252 (7) ; 
0.288 (8) ; 
0.324 (9) ; 
0.36 (10) ; 
0.396 (11)
\item Saturnian: $r^2$:0.993 Bin delimiters: 0 (0) ; 
0.064 (1) ; 
0.129 (2) ; 
0.193 (3) ; 
0.258 (4) ; 
0.322 (5) ; 
0.387 (6) ; 
0.451 (7) ; 
0.515 (8) ; 
0.58 (9)
\end{itemize}
\end{enumerate}

\subsection{Physical Characteristics}
\label{listPhysChar}
\begin{enumerate}
\setcounter{enumi}{5}
\item Density:
\begin{itemize}
\item Jovian: $r^2$:0.996 Bin delimiters: 3084.5 $kg~m^{-3}$ (0) ; 
2639 $kg~m^{-3}$ (1) ; 
2193.5 $kg~m^{-3}$ (2);  
1748 $kg~m^{-3}$ (3); 
1302.5 $kg~m^{-3}$ (4) ; 
854.3 $kg~m^{-3}$ (5) 
\item Saturnian: $r^2$:99.2 Bin delimiters: 
1880 $kg~m^{-3}$ (0) ; 
1713.6 $kg~m^{-3}$ (1) ; 
1547.3 $kg~m^{-3}$ (2);  
1380.9 $kg~m^{-3}$ (3) ; 
1214.5 $kg~m^{-3}$ (4) ; 
1048.2 $kg~m^{-3}$ (5) ;
881.8 $kg~m^{-3}$ (6) ;
715.4 $kg~m^{-3}$ (7) ;
549.1 $kg~m^{-3}$ (8) ;
382.7 $kg~m^{-3}$ (9) ;
216.3 $kg~m^{-3}$ (10) ;
48.2 $kg~m^{-3}$ (11)
\end{itemize}

\item Visual geometric albedo: 
\begin{itemize}
\item Jovian: $r^2$:0.991 Bin delimiters: 0 (0) ; 
0.09 (1) ; 
0.16 (2) ; 
0.24 (3) ; 
0.31 (4) ; 
0.38 (5) ; 
0.46 (6) ; 
0.53 (7) ; 
0.60 (8) ; 
0.68 (9)
\item Saturnian: $r^2$:0.991 Bin delimiters: 0 (0) ; 
0.13 (1) ; 
0.26 (2) ; 
0.38 (3) ; 
0.51 (4) ; 
0.63 (5) ; 
0.75 (6) ; 
0.87 (7)  
\end{itemize}
\end{enumerate}

\subsection{Compositional Characteristics}
\label{listCompChar}
\begin{enumerate}
\setcounter{enumi}{7}

\item Elemental Hydrogen (eH) Presence (0) ; Absence (1) 

\item Hydrogen (H$_2$) Presence (0) ; Absence (1) 

\item Helium (He) Presence (0) ; Absence (1) 

\item Oxygen (O$_2$) Absence (0) ; Presence (1)

\item Ozone (O$_3$) Absence (0) ; Presence (1)

\item Sodium (Na) Absence (0) ; Presence (1)

\item Potassium (K) Absence (0) ; Presence (1)

\item Carbon dioxide (CO$_2$) Absence (0) ; Presence (1)

\item Nitrogen (N$_2$) Absence (0) ; Presence (1)

\item Sulphur dioxide (SO$_2$) Absence (0) ; Presence (1)

\item Argon (Ar) Absence (0) ; Presence (1)

\item Water (H$_2$O) Absence (0) ; Presence (1)

\item Carbon monoxide (CO) Absence (0) ; Presence (1)

\item Neon (Ne) Absence (0) ; Presence (1)

\item Nitrogen oxide (NO) Absence (0) ; Presence (1)

\item Methane (CH$_4$) Absence (0) ; Presence (1)

\item Sulphuric Acid (H$_2$SO$_4$) Absence (0) ; Presence (1)

\item Iron (Fe) Absence (0) ; Presence (1)

\item Nickel (Ni) Absence (0) ; Presence (1)

\item Iron sulphide (FeS) Absence (0) ; Presence (1)

\item Iron oxide (FeO) Absence (0) ; Presence (1)

\item Silicone oxide (SiO) Absence (0) ; Presence (1)

\item Magnesium oxide (MgO) Absence (0) ; Presence (1)

\item Basalt(Bas) Absence (0) ; Presence (1)

\item Sulphur (S) Absence (0) ; Presence (1)

\item Silicates (Sil) Absence (0) ; Presence (1)

\item Carbonates (Carb) Absence (0) ; Presence (1)

\item Ammonia (NH$_4$) Absence (0) ; Presence (1)

\item Hydrocarbons (HyCarb) Absence (0) ; Presence (1)

\item Hydrogen peroxide (H$_2$O$_2$) Absence (0) ; Presence (1)

\item Tholins (Thol) Absence (0) ; Presence (1)

\end{enumerate}

 \section{Jovian Satellite matrix}
 \label{JupiterMatrix}

\begin{longrotatetable}
\begin{deluxetable}{cccccccccccccccccccccccccccccccccccccccc}
\tabletypesize{\tiny}

\tablecaption{Taxon-Character Matrix of the Jovian Satellite System }
\tablenum{3}

\tablehead{
\colhead{Identifier} & \colhead{Orb} & \colhead{Rev} & \colhead{a} & \colhead{i} & \colhead{e} & \colhead{D} & \colhead{Alb} & \colhead{eH} & \colhead{H$_2$} & \colhead{He} & \colhead{O$_2$} & \colhead{O$_3$} & \colhead{Na} & \colhead{K} & \colhead{CO$_2$} & \colhead{N$_2$} & \colhead{SO$_2$} & \colhead{Ar} & \colhead{H$_2$O} & \colhead{CO} & \colhead{Ne} & \colhead{NO} & \colhead{CH$_4$} & \colhead{H$_2$SO$_4$} & \colhead{Fe} & \colhead{Ni} & \colhead{FeS} & \colhead{FeO} & \colhead{SiO} & \colhead{MgO} & \colhead{Bas} & \colhead{S} & \colhead{Sil} & \colhead{Carb} & \colhead{NH$_3$} & \colhead{HyCarb} & \colhead{H$_2$O$_2$} & \colhead{Thol} & \colhead{Reference}\\
} 

\startdata
Sun & 0 & 0 & 0 & 0 & 0 & 0 & 0 & 0 & 0 & 0 & 0 & 0 & 0 & 0 & 0 & 0 & 0 & 0 & 0 & 0 & 0 & 0 & 0 & 0 & 0 & 0 & 0 & 0 & 0 & 0 & 0 & 0 & 0 & 0 & 0 & 0 & 0 & 0 & 1 \\
Jupiter Main Ring & 1 & 0 & 0 & 0 & 0 & ? & 0 & 1 & 1 & 1 & 0 & 0 & 0 & 0 & 0 & 0 & 0 & 0 & 1 & 0 & 0 & 0 & 0 & 0 & 0 & 0 & 0 & 0 & 0 & 0 & 0 & 0 & 1 & 0 & 0 & 0 & 0 & 0 & 2,3,4,5,6,7,8 \\
Metis & 1 & 0 & 0 & 0 & 0 & 6 & 0 & 1 & 1 & ? & ? & ? & ? & ? & ? & ? & ? & ? & 1 & ? & ? & ? & ? & ? & ? & ? & ? & ? & ? & ? & ? & ? & 1 & ? & ? & 0 & ? & ? & 9 \\
Adrastea & 1 & 0 & 0 & 0 & 0 & ? & 1 & 1 & 1 & ? & ? & ? & ? & ? & ? & ? & ? & ? & 1 & ? & ? & ? & ? & ? & ? & ? & ? & ? & ? & ? & ? & ? & 1 & ? & ? & 0 & ? & ? & 9 \\
Amalthea & 1 & 0 & 0 & 0 & 0 & 6 & 1 & 1 & 1 & 1 & 0 & 0 & 0 & 0 & 0 & 0 & 0 & 0 & 1 & 0 & 0 & 0 & ? & 0 & 0 & 0 & 0 & 0 & ? & ? & 0 & 1 & 1 & 1 & ? & 0 & 0 & ? & 9,10,11 \\
Thebe & 1 & 0 & 0 & 0 & 0 & ? & 0 & 1 & 1 & ? & ? & ? & ? & ? & ? & ? & ? & 0 & 1 & ? & ? & ? & ? & ? & ? & ? & ? & ? & ? & ? & ? & ? & 1 & ? & ? & 0 & ? & ? & 9,10 \\
Io & 1 & 0 & 0 & 0 & 0 & 0 & 8 & 1 & 1 & 1 & 0 & 0 & 0 & 0 & 0 & 0 & 1 & 0 & 1 & 0 & 0 & 0 & 0 & 0 & 1 & 0 & 0 & 0 & 0 & 1 & 0 & 1 & 1 & 0 & 0 & 0 & 0 & 0 & 3,12,13,14,15 \\
Europa & 1 & 0 & 0 & 0 & 0 & 2 & 8 & 1 & 1 & 1 & 1 & 0 & 0 & 0 & 1 & 0 & 1 & 0 & 1 & 0 & 0 & 0 & 0 & 1 & 1 & 0 & 0 & 0 & 0 & 0 & 0 & 0 & 1 & 0 & 0 & 0 & 1 & 1 & 3,12,13,14,15 \\
Ganymede & 1 & 0 & 0 & 0 & 0 & 4 & 5 & 1 & 1 & 1 & 1 & 1 & 0 & 0 & 1 & 0 & 1 & 0 & 1 & 0 & 0 & 0 & 0 & 0 & 1 & 0 & 0 & 0 & 0 & 0 & 0 & 0 & 1 & 0 & 0 & 0 & 0 & 1 & 3,12,13,14,15 \\
Callisto & 1 & 0 & 0 & 0 & 0 & 4 & 2 & 1 & 1 & 1 & 1 & 0 & 0 & 0 & 1 & 0 & 1 & 0 & 1 & 0 & 0 & 0 & 0 & 0 & 0 & 0 & 0 & 0 & 0 & 0 & 0 & 0 & 1 & 0 & 1 & 0 & 0 & 1 & 3,12,13,14,15 \\
Themisto & 1 & 0 & 2 & 2 & 6 & ? & 0 & 1 & 1 & 1 & 0 & 0 & 0 & 0 & 0 & 0 & 0 & 0 & 0 & 0 & 0 & 0 & 0 & 0 & 0 & 0 & 0 & 0 & 0 & 0 & 0 & 0 & 1 & 0 & 0 & 0 & 0 & 0 & 16,17,18 \\
Leda & 1 & 0 & 3 & 1 & 4 & ? & 0 & 1 & 1 & 1 & 0 & 0 & 0 & 0 & 0 & 0 & 0 & 0 & 0 & 0 & 0 & 0 & 0 & 0 & 0 & 0 & 0 & 0 & 0 & 0 & 0 & 0 & 1 & 0 & 0 & 0 & 0 & 0 & 16,17,19,20 \\
Himalia & 1 & 0 & 3 & 1 & 4 & 1 & 0 & 1 & 1 & 1 & 0 & 0 & 0 & 0 & 0 & 0 & 0 & 0 & 1 & 0 & 0 & 0 & 0 & 0 & 0 & 0 & 0 & 0 & 0 & 0 & 0 & 0 & 1 & 1 & 0 & 0 & 0 & 0 & 3,17,16,19,20,21,22,23 \\
Lysithea & 1 & 0 & 3 & 1 & 3 & ? & 0 & 1 & ? & ? & ? & ? & ? & ? & ? & ? & ? & ? & 0 & ? & ? & ? & ? & ? & ? & ? & ? & ? & ? & ? & ? & ? & ? & ? & ? & ? & ? & ? & 16,17,18,19,20 \\
Elara & 1 & 0 & 3 & 1 & 6 & ? & 0 & 1 & ? & ? & ? & ? & ? & ? & ? & ? & ? & ? & 0 & ? & ? & ? & ? & ? & ? & ? & ? & ? & ? & ? & ? & ? & ? & ? & ? & ? & ? & ? & 16,17,19,20,21 \\
Dia & 1 & 0 & 3 & 1 & 5 & ? & 0 & 1 & ? & ? & ? & ? & ? & ? & ? & ? & ? & ? & ? & ? & ? & ? & ? & ? & ? & ? & ? & ? & ? & ? & ? & ? & ? & ? & ? & ? & ? & ? & 16,17 \\
Carpo & 1 & 0 & 4 & 3 & 11 & ? & ? & 1 & ? & ? & ? & ? & ? & ? & ? & ? & ? & ? & ? & ? & ? & ? & ? & ? & ? & ? & ? & ? & ? & ? & ? & ? & ? & ? & ? & ? & ? & ? & 16 \\
Euporie & 1 & 1 & 5 & 8 & 3 & ? & ? & 1 & ? & ? & ? & ? & ? & ? & ? & ? & ? & ? & ? & ? & ? & ? & ? & ? & ? & ? & ? & ? & ? & ? & ? & ? & ? & ? & ? & ? & ? & ? & 16,17 \\
Orthosie & 1 & 1 & 5 & 8 & 7 & ? & ? & 1 & ? & ? & ? & ? & ? & ? & ? & ? & ? & ? & ? & ? & ? & ? & ? & ? & ? & ? & ? & ? & ? & ? & ? & ? & ? & ? & ? & ? & ? & ? & 16,17 \\
Euanthe & 1 & 1 & 5 & 8 & 6 & ? & ? & 1 & ? & ? & ? & ? & ? & ? & ? & ? & ? & ? & ? & ? & ? & ? & ? & ? & ? & ? & ? & ? & ? & ? & ? & ? & ? & ? & ? & ? & ? & ? & 16,17 \\
Thyone & 1 & 1 & 5 & 8 & 6 & ? & ? & 1 & ? & ? & ? & ? & ? & ? & ? & ? & ? & ? & ? & ? & ? & ? & ? & ? & ? & ? & ? & ? & ? & ? & ? & ? & ? & ? & ? & ? & ? & ? & 16,17 \\
Mneme & 1 & 1 & 5 & 8 & 6 & ? & ? & 1 & ? & ? & ? & ? & ? & ? & ? & ? & ? & ? & ? & ? & ? & ? & ? & ? & ? & ? & ? & ? & ? & ? & ? & ? & ? & ? & ? & ? & ? & ? & 16 \\
Harpalyke & 1 & 1 & 5 & 8 & 6 & ? & 0 & 1 & ? & ? & ? & ? & ? & ? & ? & ? & ? & ? & ? & ? & ? & ? & ? & ? & ? & ? & ? & ? & ? & ? & ? & ? & ? & ? & ? & ? & ? & ? & 16,17,18 \\
Hermippe & 1 & 1 & 5 & 9 & 5 & ? & ? & 1 & ? & ? & ? & ? & ? & ? & ? & ? & ? & ? & ? & ? & ? & ? & ? & ? & ? & ? & ? & ? & ? & ? & ? & ? & ? & ? & ? & ? & ? & ? & 16,17 \\
Praxidike & 1 & 1 & 5 & 9 & 6 & ? & 0 & 1 & ? & ? & ? & ? & ? & ? & ? & ? & ? & ? & ? & ? & ? & ? & ? & ? & ? & ? & ? & ? & ? & ? & ? & ? & ? & ? & ? & ? & ? & ? & 16,17,18,19 \\
Thelxinoe & 1 & 1 & 5 & 9 & 6 & ? & ? & 1 & ? & ? & ? & ? & ? & ? & ? & ? & ? & ? & ? & ? & ? & ? & ? & ? & ? & ? & ? & ? & ? & ? & ? & ? & ? & ? & ? & ? & ? & ? & 16 \\
Iocaste & 1 & 1 & 5 & 9 & 5 & ? & 0 & 1 & ? & ? & ? & ? & ? & ? & ? & ? & ? & ? & ? & ? & ? & ? & ? & ? & ? & ? & ? & ? & ? & ? & ? & ? & ? & ? & ? & ? & ? & ? & 16,17,18 \\
Ananke & 1 & 1 & 5 & 8 & 6 & ? & 0 & 1 & 1 & 1 & 0 & 0 & 0 & 0 & 0 & 0 & 0 & 0 & 1 & 0 & 0 & 0 & 0 & 0 & 0 & 0 & 0 & 0 & 0 & 0 & 0 & 0 & 1 & 1 & 0 & 0 & 0 & 0 & 16,17,18,19 \\
Arche & 1 & 1 & 6 & 9 & 7 & ? & ? & 1 & ? & ? & ? & ? & ? & ? & ? & ? & ? & ? & ? & ? & ? & ? & ? & ? & ? & ? & ? & ? & ? & ? & ? & ? & ? & ? & ? & ? & ? & ? & 16,17 \\
Pasithee & 1 & 1 & 6 & 9 & 7 & ? & ? & 1 & ? & ? & ? & ? & ? & ? & ? & ? & ? & ? & ? & ? & ? & ? & ? & ? & ? & ? & ? & ? & ? & ? & ? & ? & ? & ? & ? & ? & ? & ? & 16,17 \\
Herse & 1 & 1 & 6 & 9 & 5 & ? & ? & 1 & ? & ? & ? & ? & ? & ? & ? & ? & ? & ? & ? & ? & ? & ? & ? & ? & ? & ? & ? & ? & ? & ? & ? & ? & ? & ? & ? & ? & ? & ? & 16 \\
Chaldene & 1 & 1 & 6 & 9 & 6 & ? & 0 & 1 & ? & ? & ? & ? & ? & ? & ? & ? & ? & ? & ? & ? & ? & ? & ? & ? & ? & ? & ? & ? & ? & ? & ? & ? & ? & ? & ? & ? & ? & ? & 16,17 \\
Kale & 1 & 1 & 6 & 9 & 7 & ? & ? & 1 & ? & ? & ? & ? & ? & ? & ? & ? & ? & ? & ? & ? & ? & ? & ? & ? & ? & ? & ? & ? & ? & ? & ? & ? & ? & ? & ? & ? & ? & ? & 16,17 \\
Isonoe & 1 & 1 & 6 & 9 & 6 & ? & 0 & 1 & ? & ? & ? & ? & ? & ? & ? & ? & ? & ? & ? & ? & ? & ? & ? & ? & ? & ? & ? & ? & ? & ? & ? & ? & ? & ? & ? & ? & ? & ? & 16,17 \\
Aitne & 1 & 1 & 6 & 9 & 7 & ? & ? & 1 & ? & ? & ? & ? & ? & ? & ? & ? & ? & ? & ? & ? & ? & ? & ? & ? & ? & ? & ? & ? & ? & ? & ? & ? & ? & ? & ? & ? & ? & ? & 16 \\
Erinome & 1 & 1 & 6 & 9 & 7 & ? & 0 & 1 & ? & ? & ? & ? & ? & ? & ? & ? & ? & ? & ? & ? & ? & ? & ? & ? & ? & ? & ? & ? & ? & ? & ? & ? & ? & ? & ? & ? & ? & ? & 16,17 \\
Taygete & 1 & 1 & 6 & 9 & 6 & ? & 0 & 1 & ? & ? & ? & ? & ? & ? & ? & ? & ? & ? & ? & ? & ? & ? & ? & ? & ? & ? & ? & ? & ? & ? & ? & ? & ? & ? & ? & ? & ? & ? & 16,17,18 \\
Carme & 1 & 1 & 6 & 9 & 7 & ? & 0 & 1 & 1 & 1 & 0 & 0 & 0 & 0 & 0 & 0 & 0 & 0 & 0 & 0 & 0 & 0 & 0 & 0 & 0 & 0 & 0 & 0 & 0 & 0 & 0 & 0 & 1 & 1 & 0 & 0 & 0 & 0 & 16,17,18,19,21 \\
Kalyke & 1 & 1 & 6 & 9 & 6 & ? & 0 & 1 & ? & ? & ? & ? & ? & ? & ? & ? & ? & ? & ? & ? & ? & ? & ? & ? & ? & ? & ? & ? & ? & ? & ? & ? & ? & ? & ? & ? & ? & ? & 16,17,18,19 \\
Eukelade & 1 & 1 & 6 & 9 & 7 & ? & ? & 1 & ? & ? & ? & ? & ? & ? & ? & ? & ? & ? & ? & ? & ? & ? & ? & ? & ? & ? & ? & ? & ? & ? & ? & ? & ? & ? & ? & ? & ? & ? & 16 \\
Kallichore & 1 & 1 & 6 & 9 & 7 & ? & ? & 1 & ? & ? & ? & ? & ? & ? & ? & ? & ? & ? & ? & ? & ? & ? & ? & ? & ? & ? & ? & ? & ? & ? & ? & ? & ? & ? & ? & ? & ? & ? & 16 \\
Helike & 1 & 1 & 5 & 9 & 4 & ? & ? & 1 & ? & ? & ? & ? & ? & ? & ? & ? & ? & ? & ? & ? & ? & ? & ? & ? & ? & ? & ? & ? & ? & ? & ? & ? & ? & ? & ? & ? & ? & ? & 16 \\
Eurydome & 1 & 1 & 6 & 9 & 7 & ? & ? & 1 & ? & ? & ? & ? & ? & ? & ? & ? & ? & ? & ? & ? & ? & ? & ? & ? & ? & ? & ? & ? & ? & ? & ? & ? & ? & ? & ? & ? & ? & ? & 16,17 \\
Autonoe & 1 & 1 & 6 & 9 & 7 & ? & ? & 1 & ? & ? & ? & ? & ? & ? & ? & ? & ? & ? & ? & ? & ? & ? & ? & ? & ? & ? & ? & ? & ? & ? & ? & ? & ? & ? & ? & ? & ? & ? & 16,17 \\
Sponde & 1 & 1 & 6 & 9 & 11 & ? & ? & 1 & ? & ? & ? & ? & ? & ? & ? & ? & ? & ? & ? & ? & ? & ? & ? & ? & ? & ? & ? & ? & ? & ? & ? & ? & ? & ? & ? & ? & ? & ? & 16,17 \\
Pasiphae & 1 & 1 & 6 & 9 & 11 & ? & 1 & 1 & 1 & 1 & 0 & 0 & 0 & 0 & 0 & 0 & 0 & 0 & 0 & 0 & 0 & 0 & 0 & 0 & 0 & 0 & 0 & 0 & 0 & 0 & 0 & 0 & 1 & 1 & 0 & 0 & 0 & 0 & 16,17,18,19,21 \\
Megaclite & 1 & 1 & 6 & 9 & 11 & ? & 0 & 1 & ? & ? & ? & ? & ? & ? & ? & ? & ? & ? & ? & ? & ? & ? & ? & ? & ? & ? & ? & ? & ? & ? & ? & ? & ? & ? & ? & ? & ? & ? & 16,17,18 \\
Sinope & 1 & 1 & 6 & 9 & 6 & ? & 0 & 1 & 1 & 1 & 0 & 0 & 0 & 0 & 0 & 0 & 0 & 0 & 0 & 0 & 0 & 0 & 0 & 0 & 0 & 0 & 0 & 0 & 0 & 0 & 0 & 0 & 1 & 1 & 0 & 0 & 0 & 0 & 16,17,18,19,21 \\
Hegemone & 1 & 1 & 6 & 9 & 9 & ? & ? & 1 & ? & ? & ? & ? & ? & ? & ? & ? & ? & ? & ? & ? & ? & ? & ? & ? & ? & ? & ? & ? & ? & ? & ? & ? & ? & ? & ? & ? & ? & ? & 16 \\
Aoede & 1 & 1 & 6 & 9 & 1 & ? & ? & 1 & ? & ? & ? & ? & ? & ? & ? & ? & ? & ? & ? & ? & ? & ? & ? & ? & ? & ? & ? & ? & ? & ? & ? & ? & ? & ? & ? & ? & ? & ? & 16 \\
Callirrhoe & 1 & 1 & 6 & 8 & 7 & ? & 0 & 1 & 1 & 1 & 0 & 0 & 0 & 0 & 0 & 0 & 0 & 0 & 0 & 0 & 0 & 0 & 0 & 0 & 0 & 0 & 0 & 0 & 0 & 0 & 0 & 0 & 1 & 1 & 0 & 0 & 0 & 0 & 16,17,18,19 \\
Cyllene & 1 & 1 & 6 & 9 & 8 & ? & ? & 1 & ? & ? & ? & ? & ? & ? & ? & ? & ? & ? & ? & ? & ? & ? & ? & ? & ? & ? & ? & ? & ? & ? & ? & ? & ? & ? & ? & ? & ? & ? & 16 \\
Kore & 1 & 1 & 6 & 8 & 9 & ? & ? & 1 & ? & ? & ? & ? & ? & ? & ? & ? & ? & ? & ? & ? & ? & ? & ? & ? & ? & ? & ? & ? & ? & ? & ? & ? & ? & ? & ? & ? & ? & ? & 16 \\
S/2003 J2 & 1 & 1 & 7 & 9 & 10 & ? & ? & 1 & ? & ? & ? & ? & ? & ? & ? & ? & ? & ? & ? & ? & ? & ? & ? & ? & ? & ? & ? & ? & ? & ? & ? & ? & ? & ? & ? & ? & ? & ? & 16 \\
S/2003 J3 & 1 & 1 & 5 & 8 & 6 & ? & ? & 1 & ? & ? & ? & ? & ? & ? & ? & ? & ? & ? & ? & ? & ? & ? & ? & ? & ? & ? & ? & ? & ? & ? & ? & ? & ? & ? & ? & ? & ? & ? & 16 \\
S/2003 J4 & 1 & 1 & 6 & 8 & 5 & ? & ? & 1 & ? & ? & ? & ? & ? & ? & ? & ? & ? & ? & ? & ? & ? & ? & ? & ? & ? & ? & ? & ? & ? & ? & ? & ? & ? & ? & ? & ? & ? & ? & 16 \\
S/2003 J5 & 1 & 1 & 6 & 9 & 5 & ? & ? & 1 & ? & ? & ? & ? & ? & ? & ? & ? & ? & ? & ? & ? & ? & ? & ? & ? & ? & ? & ? & ? & ? & ? & ? & ? & ? & ? & ? & ? & ? & ? & 16 \\
S/2003 J9 & 1 & 1 & 6 & 9 & 7 & ? & ? & 1 & ? & ? & ? & ? & ? & ? & ? & ? & ? & ? & ? & ? & ? & ? & ? & ? & ? & ? & ? & ? & ? & ? & ? & ? & ? & ? & ? & ? & ? & ? & 16 \\
S/2003 J10 & 1 & 1 & 6 & 9 & 5 & ? & ? & 1 & ? & ? & ? & ? & ? & ? & ? & ? & ? & ? & ? & ? & ? & ? & ? & ? & ? & ? & ? & ? & ? & ? & ? & ? & ? & ? & ? & ? & ? & ? & 16 \\
S/2003 J12 & 1 & 1 & 5 & 8 & 10 & ? & ? & 1 & ? & ? & ? & ? & ? & ? & ? & ? & ? & ? & ? & ? & ? & ? & ? & ? & ? & ? & ? & ? & ? & ? & ? & ? & ? & ? & ? & ? & ? & ? & 16 \\
S/2003 J15 & 1 & 1 & 6 & 8 & 3 & ? & ? & 1 & ? & ? & ? & ? & ? & ? & ? & ? & ? & ? & ? & ? & ? & ? & ? & ? & ? & ? & ? & ? & ? & ? & ? & ? & ? & ? & ? & ? & ? & ? & 16 \\
S/2003 J16 & 1 & 1 & 5 & 8 & 7 & ? & ? & 1 & ? & ? & ? & ? & ? & ? & ? & ? & ? & ? & ? & ? & ? & ? & ? & ? & ? & ? & ? & ? & ? & ? & ? & ? & ? & ? & ? & ? & ? & ? & 16 \\
S/2003 J18 & 1 & 1 & 5 & 8 & 3 & ? & ? & 1 & ? & ? & ? & ? & ? & ? & ? & ? & ? & ? & ? & ? & ? & ? & ? & ? & ? & ? & ? & ? & ? & ? & ? & ? & ? & ? & ? & ? & ? & ? & 16 \\
S/2003 J19 & 1 & 1 & 6 & 9 & 9 & ? & ? & 1 & ? & ? & ? & ? & ? & ? & ? & ? & ? & ? & ? & ? & ? & ? & ? & ? & ? & ? & ? & ? & ? & ? & ? & ? & ? & ? & ? & ? & ? & ? & 16 \\
S/2003 J23 & 1 & 1 & 6 & 9 & 8 & ? & ? & 1 & ? & ? & ? & ? & ? & ? & ? & ? & ? & ? & ? & ? & ? & ? & ? & ? & ? & ? & ? & ? & ? & ? & ? & ? & ? & ? & ? & ? & ? & ? & 16 \\
S/2010 J1 & 1 & 1 & 6 & 9 & 8 & ? & ? & 1 & ? & ? & ? & ? & ? & ? & ? & ? & ? & ? & ? & ? & ? & ? & ? & ? & ? & ? & ? & ? & ? & ? & ? & ? & ? & ? & ? & ? & ? & ? & 16 \\
S/2010 J2 & 1 & 1 & 5 & 9 & 8 & ? & ? & 1 & ? & ? & ? & ? & ? & ? & ? & ? & ? & ? & ? & ? & ? & ? & ? & ? & ? & ? & ? & ? & ? & ? & ? & ? & ? & ? & ? & ? & ? & ? & 16 \\
S/2011 J1 & 1 & 1 & 5 & 9 & 8 & ? & ? & 1 & ? & ? & ? & ? & ? & ? & ? & ? & ? & ? & ? & ? & ? & ? & ? & ? & ? & ? & ? & ? & ? & ? & ? & ? & ? & ? & ? & ? & ? & ? & 16 \\
S/2011 J2 & 1 & 1 & 6 & 9 & 10 & ? & ? & 1 & ? & ? & ? & ? & ? & ? & ? & ? & ? & ? & ? & ? & ? & ? & ? & ? & ? & ? & ? & ? & ? & ? & ? & ? & ? & ? & ? & ? & ? & ? & 16 \\
\enddata

\tablecomments{Character Abbreviations: 
In orbit around the gas giant (Orb);
Revolution (Rev); 
Semi-major axis(a);
Orbital inclination to the plane(i);
Orbital eccentricity(e);
Density (D);
Visual geometric albedo (Alb);
Elemental Hydrogen (eH);
Hydrogen (H$_2$);
Helium (He); 
Oxygen (O$_2$);
Ozone (O$_3$);
Sodium (Na);
Potassium (K);
Carbon dioxide (CO$_2$);
Nitrogen (N$_2$);
Sulphur dioxide (SO$_2$);
Argon (Ar);
Water (H$_2$O);
Carbon monoxide (CO);
Neon (Ne);
Nitrogen oxide (NO);
Methane (CH$_4$);
Sulphuric Acid (H$_2$SO$_4$);
Iron (Fe);
Nickel (Ni);
Iron sulphide (FeS);
Iron oxide (FeO);
Silicone oxide (SiO);
Magnesium oxide (MgO);
Basalt(Bas);
Sulphur (S);
Silicates (Sil);
Carbonates (Carb;
Ammonia (NH$_4$);
Hydrocarbons (HyCarb);
Hydrogen peroxide (H$_2$O$_2$);
Tholins (Thol). 
The compositional characters eH, H$_2$ and He have absence indicated by a 1. In the remainder of compositional characteristics, a 1 is indicative of presence of the chemical species.
}

\tablerefs{(1) \citet{Lodders2003Abundances}; 
(2) \citet{Brooks2004JupiterRing};
(3) \citet{Brown2003CassiniJupiter};
(4) \citet{Burns1999JupiterRingForm};
(5) \citet{Kruger2009JupiterRingIntitu};
(6) \citet{OckertBell1999JupiterRing};
(7) \citet{Throop2004JupiterRings};
(8) \citet{Wong2006GBIRJupRingMoon};
(9) \citet{Thomas1998InnerSatJup};
(10) \citet{Cooper2006CassiniAmaltheaThebe}
(11) \citet{Takato2004Amalthea};
(12) \citet{Dalton2010IcySatComp};
(13) \citet{Dalton2010IcyMoonSpec};
(14) \citet{Greenberg2010IcyJovian};
(15) \citet{Haussmann2006SatTNO};
(16) \citet{Beauge2007IrregSatRsonance}; 
(17) \citet{Sheppard2003IrrSatNature}; 
(18) \citet{Grav2003IrregSatPhoto}; 
(19) \citet{Grav2015NEOWISEIrregulars}; 
(20) \citet{Rettig2001JovianIrrBVR}; 
(21) \citet{Sykes2000JupiterSatInfrared}; 
(22) \citet{Chamberlain2004Himalia}; 
(23) \citet{Emelyanov2005HimaliaMass};}

\end{deluxetable}
\end{longrotatetable}

\section{Saturnian Satellite matrix}
\label{SaturnMatrix}

\begin{longrotatetable}
\begin{deluxetable}{cccccccccccccccccccccccccccccccccccccccc}

\tabletypesize{\tiny}

\tablecaption{Taxon-Character Matrix of the Saturnian Satellite System }
\tablenum{4}

\tablehead{
\colhead{Identifier} & \colhead{Orb.} & \colhead{Rev.} & \colhead{a} & \colhead{i} & \colhead{e} & \colhead{D} & \colhead{Alb} & \colhead{eH} & \colhead{H$_2$} & \colhead{He} & \colhead{O$_2$} & \colhead{O$_3$} & \colhead{Na} & \colhead{K} & \colhead{CO$_2$} & \colhead{N$_2$} & \colhead{SO$_2$} & \colhead{Ar} & \colhead{H$_2$O} & \colhead{CO} & \colhead{Ne} & \colhead{NO} & \colhead{CH$_4$} & \colhead{H$_2$SO$_4$} & \colhead{Fe} & \colhead{Ni} & \colhead{FeS} & \colhead{FeO} & \colhead{SiO} & \colhead{MgO} & \colhead{Bas} & \colhead{S} & \colhead{Sil} & \colhead{Carb} & \colhead{NH$_3$} & \colhead{HyCarb} & \colhead{H$_2$O$_2$} & \colhead{Thol} & \colhead{Reference}\\
} 
\startdata
Sun & 0 & 0 & 0 & 0 & 0 & 0 & 0 & 0 & 0 & 0 & 0 & 0 & 0 & 0 & 0 & 0 & 0 & 0 & 0 & 0 & 0 & 0 & 0 & 0 & 0 & 0 & 0 & 0 & 0 & 0 & 0 & 0 & 0 & 0 & 0 & 0 & 0 & 0 & 1 \\
D ring & 1 & 0 & 0 & 0 & 0 & 9 & 4 & 1 & 1 & 1 & 0 & 0 & 0 & 0 & 0 & 0 & 0 & 0 & 1 & 0 & 0 & 0 & 0 & 0 & 0 & 0 & 0 & 0 & 0 & 0 & 0 & 0 & 0 & 0 & 0 & 0 & 0 & 0 & 2 \\
C Ring & 1 & 0 & 0 & 0 & 0 & 11 & 2 & 1 & 1 & 1 & 0 & 0 & 0 & 0 & 0 & 0 & 0 & 0 & 1 & 0 & 0 & 0 & 0 & 0 & 0 & 0 & 0 & 0 & 0 & 0 & 0 & 0 & 0 & 1 & 0 & 0 & 0 & 0 & 3,4 \\
B Ring & 1 & 0 & 0 & 0 & 0 & 9 & 3 & 1 & 1 & 1 & 0 & 0 & 0 & 0 & 0 & 0 & 0 & 0 & 1 & 0 & 0 & 0 & 0 & 0 & 0 & 0 & 0 & 0 & 0 & 0 & 0 & 0 & 0 & 0 & 0 & 0 & 0 & 0 & 3,4 \\
Cassini Division & 1 & 0 & 0 & 0 & 0 & 10 & 3 & 1 & 1 & 1 & 0 & 0 & 0 & 0 & 0 & 0 & 0 & 0 & 1 & 0 & 0 & 0 & 0 & 0 & 0 & 0 & 0 & 0 & 0 & 0 & 0 & 0 & 0 & 1 & 0 & 0 & 0 & 0 & 3,4 \\
A Ring & 1 & 0 & 0 & 0 & 0 & 9 & 6 & 1 & 1 & 1 & 0 & 0 & 0 & 0 & 0 & 0 & 0 & 0 & 1 & 0 & 0 & 0 & 0 & 0 & 0 & 0 & 0 & 0 & 0 & 0 & 0 & 0 & 0 & 0 & 0 & 0 & 0 & 0 & 3,4 \\
F Ring & 1 & 0 & 0 & 0 & 0 & ? & 4 & 1 & 1 & 1 & 0 & 0 & 0 & 0 & 0 & 0 & 0 & 0 & 1 & 0 & 0 & 0 & 0 & 0 & 0 & 0 & 0 & 0 & 0 & 0 & 0 & 0 & 0 & 0 & 0 & 0 & 0 & 0 & 3,4 \\
S/2009 S1 & 1 & 0 & 0 & 0 & 0 & ? & ? & 1 & ? & ? & ? & ? & ? & ? & ? & ? & ? & ? & ? & ? & ? & ? & ? & ? & ? & ? & ? & ? & ? & ? & ? & ? & ? & ? & ? & ? & ? & ? & 5 \\
Aegaeon & 1 & 0 & 0 & 0 & 0 & 9 & 1 & 1 & ? & ? & ? & ? & ? & ? & ? & ? & ? & ? & ? & ? & ? & ? & ? & ? & ? & ? & ? & ? & ? & ? & ? & ? & ? & ? & ? & ? & ? & ? & 6,7 \\
Pan & 1 & 0 & 0 & 0 & 0 & 9 & 3 & 1 & 1 & 1 & 0 & 0 & 0 & 0 & 0 & 0 & 0 & 0 & 1 & 0 & 0 & 0 & 0 & 0 & 0 & 0 & 0 & 0 & 0 & 0 & 0 & 0 & 0 & 0 & 0 & 0 & 0 & 0 & 7,8 \\
Daphnis & 1 & 0 & 0 & 0 & 0 & 10 & ? & 1 & ? & ? & ? & ? & ? & ? & ? & ? & ? & ? & ? & ? & ? & ? & ? & ? & ? & ? & ? & ? & ? & ? & ? & ? & ? & ? & ? & ? & ? & ? & 7 \\
Atlas & 1 & 0 & 0 & 0 & 0 & 9 & 6 & 1 & 1 & 1 & 0 & 0 & 0 & 0 & 0 & 0 & 0 & 0 & 1 & 0 & 0 & 0 & 0 & 0 & 0 & 0 & 0 & 0 & 0 & 0 & 0 & 0 & 0 & 0 & 0 & 0 & 0 & 0 & 7,8 \\
Prometheus & 1 & 0 & 0 & 0 & 0 & 9 & 3 & 1 & 1 & 1 & 0 & 0 & 0 & 0 & 0 & 0 & 0 & 0 & 1 & 0 & 0 & 0 & 0 & 0 & 0 & 0 & 0 & 0 & 0 & 0 & 0 & 0 & 0 & 0 & 0 & 0 & 0 & 0 & 7 \\
Pandora & 1 & 0 & 0 & 0 & 0 & 9 & 5 & 1 & 1 & 1 & 0 & 0 & 0 & 0 & 0 & 0 & 0 & 0 & 1 & 0 & 0 & 0 & 0 & 0 & 0 & 0 & 0 & 0 & 0 & 0 & 0 & 0 & 0 & 0 & 0 & 0 & 0 & 0 & 7,8 \\
Epimetheus & 1 & 0 & 0 & 0 & 0 & 8 & 6 & 1 & 1 & 1 & 0 & 0 & 0 & 0 & 0 & 0 & 0 & 0 & 1 & 0 & 0 & 0 & 0 & 0 & 0 & 0 & 0 & 0 & 0 & 0 & 0 & 0 & 0 & 0 & 0 & 0 & 0 & 0 & 7,8 \\
Janus & 1 & 0 & 0 & 0 & 0 & 8 & 7 & 1 & 1 & 1 & 0 & 0 & 0 & 0 & 0 & 0 & 0 & 0 & 1 & 0 & 0 & 0 & 0 & 0 & 0 & 0 & 0 & 0 & 0 & 0 & 0 & 0 & 0 & 0 & 0 & 0 & 0 & 0 & 7,8 \\
Janus/Epimetheus Ring & 1 & 0 & 0 & 0 & 0 & ? & ? & 1 & ? & ? & ? & ? & ? & ? & ? & ? & ? & ? & ? & ? & ? & ? & ? & ? & ? & ? & ? & ? & ? & ? & ? & ? & ? & ? & ? & ? & ? & ? & 7,9 \\
G Ring & 1 & 0 & 0 & 0 & 0 & ? & ? & 1 & 1 & 1 & 0 & 0 & 0 & 0 & 0 & 0 & 0 & 0 & 1 & 0 & 0 & 0 & 0 & 0 & 0 & 0 & 0 & 0 & 0 & 0 & 0 & 0 & 0 & 0 & 0 & 0 & 0 & 0 & 10 \\
E Ring & 1 & 0 & 0 & 0 & 0 & ? & ? & 1 & 1 & 1 & 0 & 0 & 0 & 0 & 1 & 1 & 0 & 0 & 1 & 1 & 0 & 0 & 0 & 0 & 0 & 0 & 0 & 0 & 0 & 0 & 0 & 0 & 1 & 0 & 1 & 1 & 0 & 0 & 11,12 \\
Phoebe Ring & 1 & 0 & 2 & 5 & ? & ? & 1 & 1 & ? & ? & ? & ? & ? & ? & ? & ? & ? & ? & ? & ? & ? & ? & ? & ? & ? & ? & ? & ? & ? & ? & ? & ? & ? & ? & ? & ? & ? & ? & 13,14 \\
Methone Ring Arc & 1 & 0 & 0 & 0 & 0 & ? & ? & 1 & ? & ? & ? & ? & ? & ? & ? & ? & ? & ? & ? & ? & ? & ? & ? & ? & ? & ? & ? & ? & ? & ? & ? & ? & ? & ? & ? & ? & ? & ? & 15 \\
Anthe Ring Arc & 1 & 0 & 0 & 0 & 0 & ? & ? & 1 & ? & ? & ? & ? & ? & ? & ? & ? & ? & ? & ? & ? & ? & ? & ? & ? & ? & ? & ? & ? & ? & ? & ? & ? & ? & ? & ? & ? & ? & ? & 15 \\
Pallene Ring Arc & 1 & 0 & 0 & 0 & 0 & ? & ? & 1 & ? & ? & ? & ? & ? & ? & ? & ? & ? & ? & ? & ? & ? & ? & ? & ? & ? & ? & ? & ? & ? & ? & ? & ? & ? & ? & ? & ? & ? & ? & 15 \\
Mimas & 1 & 0 & 0 & 0 & 0 & 5 & 4 & 1 & 1 & 1 & 0 & 0 & 0 & 0 & 1 & 0 & 0 & 0 & 1 & 0 & 0 & 0 & 0 & 0 & 0 & 0 & 0 & 0 & 0 & 0 & 0 & 0 & 1 & 0 & 0 & 0 & 0 & 0 & 16,17,18,19 \\
Enceladus & 1 & 0 & 0 & 0 & 0 & 2 & 7 & 1 & 1 & 1 & 0 & 0 & 0 & 0 & 1 & 1 & 0 & 0 & 1 & 1 & 0 & 0 & 1 & 0 & 0 & 0 & 0 & 0 & 0 & 0 & 0 & 0 & 1 & 0 & 1 & 1 & 1 & 0 & 16,17,18,19,20 \\
Tethys & 1 & 0 & 0 & 0 & 0 & 6 & 6 & 1 & 1 & 1 & 0 & 0 & 0 & 0 & 1 & 0 & 0 & 0 & 1 & 0 & 0 & 0 & 0 & 0 & 0 & 0 & 0 & 0 & 0 & 0 & 0 & 0 & 1 & 0 & 1 & 0 & 0 & 0 & 16,17,18,19 \\
Dione & 1 & 0 & 0 & 0 & 0 & 3 & 5 & 1 & 1 & 1 & 1 & 1 & 0 & 0 & 1 & 0 & 0 & 0 & 1 & 0 & 0 & 0 & 0 & 0 & 0 & 0 & 0 & 0 & 0 & 0 & 0 & 0 & 1 & 0 & 1 & 1 & 0 & 0 & 16,17,18,19 \\
Methone & 1 & 0 & 0 & 0 & 0 & ? & ? & 1 & ? & ? & ? & ? & ? & ? & ? & ? & ? & ? & ? & ? & ? & ? & ? & ? & ? & ? & ? & ? & ? & ? & ? & ? & ? & ? & ? & ? & ? & ? & 15 \\
Anthe & 1 & 0 & 0 & 0 & 0 & ? & ? & 1 & ? & ? & ? & ? & ? & ? & ? & ? & ? & ? & ? & ? & ? & ? & ? & ? & ? & ? & ? & ? & ? & ? & ? & ? & ? & ? & ? & ? & ? & ? & 15 \\
Pallene & 1 & 0 & 0 & 0 & 0 & ? & 3 & 1 & ? & ? & ? & ? & ? & ? & ? & ? & ? & ? & ? & ? & ? & ? & ? & ? & ? & ? & ? & ? & ? & ? & ? & ? & ? & ? & ? & ? & ? & ? & 15 \\
Telesto & 1 & 0 & 0 & 0 & 0 & 6 & 7 & 1 & 1 & 1 & 0 & 0 & 0 & 0 & 0 & 0 & 0 & 0 & 1 & 0 & 0 & 0 & 0 & 0 & 0 & 0 & 0 & 0 & 0 & 0 & 0 & 0 & 0 & 0 & 0 & 0 & 0 & 0 & 7,8 \\
Calypso & 1 & 0 & 0 & 0 & 0 & 6 & 7 & 1 & 1 & 1 & 0 & 0 & 0 & 0 & 0 & 0 & 0 & 0 & 1 & 0 & 0 & 0 & 0 & 0 & 0 & 0 & 0 & 0 & 0 & 0 & 0 & 0 & 0 & 0 & 0 & 0 & 0 & 0 & 7,8 \\
Polydeuces & 1 & 0 & 0 & 0 & 0 & ? & ? & 1 & ? & ? & ? & ? & ? & ? & ? & ? & ? & ? & ? & ? & ? & ? & ? & ? & ? & ? & ? & ? & ? & ? & ? & ? & ? & ? & ? & ? & ? & ? & 7 \\
Helene & 1 & 0 & 0 & 0 & 0 & 4 & 5 & 1 & 1 & 1 & 1 & 1 & 0 & 0 & 1 & 0 & 0 & 0 & 1 & 0 & 0 & 0 & 0 & 0 & 0 & 0 & 0 & 0 & 0 & 0 & 0 & 0 & 1 & 0 & 1 & 1 & 0 & 0 & 7 \\
Rhea & 1 & 0 & 0 & 0 & 0 & 4 & 5 & 1 & 1 & 1 & 0 & 1 & 0 & 0 & 1 & 0 & 0 & 0 & 1 & 0 & 0 & 0 & 1 & 0 & 0 & 0 & 0 & 0 & 0 & 0 & 0 & 0 & 1 & 0 & 0 & 1 & 0 & 1 & 16,17,18,19 \\
Titan & 1 & 0 & 0 & 0 & 0 & 0 & 1 & 1 & 1 & 1 & 0 & 0 & 0 & 0 & 1 & 0 & 0 & 0 & 1 & 1 & 0 & 0 & 1 & 0 & 0 & 0 & 0 & 0 & 0 & 0 & 0 & 0 & 1 & 0 & 1 & 1 & 0 & 1 & 21,22,23 \\
Hyperion & 1 & 0 & 0 & 0 & 0 & 8 & 2 & 1 & 1 & 1 & 0 & 0 & 0 & 0 & 1 & 0 & 0 & 0 & 1 & 0 & 0 & 0 & 0 & 0 & 0 & 0 & 0 & 0 & 0 & 0 & 0 & 0 & 1 & 0 & 0 & 1 & 0 & 1 & 16,17,18,19 \\
Iapetus & 1 & 0 & 0 & 0 & 0 & 5 & 3 & 1 & 1 & 1 & 0 & 0 & 0 & 0 & 1 & 0 & 0 & 0 & 1 & 0 & 0 & 0 & 0 & 0 & 0 & 0 & 0 & 0 & 0 & 0 & 0 & 0 & 1 & 0 & 1 & 1 & 0 & 1 & 16,17,18,19 \\
Kiviuq & 1 & 0 & 3 & 1 & 5 & ? & 0 & 1 & ? & ? & ? & ? & ? & ? & ? & ? & ? & ? & ? & ? & ? & ? & ? & ? & ? & ? & ? & ? & ? & ? & ? & ? & ? & ? & ? & ? & ? & ? & 24,25,27 \\
Ijiraq & 1 & 0 & 3 & 1 & 4 & ? & 0 & 1 & ? & ? & ? & ? & ? & ? & ? & ? & ? & ? & ? & ? & ? & ? & ? & ? & ? & ? & ? & ? & ? & ? & ? & ? & ? & ? & ? & ? & ? & ? & 24,25,27 \\
Siarnaq & 1 & 0 & 4 & 1 & 4 & ? & 0 & 1 & ? & ? & ? & ? & ? & ? & ? & ? & ? & ? & ? & ? & ? & ? & ? & ? & ? & ? & ? & ? & ? & ? & ? & ? & ? & ? & ? & ? & ? & ? & 24,25,27 \\
Tarqeq & 1 & 0 & 5 & 1 & 2 & ? & ? & 1 & ? & ? & ? & ? & ? & ? & ? & ? & ? & ? & ? & ? & ? & ? & ? & ? & ? & ? & ? & ? & ? & ? & ? & ? & ? & ? & ? & ? & ? & ? & 24 \\
Paaliaq & 1 & 0 & 4 & 1 & 5 & ? & ? & 1 & ? & ? & ? & ? & ? & ? & ? & ? & ? & ? & ? & ? & ? & ? & ? & ? & ? & ? & ? & ? & ? & ? & ? & ? & ? & ? & ? & ? & ? & ? & 24,25,27 \\
Albiorix & 1 & 0 & 4 & 1 & 7 & ? & 0 & 1 & ? & ? & ? & ? & ? & ? & ? & ? & ? & ? & ? & ? & ? & ? & ? & ? & ? & ? & ? & ? & ? & ? & ? & ? & ? & ? & ? & ? & ? & ? & 24,25,27 \\
Bebhionn & 1 & 0 & 4 & 1 & 7 & ? & ? & 1 & ? & ? & ? & ? & ? & ? & ? & ? & ? & ? & ? & ? & ? & ? & ? & ? & ? & ? & ? & ? & ? & ? & ? & ? & ? & ? & ? & ? & ? & ? & 24,25,27 \\
Erriapus & 1 & 0 & 4 & 1 & 7 & ? & 0 & 1 & ? & ? & ? & ? & ? & ? & ? & ? & ? & ? & ? & ? & ? & ? & ? & ? & ? & ? & ? & ? & ? & ? & ? & ? & ? & ? & ? & ? & ? & ? & 24,25,27 \\
Tarvos & 1 & 0 & 5 & 1 & 8 & ? & 0 & 1 & ? & ? & ? & ? & ? & ? & ? & ? & ? & ? & ? & ? & ? & ? & ? & ? & ? & ? & ? & ? & ? & ? & ? & ? & ? & ? & ? & ? & ? & ? & 24,25,27 \\
Phoebe & 1 & 1 & 3 & 5 & 2 & 2 & 0 & 1 & 1 & 1 & 0 & 0 & 0 & 0 & 1 & 0 & 0 & 0 & 1 & 0 & 0 & 0 & 0 & 0 & 1 & 0 & 0 & 0 & 0 & 0 & 0 & 0 & 1 & 0 & 1 & 1 & 0 & 1 & 16,17,18,19,24,25,26,27 \\
Skathi & 1 & 1 & 4 & 5 & 4 & ? & 0 & 1 & ? & ? & ? & ? & ? & ? & ? & ? & ? & ? & ? & ? & ? & ? & ? & ? & ? & ? & ? & ? & ? & ? & ? & ? & ? & ? & ? & ? & ? & ? & 24,25,27 \\
Skoll & 1 & 1 & 4 & 5 & 7 & ? & ? & 1 & ? & ? & ? & ? & ? & ? & ? & ? & ? & ? & ? & ? & ? & ? & ? & ? & ? & ? & ? & ? & ? & ? & ? & ? & ? & ? & ? & ? & ? & ? & 24 \\
Greip & 1 & 1 & 5 & 5 & 5 & ? & ? & 1 & ? & ? & ? & ? & ? & ? & ? & ? & ? & ? & ? & ? & ? & ? & ? & ? & ? & ? & ? & ? & ? & ? & ? & ? & ? & ? & ? & ? & ? & ? & 24 \\
Hyrrokkin & 1 & 1 & 5 & 5 & 5 & ? & ? & 1 & ? & ? & ? & ? & ? & ? & ? & ? & ? & ? & ? & ? & ? & ? & ? & ? & ? & ? & ? & ? & ? & ? & ? & ? & ? & ? & ? & ? & ? & ? & 24 \\
Mundilfari & 1 & 1 & 5 & 5 & 3 & ? & ? & 1 & ? & ? & ? & ? & ? & ? & ? & ? & ? & ? & ? & ? & ? & ? & ? & ? & ? & ? & ? & ? & ? & ? & ? & ? & ? & ? & ? & ? & ? & ? & 24,25,27 \\
Jarnsaxa & 1 & 1 & 5 & 5 & 3 & ? & ? & 1 & ? & ? & ? & ? & ? & ? & ? & ? & ? & ? & ? & ? & ? & ? & ? & ? & ? & ? & ? & ? & ? & ? & ? & ? & ? & ? & ? & ? & ? & ? & 24 \\
Narvi & 1 & 1 & 5 & 4 & 6 & ? & ? & 1 & ? & ? & ? & ? & ? & ? & ? & ? & ? & ? & ? & ? & ? & ? & ? & ? & ? & ? & ? & ? & ? & ? & ? & ? & ? & ? & ? & ? & ? & ? & 24 \\
Bergelmir & 1 & 1 & 5 & 5 & 2 & ? & ? & 1 & ? & ? & ? & ? & ? & ? & ? & ? & ? & ? & ? & ? & ? & ? & ? & ? & ? & ? & ? & ? & ? & ? & ? & ? & ? & ? & ? & ? & ? & ? & 24 \\
Suttungr & 1 & 1 & 5 & 5 & 1 & ? & 0 & 1 & ? & ? & ? & ? & ? & ? & ? & ? & ? & ? & ? & ? & ? & ? & ? & ? & ? & ? & ? & ? & ? & ? & ? & ? & ? & ? & ? & ? & ? & ? & 24,25,27 \\
Hati & 1 & 1 & 5 & 5 & 5 & ? & ? & 1 & ? & ? & ? & ? & ? & ? & ? & ? & ? & ? & ? & ? & ? & ? & ? & ? & ? & ? & ? & ? & ? & ? & ? & ? & ? & ? & ? & ? & ? & ? & 24 \\
Bestla & 1 & 1 & 5 & 4 & 8 & ? & ? & 1 & ? & ? & ? & ? & ? & ? & ? & ? & ? & ? & ? & ? & ? & ? & ? & ? & ? & ? & ? & ? & ? & ? & ? & ? & ? & ? & ? & ? & ? & ? & 24 \\
Farbauti & 1 & 1 & 5 & 5 & 3 & ? & ? & 1 & ? & ? & ? & ? & ? & ? & ? & ? & ? & ? & ? & ? & ? & ? & ? & ? & ? & ? & ? & ? & ? & ? & ? & ? & ? & ? & ? & ? & ? & ? & 24 \\
Thrymr & 1 & 1 & 5 & 5 & 7 & ? & 0 & 1 & ? & ? & ? & ? & ? & ? & ? & ? & ? & ? & ? & ? & ? & ? & ? & ? & ? & ? & ? & ? & ? & ? & ? & ? & ? & ? & ? & ? & ? & ? & 24,25,27 \\
Aegir & 1 & 1 & 5 & 5 & 3 & ? & ? & 1 & ? & ? & ? & ? & ? & ? & ? & ? & ? & ? & ? & ? & ? & ? & ? & ? & ? & ? & ? & ? & ? & ? & ? & ? & ? & ? & ? & ? & ? & ? & 24 \\
Kari & 1 & 1 & 6 & 5 & 7 & ? & ? & 1 & ? & ? & ? & ? & ? & ? & ? & ? & ? & ? & ? & ? & ? & ? & ? & ? & ? & ? & ? & ? & ? & ? & ? & ? & ? & ? & ? & ? & ? & ? & 24 \\
Fenrir & 1 & 1 & 6 & 5 & 2 & ? & ? & 1 & ? & ? & ? & ? & ? & ? & ? & ? & ? & ? & ? & ? & ? & ? & ? & ? & ? & ? & ? & ? & ? & ? & ? & ? & ? & ? & ? & ? & ? & ? & 24 \\
Surtur & 1 & 1 & 6 & 5 & 6 & ? & ? & 1 & ? & ? & ? & ? & ? & ? & ? & ? & ? & ? & ? & ? & ? & ? & ? & ? & ? & ? & ? & ? & ? & ? & ? & ? & ? & ? & ? & ? & ? & ? & 24 \\
Ymir & 1 & 1 & 6 & 5 & 5 & ? & 0 & 1 & ? & ? & ? & ? & ? & ? & ? & ? & ? & ? & ? & ? & ? & ? & ? & ? & ? & ? & ? & ? & ? & ? & ? & ? & ? & ? & ? & ? & ? & ? & 24,25,27 \\
Loge & 1 & 1 & 6 & 5 & 2 & ? & ? & 1 & ? & ? & ? & ? & ? & ? & ? & ? & ? & ? & ? & ? & ? & ? & ? & ? & ? & ? & ? & ? & ? & ? & ? & ? & ? & ? & ? & ? & ? & ? & 24 \\
Fornjot & 1 & 1 & 6 & 5 & 3 & ? & ? & 1 & ? & ? & ? & ? & ? & ? & ? & ? & ? & ? & ? & ? & ? & ? & ? & ? & ? & ? & ? & ? & ? & ? & ? & ? & ? & ? & ? & ? & ? & ? & 24 \\
S/2004 S07 & 1 & 1 & 5 & 5 & 8 & ? & ? & 1 & ? & ? & ? & ? & ? & ? & ? & ? & ? & ? & ? & ? & ? & ? & ? & ? & ? & ? & ? & ? & ? & ? & ? & ? & ? & ? & ? & ? & ? & ? & 24 \\
S/2004 S12 & 1 & 1 & 5 & 5 & 6 & ? & ? & 1 & ? & ? & ? & ? & ? & ? & ? & ? & ? & ? & ? & ? & ? & ? & ? & ? & ? & ? & ? & ? & ? & ? & ? & ? & ? & ? & ? & ? & ? & ? & 24 \\
S/2004 S13 & 1 & 1 & 5 & 5 & 4 & ? & ? & 1 & ? & ? & ? & ? & ? & ? & ? & ? & ? & ? & ? & ? & ? & ? & ? & ? & ? & ? & ? & ? & ? & ? & ? & ? & ? & ? & ? & ? & ? & ? & 24 \\
S/2004 S17 & 1 & 1 & 5 & 5 & 4 & ? & ? & 1 & ? & ? & ? & ? & ? & ? & ? & ? & ? & ? & ? & ? & ? & ? & ? & ? & ? & ? & ? & ? & ? & ? & ? & ? & ? & ? & ? & ? & ? & ? & 24 \\
S/2006 S1 & 1 & 1 & 5 & 5 & 2 & ? & ? & 1 & ? & ? & ? & ? & ? & ? & ? & ? & ? & ? & ? & ? & ? & ? & ? & ? & ? & ? & ? & ? & ? & ? & ? & ? & ? & ? & ? & ? & ? & ? & 24 \\
S/2006 S3 & 1 & 1 & 5 & 5 & 7 & ? & ? & 1 & ? & ? & ? & ? & ? & ? & ? & ? & ? & ? & ? & ? & ? & ? & ? & ? & ? & ? & ? & ? & ? & ? & ? & ? & ? & ? & ? & ? & ? & ? & 24 \\
S/2007 S2 & 1 & 1 & 4 & 5 & 3 & ? & ? & 1 & ? & ? & ? & ? & ? & ? & ? & ? & ? & ? & ? & ? & ? & ? & ? & ? & ? & ? & ? & ? & ? & ? & ? & ? & ? & ? & ? & ? & ? & ? & 24 \\
S/2007 S3 & 1 & 1 & 5 & 5 & 2 & ? & ? & 1 & ? & ? & ? & ? & ? & ? & ? & ? & ? & ? & ? & ? & ? & ? & ? & ? & ? & ? & ? & ? & ? & ? & ? & ? & ? & ? & ? & ? & ? & ? & 24 \\
\enddata

\tablecomments{Character Abbreviations: 
In orbit around the gas giant (Orb.);
Revolution (Rev.); 
Semi-major axis(a);
Orbital inclination to the plane(i);
Orbital eccentricity(e);
Density (D);
Visual geometric albedo (Alb);
Elemental Hydrogen (eH);
Hydrogen (H$_2$);
Helium (He); 
Oxygen (O$_2$);
Ozone (O$_3$);
Sodium (Na);
Potassium (K);
Carbon dioxide (CO$_2$);
Nitrogen (N$_2$);
Sulphur dioxide (SO$_2$);
Argon (Ar);
Water (H$_2$O);
Carbon monoxide (CO);
Neon (Ne);
Nitrogen oxide (NO);
Methane (CH$_4$);
Sulphuric Acid (H$_2$SO$_4$);
Iron (Fe);
Nickel (Ni);
Iron sulphide (FeS);
Iron oxide (FeO);
Silicone oxide (SiO);
Magnesium oxide (MgO);
Basalt(Bas);
Sulphur (S);
Silicates (Sil);
Carbonates (Carb;
Ammonia (NH$_4$);
Hydrocarbons (HyCarb);
Hydrogen peroxide (H$_2$O$_2$);
Tholins (Thol). 
The compositional characters eH, H$_2$ and He have absence indicated by a 1. In the remainder of compositional characteristics, a 1 is indicative of presence of the chemical species.
}

\tablerefs{(1) \citet{Lodders2003Abundances};
(2) \citet{Hedman2007Dring};
(3) \citet{Nicholson2008VIMSRings};
(4) \citet{Filacchione2014VIMSrings};
(5) \citet{Spitale2012s2009s1};
(6) \citet{Hedman2010Aegaeon};
(7) \citet{Thomas2013InnerSatSatu};
(8) \citet{Buratti2010SatInnerSat};
(9) \citet{Winter2016JanusEpiRing};
(10) \citet{Hedman2007Gring};
(11) \citet{Hedman2012EringStruc};
(12) \citet{Hillier2007EringComp};
(13) \citet{Tamayo2014PhoRingOptic};
(14) \citet{Verbiscer2009SatLarRing};
(15) \citet{Hedman2009SatRingArcs};
(16) \citet{Filacchione2010VIMS2};
(17) \citet{Filacchione2012VIMS3};
(18) \citet{Haussmann2006SatTNO};
(19) \citet{Matson2009SaturnSat};
(20) \citet{Spencer2013EnceladusRev};
(21) \citet{Hirtzig2013VIMSTitan};
(22) \citet{Hemingway2013TitanIceshell};
(23) \citet{Niemann2005TitanAtmos};
(24) \citet{Beauge2007IrregSatRsonance};
(25) \citet{Gladman200112Sat};
(26) \citet{Jewitt2007IrregularSats};
(27) \citet{Grav2007IrregSatCol};}

\end{deluxetable}
\end{longrotatetable}

\end{document}